\documentclass[10pt]{ijnam}
\usepackage{float}
\usepackage{caption}
\usepackage{subcaption}
\usepackage{afterpage}
\usepackage{graphicx}
\usepackage[margin=38.75mm]{geometry}
\begin{document}

\title[Numerical Analysis of Flow using ALE CFR scheme]
{Numerical Analysis of flow past an oscillating circular cylinder using Arbitrary Lagrangian Eulerian Consistent Flux Reconstruction Scheme}

\author[K. Bali]{Kartik Bali}
\address{
  Department of Mechanical Engineering,
  Birla Institute of Technology and Science Pilani,
  Hyderabad, 380004, India
}
\email{f20150851h@alumni.bits-pilani.ac.in}

\author[K Supradeepan]{Katiresan Supradeepan}
\address{
  Department of Mechanical Engineering,
  Birla Institute of Technology and Science Pilani,
  Hyderabad, 380004, India
}
\email{supradeepan@hyderabad.bits-pilani.ac.in}






\abstract{This paper proposes an Arbitrary Lagrangian-Eulerian incompressible finite volume scheme based on Consistent Flux Reconstruction method (ALE-CFR) on deforming two-dimensional unstructured triangular grids. The scheme is further applied to solve for a problem of two-dimensional incompressible viscous flow past a rotationally oscillating circular cylinder when subjected to forced transverse oscillations at Re = 100. The paper first focuses on the mathematical development and validation of the ALE-CFR scheme and then its application on the  problem. The rotational oscillation frequencies were taken corresponding to the four primary vortex shedding mode frequencies discovered from the studies conducted by Tokumaru and Dimotakis. The transverse oscillation frequency was varied for a frequency ratio $f_{tv}$ with respect to natural strouhal frequency for flow past a stationary circular cylinder from $0.9 \leq f_{tv} \leq 3.0$. The amplitude ratio for transverse oscillations was taken as 0.2 with respect to the cylinder diameter. The distinct effects on the vortex shedding patterns and the contributions of such rotational and transverse oscillations on lift and drag characteristics of the circular cylinder are studied using the newly developed scheme.}


\maketitle

\section{Introduction}
Lagrangian algorithms are popular for modeling phenomenon where the computational mesh moves with the continuum. They can track material interfaces accurately and are widely used in computational solid mechanics. However, the reason behind their less frequent use in fluid mechanics is that they cause the computational mesh to distort while the mesh follows the fluid particle motion. Eulerian algorithms have, on the other hand, become increasingly popular in the past for fluid mechanics as the computational mesh remains fixed and independent of fluid motion. While dealing with fluid flow where geometry changes dynamically, however, fixing the computational mesh may not be sufficient. The mesh must be dynamic in order to accommodate varied configurations of the flow domain. Neither purely Lagrangian nor Eulerian formulations are sufficient for dealing with such problems and thus Arbitrary Lagrangian-Eulerian algorithms have gained wide popularity. The latter methods use a pseudo-Eulerian computational mesh for computing fluid flow which moves in order to accommodate the solid displacement (or track solid-fluid interfaces in a Lagrangian manner), thus giving rise to a mesh dynamic in space and time. Problems ranging in fields like aerodynamics, marine engineering, and biomedical engineering which involve dynamically changing geometries and fluid-solid interaction can be modeled robustly using ALE algorithms. For such ALE algorithms, an additional law of geometric conservation needs to be introduced in order to ensure mesh velocities are computed in accordance with the changing control volumes of the computational mesh. In this paper an ALE algorithm is incorporated in a Consistent Flux Reconstruction finite volume solution scheme. The scheme is second-order accurate in space and uses a collocated grid arrangement applicable to unstructured 2D dynamic triangular grid.

The problem of flow past a circular cylinder has gained wide popularity over the past few decades, owing to the various complex flow features, vortex shedding mechanisms and lift-drag characteristics that provide valuable insight into the flow behavior. The understanding of flow by the study of these phenomena greatly aids in engineering design. Especially in marine and aerodynamic engineering applications due to use of equipment with similar geometries (for ex. risers in marine engineering and bluff bodies in aerodynamics)  where better power efficiencies are desired and structural failure needs to be avoided. Out of all such fluid phenomena, vortex shedding remains one of the most important as it explains the unsteady features of flow and the cause behind varying lift and drag forces. In certain cases, it can even trigger structural failure. Thus to mitigate such adverse effects, many methods for vortex suppression have been explored by researchers in the past. These include either passive control methods that involve a change in the geometry, like the use of splitter plates \cite{Kwon}, control cylinders \cite{Mittal}, or active control methods like imposing transverse or rotary oscillations \cite{Choi}. 
    Rotary oscillation is a popular active flow control method that is used to alter the wake behind a circular cylinder in order to minimize vortex shedding thus resulting in lesser aerodynamic/hydrodynamic force fluctuations, as studied by Williamson CHK \cite{Williamson}.  
    Many researchers in the past have investigated the effects of imposing forced rotary oscillations on the circular cylinder to study wake structures and the associated aerodynamic coefficients. 
    Taneda \cite{Taneda} showed that even for low Re in range $30\leq Re \leq 300$ and rotational oscillation frequencies $St_{f}$ in range $0\leq St_{f} \leq 55$, the vortex shedding disappears completely for higher frequencies.
    Tokumaru and Dimotakis \cite{Tokumaru} tested the effectiveness of forced rotary oscillations experimentally at high $Re = 15000$ with an amplitude of rotation $0\leq A_{r} \leq 16$ and forced rotary oscillation frequency $ 0\leq St_{f} \leq 3.3$, to examine the effect on the unsteady wake.
    Shiels and Leonard \cite{Shiels} also verified similar findings on drag reduction by carrying out numerical simulations using a 2D high-resolution viscous method for Re between 150 and 15000. According to their study multi-polar vortices generated in the wake were responsible for drag reduction at high Re, whereas for low Re this effect was dampened by the high viscous effects.
    
    It is also known that in these scenarios, fluid-structure interactions must also be accounted for an accurate depiction of the flow physics. Due to vortex shedding the cylinder is subjected to periodic lift and drag forces that lead to its transverse oscillatory motion.
    
    A common strategy to simulate this vortex-induced motion is to force the body to oscillate with a predefined motion, as described by Singh \cite{Singh}, which has been adopted in the present study.
    Many studies have been done on cylinders that undergo vortex-induced transverse oscillations. Ongoren and Rockwell \cite{Ongoren} carried out a hydrogen bubble visualization experiment to find the position of the vortex switch. It was concluded from this study that for forced transverse oscillation frequencies greater and lesser than the natural strouhal number ($St_{n}$), the vortices formed on one side shed on the opposite and same sides respectively at maximum amplitudes. Another numerical study conducted by Anagnostopoulos            \cite{Anagnostopoulos} for flow past an oscillating circular cylinder at $Re = 106$ also confirmed switching in vortex shedding as seen in experiments conducted by the same author. Numerical studies conducted by Pham \cite{Pham} on the same problem used the immersed boundary method to illustrate the vortex shedding patterns and aerodynamic characteristics at different transverse oscillation amplitudes and frequencies.
    
    
    The numerical studies on flow past a cylinder subjected to rotary or transverse oscillations, each imposed separately, have been quite popular in the past. However, there haven't been many comprehensive studies on the combined effect of rotary and transverse oscillations imposed on a cylinder exposed to open channel fluid flow to the best knowledge of the authors.
    In this study the effects on the vortex shedding and aerodynamic characteristics that arise as a result of both transverse and rotational oscillations imposed on the circular cylinder at Re = 100, have been investigated using the newly developed ALE-CFR finite volume scheme.
    
    \section{Governing Equations}
    The primary governing equations taken for this work are that of mass and momentum conservation for incompressible fluid flow in the ALE frame. In order to compute flows on grids with an ALE kinematic description, an additional equation aids in computing the flow over the dynamic mesh known as the Space Conservation Law (SCL) \cite{Demirdzic1,Demirdzic2}. The conservation laws in their integral non-dimensional forms for incompressible flows without considering body forces are described below.
    \\Space Conservation Law:
    \begin{equation}
    \frac{d}{dt} \left(\int\limits_\Omega d\Omega\right) -\int\limits_S\overrightarrow{{v}_{b}}.\overrightarrow{n}dS = 0 
    \label{eq:SCL}
    \end{equation}
    \\Mass Conservation Law:    
    \begin{equation}
      \frac{d}{dt} \left(\int\limits_\Omega\rho d\Omega\right) + \int\limits_S\rho\left(\overrightarrow{v}-\overrightarrow{{v}_{b}}\right).\overrightarrow{n}dS = 0   
      \label{eq:Mass Conservation eq}
    \end{equation}
     \\Momentum Conservation Law:
    \begin{equation}
       \frac{d}{dt}\left(\int\limits_\Omega {u}_{i}d\Omega\right) + \int\limits_S {u}_{i}\left(\overrightarrow{v}-\overrightarrow{{v}_{b}}\right).\overrightarrow{n}dS = -\int\limits_S p\overrightarrow{{i}_{i}}.\overrightarrow{n}dS + \frac{1}{Re}\int\limits_S\nabla {u}_{i}.\overrightarrow{n}dS
       \label{eq:Momentum Conservation Eq}
    \end{equation}

    
    where $i$ takes 1, 2 for $x$, $y$ components of velocities respectively.
    
    The expression $\overrightarrow{{v}_{b}}$ denotes the mesh velocities of the faces for the moving control volume. In an ALE framework, the net mass flux convecting inside the moving control volume is accounted for by taking fluid convection velocity relative to the moving control volume faces in the convective terms of the Navier-Stokes Equations. Eq.(\ref{eq:SCL}) ensures that the mesh velocities are in accordance with the change in the control volumes when the mesh is deforming. 
    For an incompressible flow, as the density remains constant,  Eq.(\ref{eq:SCL}) is reduced to,
    \begin{equation}
    \int\limits_S \overrightarrow{v}.\overrightarrow{n}dS = 0
    \label{eq:Simplified Mass Conservation Eq}
    \end{equation}
     
    \newpage
    \section{Boundary Conditions}
\begin{figure}
        \centering
        \includegraphics[scale=0.025]{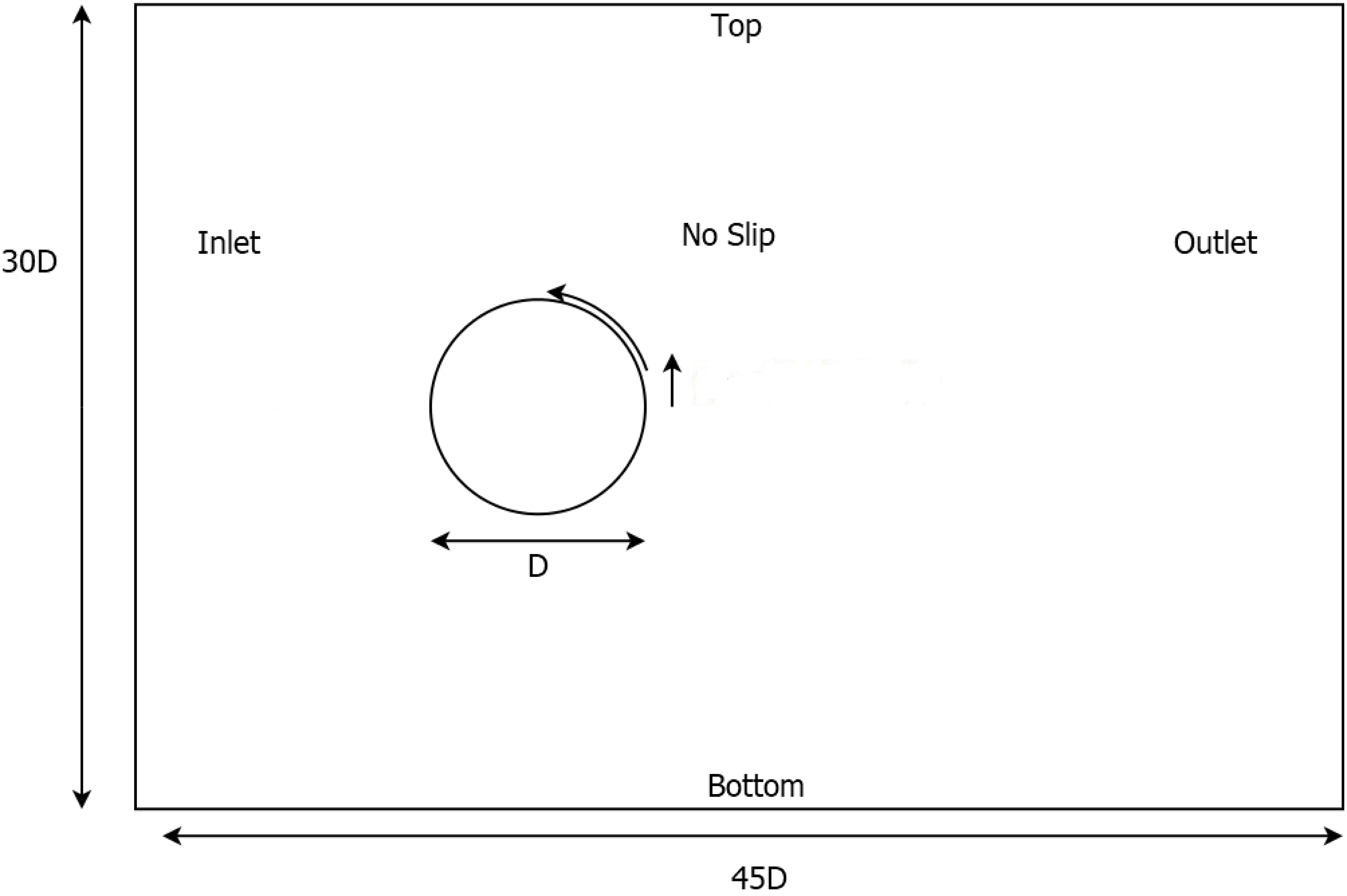}
        \caption{Boundary conditions and flow domain}
        \label{fig:Flow Domain}
    \end{figure}

    From the schematic shown in Fig.(\ref{fig:Flow Domain}), dirichlet boundary conditions have been given at the inlet with $u = {U}_{\infty}$ and $v = 0$. Zero shear boundary condition has been imposed on the top and bottom with $\frac{\partial u}{\partial y}=0$ and $v=0$. At the outlet, convective boundary condition have been imposed with $\frac{\partial \phi}{\partial t}+U_{c}\frac{\partial \phi}{\partial x}=0$ where $\phi=u,v$ and $U_{c} = U_{\infty}$. On the cylinder surface no slip boundary condition for forced rotary and transverse oscillations has been imposed with $\dot{\theta}=A_{r}\cos(2\pi St_{r}t)$ as the rotational oscillation speed and $Y=A_{t}\sin{(2\pi St_{tv}t)}$ as the transverse position at any instant of time where $ St_{r}$ and $ St_{tv}$ represent the rotational oscillation and transverse oscillation frequencies respectively. 
    
    \section{Numerical Scheme}
\begin{figure}[h]
        \centering
        \includegraphics[height=4.5 cm,width=9.5 cm]{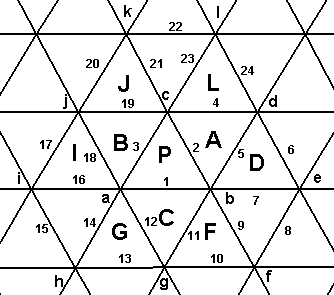}
        \caption{Main Grid Stencil}
        \label{fig:Main Grid Stencil}
        \end{figure}

    The Consistent Flux Reconstruction (CFR) scheme \cite{Bandhyopadhyay,Harichandan} was adopted in this work to accommodate flows on deformable grids and solve for a larger class of problems involving moving bodies and their effect, in turn, on the fluid flow and vice-versa.	Based on this development, the scheme is hereon referred to as \textbf{ALE-CFR}, short for Arbitrary Lagrangian Eulerian Consistent Flux Reconstruction scheme. The scheme employs a collocated grid arrangement and is second-order accurate in space and first-order accurate in time. For achieving second-order accuracy the momentum equations are solved for both the original and the reconstructed cells. This reconstructed cell consists of the main cell and its neighbor, both sharing the face where the velocity needs to be computed. Appropriate neighboring cells to the main cell are chosen for each face such that the shared face acts as the center of the reconstructed cell.          
The cell-centered flow variables (pressure and velocity) at the $(n-1)^{th}$ time level are linearly interpolated to calculate face-centered flow variables.\\The face-centered variables are interpolated from the cell-centered variables using weighted averaging \cite{Harichandan}, as shown for pressure in Eq.(\ref{eq:Interpolation}). 
    \begin{equation}
        p_{1}=\frac{p_{p}a_{c}+p_{c}a_{p}}{a_{c}+a_{p}}
	\label{eq:Interpolation}
    \end{equation}
After grid deformation, the momentum equation is then applied on the reconstructed cell explicitly in order to construct expressions of face-centered velocities at the $n^{th}$ time level in terms of the face-centered pressures. The convective, diffusive and ALE flux expressions in the momentum equation for the reconstructed cells use the interpolated face-centered variables from old cell-centered variables. These expressions of velocities at the $n^{th}$ time level are substituted in the continuity relation for cell P, given by Eq.(\ref{eq:Discret. Cont. Eq}) to construct the Pressure Poisson Equation (PPE).  
    The PPE is solved for all cells to compute the cell-centered pressure field for the new mesh.
    \\The discretized integral form of the continuity equation for a cell $P$ with vertices $abc$ in the grid shown in Fig.(\ref{fig:Main Grid Stencil}) at the ${n+1}^{th}$ time level is,
    
    \begin{equation}
    \begin{aligned}
        \int \overrightarrow{v}.\overrightarrow{n}dS =&\left(u_{1}^{n+1}\triangle x_{ab}^{n+1}-v_{1}^{n+1}\triangle y_{ab}^{n+1}\right)+ \left(u_{2}^{n+1}\triangle x_{bc}^{n+1}-v_{2}^{n+1}\triangle y_{bc}^{n+1}\right)+\\  &\left(u_{3}^{n+1}\triangle x_{ca}^{n+1}-v_{3}^{n+1}\triangle y_{ca}^{n+1}\right)=0
    \label{eq:Discret. Cont. Eq}
    \end{aligned}
    \end{equation}
    
\begin{figure}[ht]
    \centering
    \includegraphics[scale=0.15]{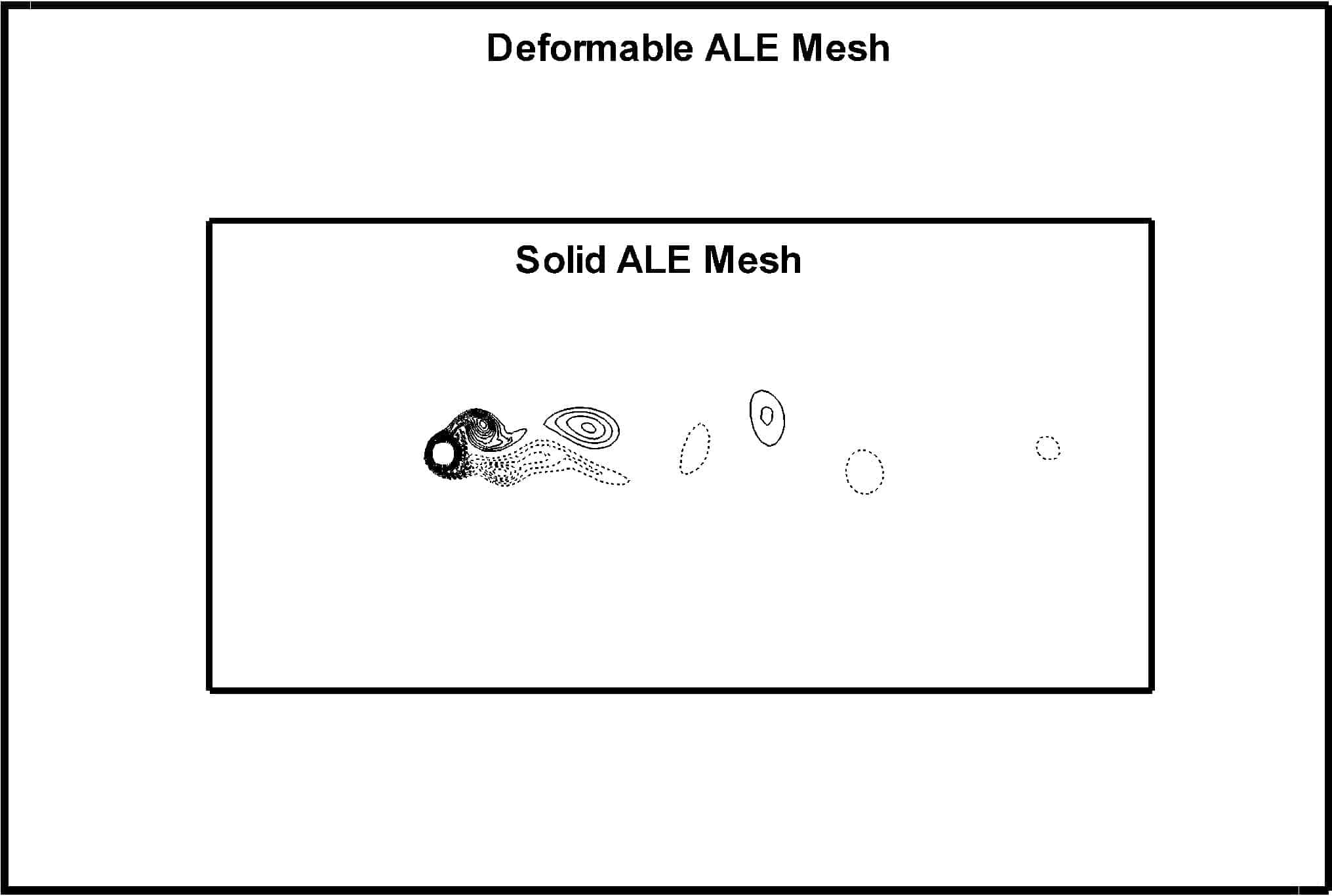}
    \caption{The Solid ALE Mesh region moves rigidly while the mesh elements outside deform}
    \label{fig:Mesh}
    \end{figure}

    where ${\triangle x}_{ab}$ denotes  length vector given as ${x}_{b}-{x}_{a}$.
    After the velocities are substituted, Eq.({\ref{eq:Discret. Cont. Eq}}) is thus solved for the new pressure field.
    The grid deforming scheme used in this work is based on linear spring analogy \cite{Zheng}. In order to sufficiently preserve the shear layers near the cylinder surface and capture flow structures accurately in the wake, the outlined region as given in Fig ({\ref{fig:Mesh}}) was made to move rigidly with the cylinder. The remaining exterior region contains deformable mesh elements that accommodate the new position of the marked region containing elements surrounding the cylinder.
    
    The ALE flux terms are computed after mesh deformation and the discretized PPE resulting from Eq.({\ref{eq:Discret. Cont. Eq}}) is solved for the new cell-centered pressures. Discretization of the unsteady term is done considering the change in both the cell-centered field variable and the dimensions of the CV:
    \begin{equation}
        \frac{d}{dt}\left(\int \phi d\Omega\right) \approx \frac{\left(\phi \Omega\right)^{n+1}-\left(\phi \Omega \right)^{n}}{\triangle t}
    \end{equation}

    The face-centered velocities ${u}_{1}$ and ${v}_{1}$ upon first order Eulerian discretization of Eq.({\ref{eq:Momentum Conservation Eq}}) are given by,
    \begin{equation}
        \begin{aligned}
        u_{1}^{n+1} = u_{1}^{n}\frac{A_{pc}^{n}}{A_{pc}^{n+1}} + \frac{\triangle t}{A_{pc}^{n+1}}(-{XCFLUX_{1}}^{n}-&{XPFLUX_{1}}^{n+1}+\frac{1}{Re}{XDFLUX_{1}}^{n}+ \\ &{XALEFLUX_{1}}^{n}) 
        \label{eq:u1}
        \end{aligned}
    \end{equation}
   
\begin{equation}
    \begin{aligned}
        v_{1}^{n+1} = v_{1}^{n}\frac{A_{pc}^{n}}{A_{pc}^{n+1}} + \frac{\triangle t}{A_{pc}^{n+1}}(-YCFLUX_{1}^{n}-&YPFLUX_{1}^{n+1}+\frac{1}{Re}YDFLUX_{1}^{n}+ \\ &YALEFLUX_{1}^{n})
        \label{eq:v1}
    \end{aligned}
\end{equation}
    
    ${A_{pc}^{n+1}}$ is the area of the reconstructed cell consisting of original cell P and neighboring cell C at the ${n+1}^{th}$ level (after grid deformation). 
    All the above flux terms are computed based on the mesh stencil given in Fig ({\ref{fig:Main Grid Stencil}}).
    
    The flux terms used in expressions for $u_{1}$ and $v_{1}$  are expressed as\\
    Convective Flux:
    \begin{equation}
    \begin{aligned}
    XCFLUX_{1}^{n} = \int u\overrightarrow{v}.\overrightarrow{n}dS = 
    u_{2}^{n}\left(u_{2}^{n}\triangle y_{bc}^{n}-v_{2}^{n}\triangle  x_{bc}^{n}\right)+u_{3}^{n}\left(u_{3}^{n}\triangle  y_{ca}^{n}-v_{3}^{n}\triangle  x_{ca}^{n}\right)+\\ u_{12}^{n}\left(u_{12}^{n}\triangle  y_{ag}^{n}-v_{12}^{n}\triangle x_{ag}^{n}\right)+u_{11}^{n}\left(u_{11}^{n}\triangle  y_{gb}^{n}-v_{11}^{n}\triangle x_{gb}^{n}\right)
    \label{eq:XCFLUX1}
    \end{aligned}
    \end{equation}
    
    \begin{equation}
    \begin{aligned}
    YCFLUX_{1}^{n}= \int v\overrightarrow{v}.\overrightarrow{n}dS =
    v_{2}^{n}\left(u_{2}^{n}\triangle y_{bc}^{n}-v_{2}^{n}\triangle x_{bc}^{n}\right)+v_{3}^{n}\left(u_{3}^{n}\triangle y_{ca}^{n}-v_{3}^{n}\triangle x_{ca}^{n}\right)+\\ v_{12}^{n}\left(u_{12}^{n}\triangle y_{ag}^{n}-v_{12}^{n}\triangle x_{ag}^{n}\right)+v_{11}^{n}\left(u_{11}^{n}\triangle y_{gb}^{n}-v_{11}^{n}\triangle x_{gb}^{n}\right)
    \label{eq:YCFLUX1}
        \end{aligned}
    \end{equation}
    \\
    Diffusive Flux:
    
    $XDFLUX_{1}^{n} = \int\nabla u\overrightarrow{.n}dS$ 
    \begin{equation}
    \begin{aligned}
     = \left[\left(\frac{ \partial u}{ \partial x}\right)_{2}^{n}\triangle y_{bc}^{n}-\left(\frac{\partial u}{\partial y}\right)_{2}^{n}\triangle x_{bc}^{n}\right]+ \left[\left(\frac{ \partial u}{ \partial x}\right)_{3}^{n}\triangle y_{ca}^{n}-\left(\frac{ \partial u}{ \partial y}\right)_{3}^{n}\triangle x_{ca}^{n}\right]+\\ \left[\left(\frac{ \partial u}{\partial x}\right)_{12}^{n}\triangle y_{ag}^{n}-\left(\frac{ \partial u}{ \partial y}\right)_{12}^{n}\triangle x_{ag}^{n}\right]+\left[\left(\frac{ \partial u}{ \partial x}\right)_{11}^{n}\triangle y_{gb}^{n}-\left(\frac{ \partial u}{ \partial y}\right)_{11}^{n}\triangle x_{gb}^{n}\right] \\
     \label{eq:XDFLUX1}
    \end{aligned}
    \end{equation}

    $YDFLUX_{1}^{n}= \int\nabla v\overrightarrow{.n}dS$
    \begin{equation}
    \begin{aligned}
    =\left[\left(\frac{\partial v}{ \partial x}\right)_{2}^{n}\triangle y_{bc}^{n}-\left(\frac{ \partial v}{ \partial y}\right)_{2}^{n}\triangle x_{bc}^{n}\right]+\left[\left(\frac{ \partial v}{ \partial x}\right)_{3}^{n}\triangle y_{ca}^{n}-\left(\frac{ \partial v}{ \partial y}\right)_{3}^{n}\triangle x_{ca}^{n}\right]+ \\ \left[\left(\frac{ \partial v}{ \partial x}\right)_{12}^{n}\triangle y_{ag}^{n}-\left(\frac{ \partial v}{ \partial y}\right)_{12}^{n}\triangle x_{ag}^{n}\right]+\left[\left(\frac{ \partial v}{ \partial x}\right)_{11}^{n}\triangle y_{gb}^{n}-\left(\frac{ \partial v}{ \partial y}\right)_{11}^{n}\triangle x_{gb}^{n}\right]
    \label{eq:YDFLUX1}
    \end{aligned}
    \end{equation}
    \\ALE Flux:
    \begin{equation}
    \begin{aligned}
        XALEFLUX_{1}^{n} = \int u\overrightarrow{{v}_{b}}.\overrightarrow{n}dS = u_{2}^{n}\left(\frac{ d \Omega}{ d t}\right)_{2}^{n+1}+u_{3}^{n}\left(\frac{ d \Omega}{ d t}\right)_{3}^{n+1}+ \\ u_{12}^{n}\left(\frac{ d \Omega}{ d t}\right)_{12}^{n+1}+u_{11}^{n}\left(\frac{ d \Omega}{d t}\right)_{11}^{n+1}
        \label{eq:XALEFLUX1}
        \end{aligned}
    \end{equation}
    
    \begin{equation}
    \begin{aligned}
        YALEFLUX_{1}^{n}= \int v\overrightarrow{{v}_{b}}.\overrightarrow{n}dS = v_{2}^{n}\left(\frac{ d \Omega}{ d t}\right)_{2}^{n+1}+v_{3}^{n}\left(\frac{ d \Omega}{ d t}\right)_{3}^{n+1}+ \\ v_{12}^{n}\left(\frac{ d \Omega}{ d t}\right)_{12}^{n+1}+v_{11}^{n}\left(\frac{ d \Omega}{ d t}\right)_{11}^{n+1}
        \label{eq:YALEFLUX1}
    \end{aligned}
    \end{equation}
    
    Pressure Flux:
    \begin{equation}
    \begin{aligned}
        XPFLUX_{1}^{n+1}= \int p\overrightarrow{i}.\overrightarrow{n}dS = p_{2}^{n+1}\triangle y_{bc}^{n+1}+p_{3}^{n+1}\triangle y_{ca}^{n+1}+\\ p_{12}^{n+1}\triangle y_{ag}^{n+1}+p_{11}^{n+1}\triangle y_{gb}^{n+1}
        \label{eq:XPFLUX1}
    \end{aligned}
    \end{equation}
    
    \begin{equation}
    \begin{aligned}
        YPFLUX_{1}^{n+1}= \int p\overrightarrow{j}.\overrightarrow{n}dS = -p_{2}^{n+1}\triangle x_{bc}^{n+1}-p_{3}^{n+1}\triangle x_{ca}^{n+1}- \\ p_{12}^{n+1}\triangle x_{ag}^{n+1}-p_{11}^{n+1}\triangle x_{gb}^{n+1}
        \label{eq:YPFLUX1}
    \end{aligned}
    \end{equation}
	Where the spatial derivatives appearing in diffusive fluxes are approximated using Taylor Series approximation.
    \begin{equation}
        \left(\frac{ \partial \phi}{ \partial y}\right)_{1} \approx \frac{(\phi_{P}-\phi_{C})\triangle x_{ab}+(\phi_{a}-\phi_{b})\triangle x_{CP}}{\triangle y_{CP}\triangle x_{ab}-\triangle y_{ab}\triangle x_{CP}}
        \label{eq:Pressure Interpolation}
    \end{equation}

\begin{figure}[h!]
        \centering
        \includegraphics[scale=0.35]{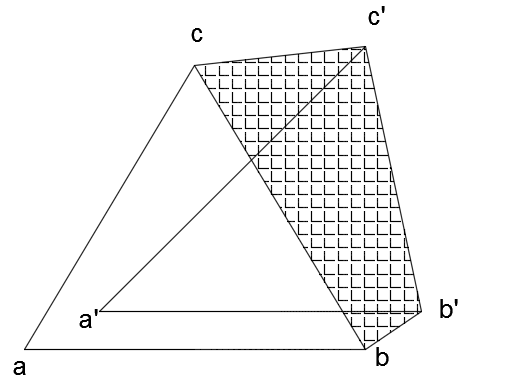}
        \caption{Volume traced by face bc during grid deformation}
        \label{fig:my_label3}
    \end{figure}   

 The derivatives occurring in the ALE flux terms are computed as follows,   

\begin{equation}
\begin{aligned}
\left(\frac{ d \Omega}{ d t}\right)_{bc}^{n+1}=\left(\frac{ d \Omega}{ d t}\right)_{2}^{n+1}=\frac{1}{2\triangle t}(x_{c}y_{b}+x_{b}y_{b'}+x_{b'}y_{c'}+x_{c'}y_{c}\\-x_{b}y_{c}-x_{b'}y_{b}-x_{c'}y_{b'}-x_{c}y_{c'})    
\label{eq:ALE Derivative}
\end{aligned}
\end{equation}
    
Similar discretization procedure is adopted for the rest of the faces of the cell. The discretized face-centered velocities (like in Eq.(\ref{eq:u1}) and Eq.(\ref{eq:v1})) are substituted into the discrete continuity equation (\ref{eq:Discret. Cont. Eq}) to obtain the PPE (\ref{eq:PPE}).
  
    \begin{equation}
    \begin{aligned}
        SOURCE = K_{P}p_{P}+K_{A}p_{A}+K_{B}p_{B}+K_{C}p_{C}+K_{D}p_{D}+K_{G}p_{G}+\\K_{F}p_{F}+K_{I}p_{I}+ K_{J}p_{J}+K_{L}p_{L}
        \label{eq:PPE}
    \end{aligned}
    \end{equation}
    Where the $K_{i}$ denote geometrical coefficients that appear as a result of discretizing Eq(\ref{eq:Discret. Cont. Eq}) and the $SOURCE$ term consists of the remaining fluxes and cell-centered velocities known at ${n}^{th}$ time level.
    
    \begin{equation}
    \begin{aligned}
        &SOURCE =\frac{A_{pc}^{n}}{A_{pc}^{n+1}\triangle t}(u_{1}^{n}\triangle y_{ab}^{n+1}-v_{1}^{n}\triangle x_{ab}^{n+1})+\frac{A_{pa}^{n}}{A_{pa}^{n+1}\triangle t}(u_{2}^{n}\triangle y_{bc}^{n+1}-v_{2}^{n}\triangle x_{bc}^{n+1}) + \\ &\frac{A_{pb}^{n}}{A_{pb}^{n+1}\triangle t}(u_{3}^{n}\triangle y_{ca}^{n+1}-v_{3}^{n}\triangle x_{ca}^{n+1})+\\ &\frac{\triangle y_{ab}^{n+1}}{A_{pc}^{n+1}}(-XCFLUX_{1}^n+\frac{1}{Re}XDFLUX_{1}^n+XALEFLUX_{1}^n)+ \\ &\frac{\triangle y_{bc}^{n+1}}{A_{pa}^{n+1}}(-XCFLUX_{2}^n+\frac{1}{Re}XDFLUX_{2}^n+XALEFLUX_{2}^n)+ \\ &\frac{\triangle y_{ca}^{n+1}}{A_{pb}^{n+1}}(-XCFLUX_{3}^n+\frac{1}{Re}XDFLUX_{3}^n+XALEFLUX_{3}^n)- \\ &\frac{\triangle x_{ab}^{n+1}}{A_{pc}^{n+1}}(-YCFLUX_{1}^n+\frac{1}{Re}YDFLUX_{1}^n+YALEFLUX_{1}^n)- \\ &\frac{\triangle x_{bc}^{n+1}}{A_{pa}^{n+1}}(-YCFLUX_{2}^n+\frac{1}{Re}YDFLUX_{2}^n+YALEFLUX_{2}^n)- \\ &\frac{\triangle x_{ca}^{n+1}}{A_{pb}^{n+1}}(-YCFLUX_{3}^n+\frac{1}{Re}YDFLUX_{3}^n+YALEFLUX_{3}^n)
        \label{eq:SOURCE}
    \end{aligned}
    \end{equation}
    
    Solving subsequently for the cell-centered pressures, face-centered velocities are computed further using (Eq.{\ref{eq:u1}}) and Eq.({\ref{eq:v1}}).
    For each cell, all fluxes are computed from the newly computed face-centered velocities and pressures and the velocity field is computed at the cell centers.
    \begin{equation}
    \begin{aligned}
        u_{iP}^{n+1}=u_{iP}^{n}\frac{a_{P}^{n}}{a_{P}^{n+1}}+\frac{\triangle t}{a_{p}^{n+1}}(-{CFLUX_i}^{n+1}-{PFLUX_i}^{n+1}+\frac{1}{Re}{DFLUX_i}^{n+1}+ \\ {ALEFLUX_i}^{n+1})
        \label{eq:CCV}
        \end{aligned}
    \end{equation}
    
    The flux terms appearing in the momentum equation above (for velocities $i=1,2$ as u,v) are computed from the new face-centered variables using a similar approach for the original cell P.
\\
\\The various steps involved in the solution procedure can be summarized as follows:
\\
\\(1) The cell-centered velocity and pressure field $u^{n}$, $v^{n}$ and $p^{n}$ are initialized. This could either be available from the past flow data or from the prescribed initial conditions.
\\(2) The face-centered velocity and pressure field  $u^{n}$, $v^{n}$ and $p^{n}$ are computed upon interpolation from the cell-centered data at $n^{th}$ level. 
\\(3) Grid is deformed to accommodate the new position of the object.
\\(4) Fluxes appearing in the momentum equation (Eq.(\ref{eq:XCFLUX1}) to Eq.(\ref{eq:YPFLUX1})) are computed using the interpolated velocity field at $n^{th}$ level and mesh data at the current level.
\\(5) The new face-centered velocity field expressions are computed by solving momentum equations for the reconstructed cells.
\\(6) These face-centered velocity field expressions are substituted in the continuity equation Eq.(\ref{eq:Discret. Cont. Eq}) and the resulting PPE is solved to compute the cell-centered pressure field $p^{n+1}$. 
\\(7) The cell-centered pressure field $p^{n+1}$ is interpolated to compute face-centered pressure field $p^{n+1}$.
\\(8) The pressure flux terms (Eq.(\ref{eq:XPFLUX1}) and Eq.(\ref{eq:YPFLUX1})) are updated and the face-centered velocities $u^{n+1}$ and $v^{n+1}$ are computed by solving the momentum equations for the reconstructed cells using fluxes computed from velocity and pressure fields $u^{n}$, $v^{n}$ and $p^{n+1}$ .
\\(9) The momentum equations are then solved for each cell by computing fluxes from the face-centered velocities and pressures $u^{n+1}$, $v^{n+1}$ and $p^{n+1}$ to compute the cell-centered velocity and pressure field $u^{n+1}$, $v^{n+1}$ and $p^{n+1}$.

The solution procedure is repeated for the next cycle and is continued until sufficient convergence is achieved.         

 \section{Results and Discussion}

 In the present study for flow past a circular cylinder subjected to rotary and transverse oscillations, its wake structure  and aerodynamic characteristics have been investigated.
 The rotary oscillation parametric space corresponds to four vortex shedding modes primarily \cite{Tokumaru}.
 The amplitude ratios with the diameter(D) of the cylinder for rotary and transverse oscillations have been taken as $A_{r} = 2.0$ and $A_{tv} = 0.2$ respectively. The rotary forced frequency of oscillation, denoted by $St_{r}$ has been varied as $0.4\leq St_{r} \leq 2.0$ and the transverse forced oscillation frequency ratio denoted by $f_{tr} = St_{tv}/{St_{n}}$, is varied as $0.9\leq f_{tr}\leq 2.0$.      
In the following sections, we present the aerodynamic and vortex shedding characteristics for each of the vortex shedding modes when subjected to transverse oscillations at Re=100.

\begin{table}[h]
\centering
\caption{Distinct vortex shedding modes}
\begin{tabular}{c  c  c  c  c}
\hline
 & Mode 1 & Mode 2 & Mode 3 & Mode 4\\ 
\hline
$A_{r}$ & 2.0 & 2.0 & 2.0 & 0.45\\ 
$St_{r}$ & 0.165 & 0.4 & 0.8 & 0.8\\ 
\hline
\end{tabular}
\label{tab:VSModes}
\end{table}

 \subsection{Validation}
~\\
The Solver based on ALE-CFR was validated for both transverse and rotational oscillation problems of simulating flow past an oscillating circular cylinder. A systematic grid independence study was carried out to test the grid convergence of the new ALE scheme. The study was performed on flow past a transversely oscillating cylinder with amplitude and frequency of transverse oscillations were taken as 0.5D and $f=0.192$ respectively at $Re=185$. The difference in the grids is based on the number of nodes on the cylinder surface. Coefficient of drag was chosen for comparison since it is one of the most sensitive numerical parameters to change in grid size. Table \ref{tab:Grid Independence Study} gives the details of the grid independence test based on which grid number 3 was chosen which comprising of 57,962 triangular elements and 29,156 nodes out of which 200 nodes were taken on the body surface. Further studies, using the same grid at $Re=185$, were taken as reference for validating the former \cite{Pham}, and the latter problem \cite{Choi} was validated at $Re=100$. The coefficients $\overline{c}_{d}$ and $c_{l,rms}$ have been plotted for frequency ratios $0.8\leq f_{r} \leq 1.2$ and amplitude ratio $A/D=0.5$ in Fig \ref{fig:Cd,avg_tv} and \ref{fig:Cl,rms_tv} for the transverse oscillation problem. For the rotational oscillation problem $\overline{c}_{d}$ and $c_{l,amp}$ have been plotted for frequency ratios $0.9\leq f_{r} \leq 1.2$ and amplitude ratio $A/D=2.0$ in Fig \ref{fig:Cd,avg_r} and \ref{fig:Cl,amp_r}.
    
    \begin{table}[h]
    \centering
    \caption{Grid independence test carried out at $Re=185$ for transversely oscillating cylinder}
    \begin{tabular}{c  c  c}
    \hline
        Grid no. & Number of nodes on the body & Drag coefficient \\ 
    \hline
    1 & 160 & $1.754\pm0.49$\\ 
    2 & 180 & $1.813\pm0.52$\\
    3 & 200 & $1.878\pm0.53$\\
    4 & 220 & $1.887\pm0.53$\\
    \hline
    \end{tabular}
    \label{tab:Grid Independence Study}
    \end{table}
    
    \begin{figure}[h!]
        \begin{subfigure}[b]{6.5cm}
        \includegraphics[scale=0.04]{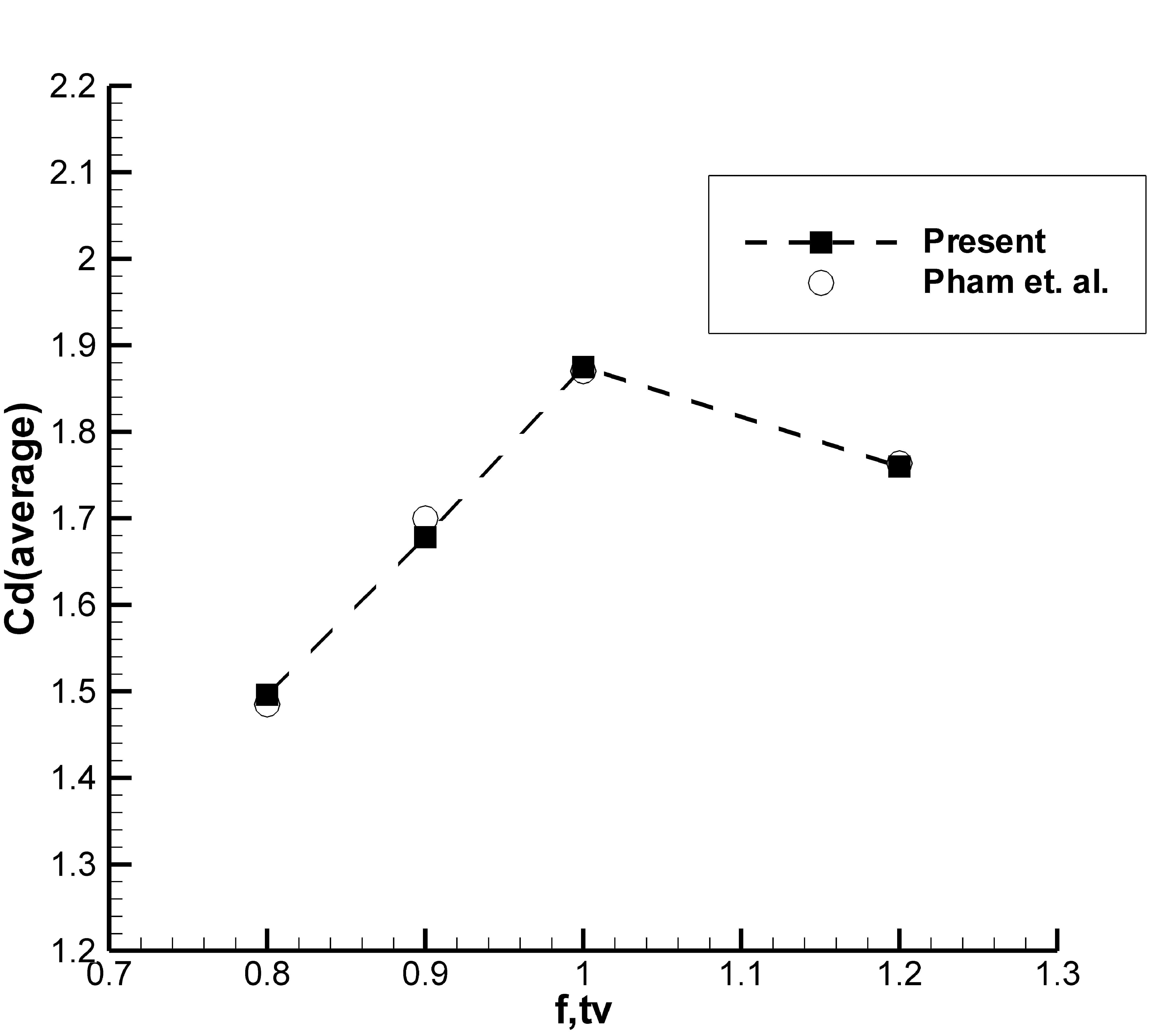}
        \caption{\normalsize{Time Averaged $C_{D}$ values for $A/D=0.5$}}
        \label{fig:Cd,avg_tv}
        \end{subfigure}
        \hfill
        \begin{subfigure}[b]{6cm}
        \includegraphics[scale=0.04]{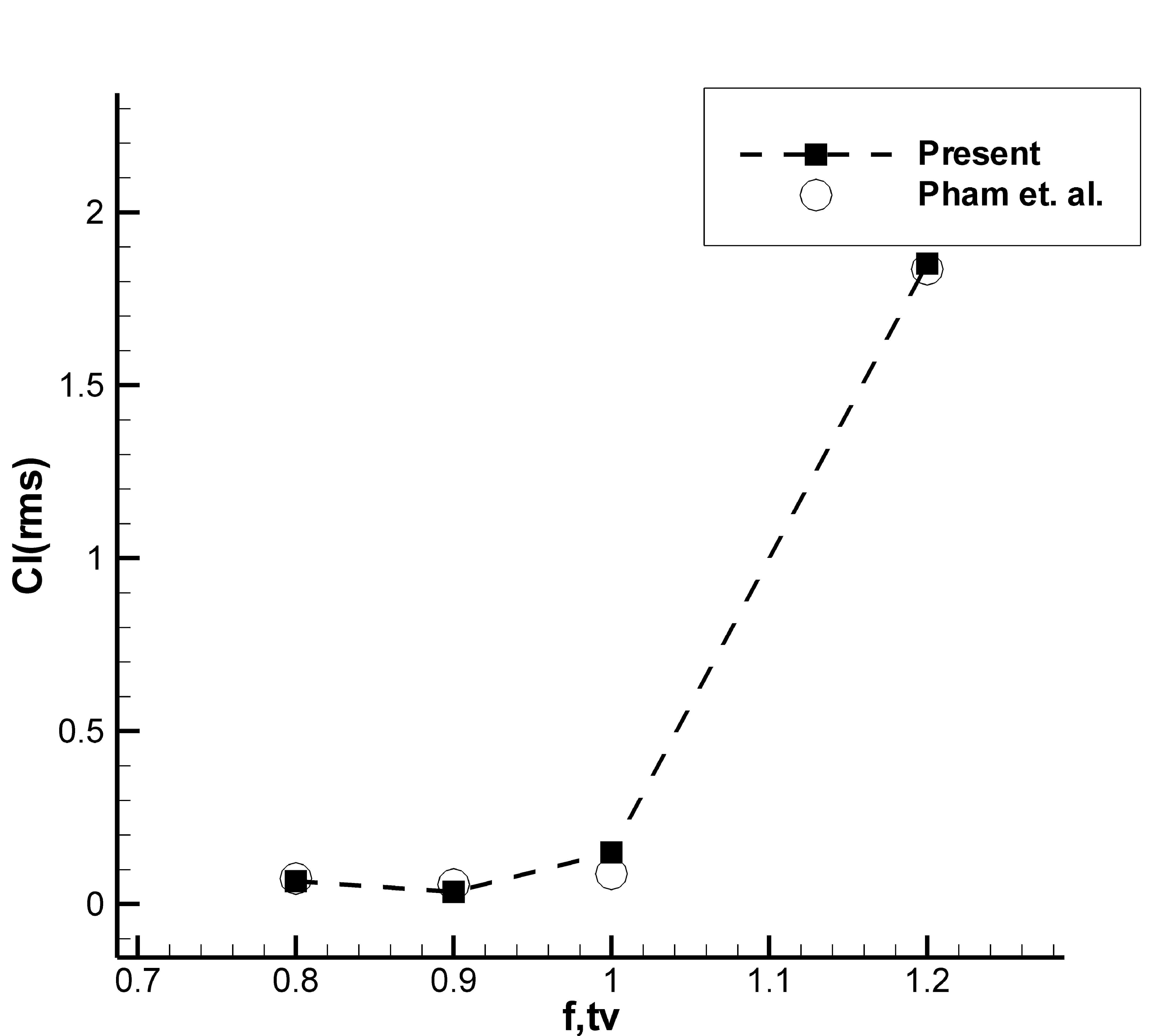}
        \caption{\normalsize{Root-Mean-Squared $C_{L}$ values for $A/D=0.5$}}
        \label{fig:Cl,rms_tv}
        \end{subfigure}
        \caption{\normalsize{Aerodynamic coefficients for transverse oscillation frequency ratios at $Re=185$}}
    \end{figure}
    
    \begin{figure}[h!]
        \begin{subfigure}[b]{6cm}
        \includegraphics[scale=0.04]{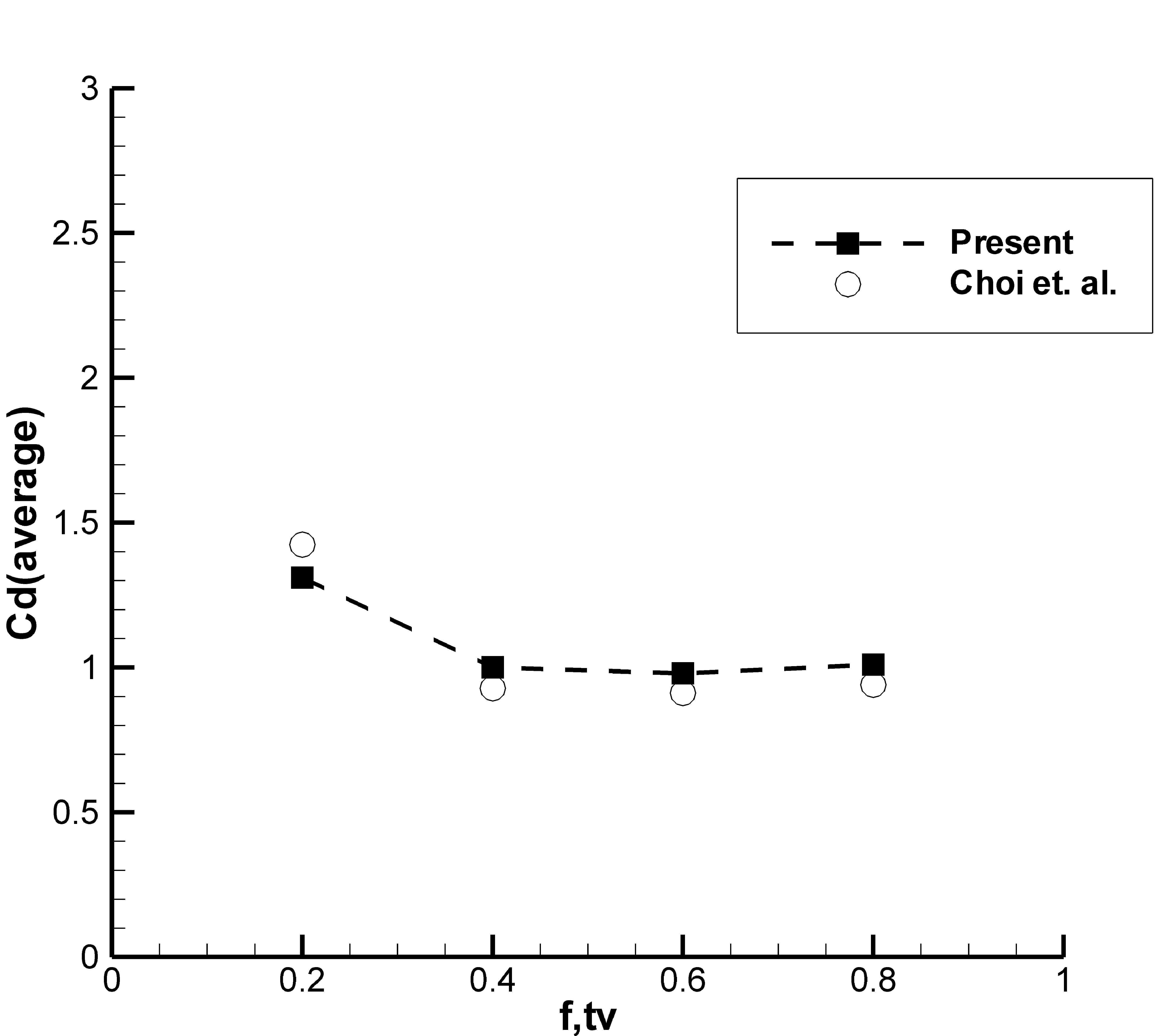}
        \caption{\normalsize{Time Averaged $C_{D}$ values for A/D = 2.0}}
        \label{fig:Cd,avg_r}
        \end{subfigure}
        \hfill
        \begin{subfigure}[b]{6cm}
        \includegraphics[scale=0.04]{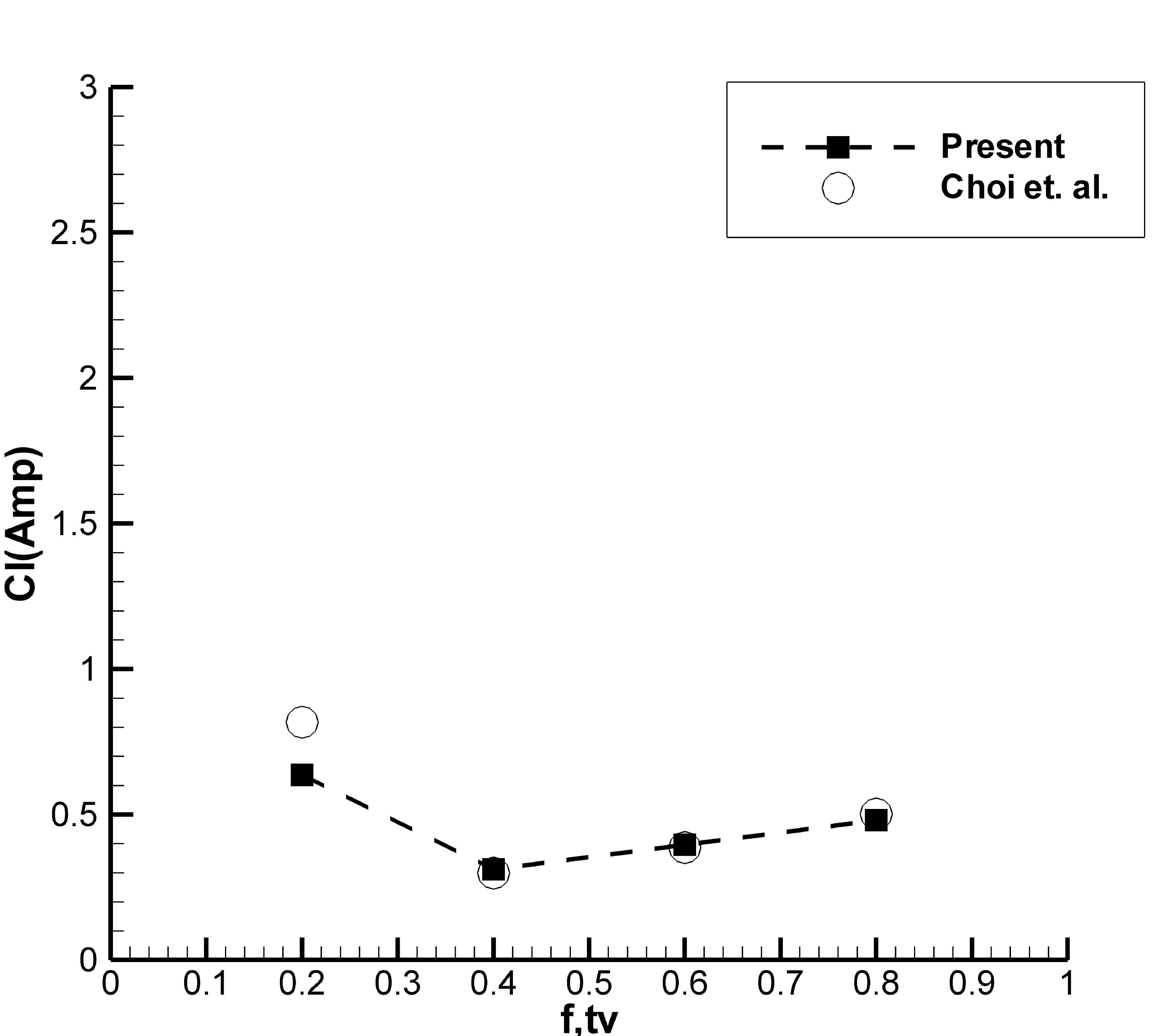}
        \caption{\normalsize{Max Fluctuation Amplitude of $C_{L}$ for A/D = 2.0}}
        \label{fig:Cl,amp_r}
        \end{subfigure}
        \caption{\normalsize{Aerodynamic coefficients for rotational oscillation frequencies at $Re=100$}}
    \end{figure}
    
 \newpage
 \subsection{Vortex shedding characteristics and contribution of rotational and transverse oscillations to lift generation}
~\\
    This vortex shedding mode 1 corresponding to $St_{r}=0.165$ and $A_{r}=2.0$ can be characterized by the generation of two like signed vortices in one-half cycle \cite{Tokumaru}, however, this was not observed in our numerical study at $Re = 100$ \cite{Choi}. Nonetheless, the characteristic feature of this mode of vortex shedding was observed to be the symmetric shedding of oppositely signed vortices, as can be seen in Fig \ref{fig:VSMode1}. The transverse oscillation frequency ratio is varied as $0.9\leq f_{tr}\leq 2.0$ for an amplitude ratio $A_{tv}=0.2$. Since this mode is lock-in with respect to rotary oscillations ($St_{r} = St_{n}$), the peak corresponding to $St_{r}$ in the frequency characteristic plot for lift was observed to be dominant. 
    For $f_{tr}=0.9$ and $f_{tr}=0.95$, two nearby peaks are observed from the frequency characteristics referring from Fig \ref{fig:_1,3_} and \ref{fig:_1,4_}, the primary and secondary ones corresponding to rotary and transverse oscillation frequencies respectively. The primary reason for only two peaks being visibly dominant in the frequency plot was due to the lock-in established between the forced rotary oscillation and vortex shedding frequency ($St_{vs} \approx St_{r}$ due to lock-in at the natural frequency of the system).
    
\begin{figure}[hp]
        \begin{subfigure}[b]{6.2cm}
        \includegraphics[scale=0.055]{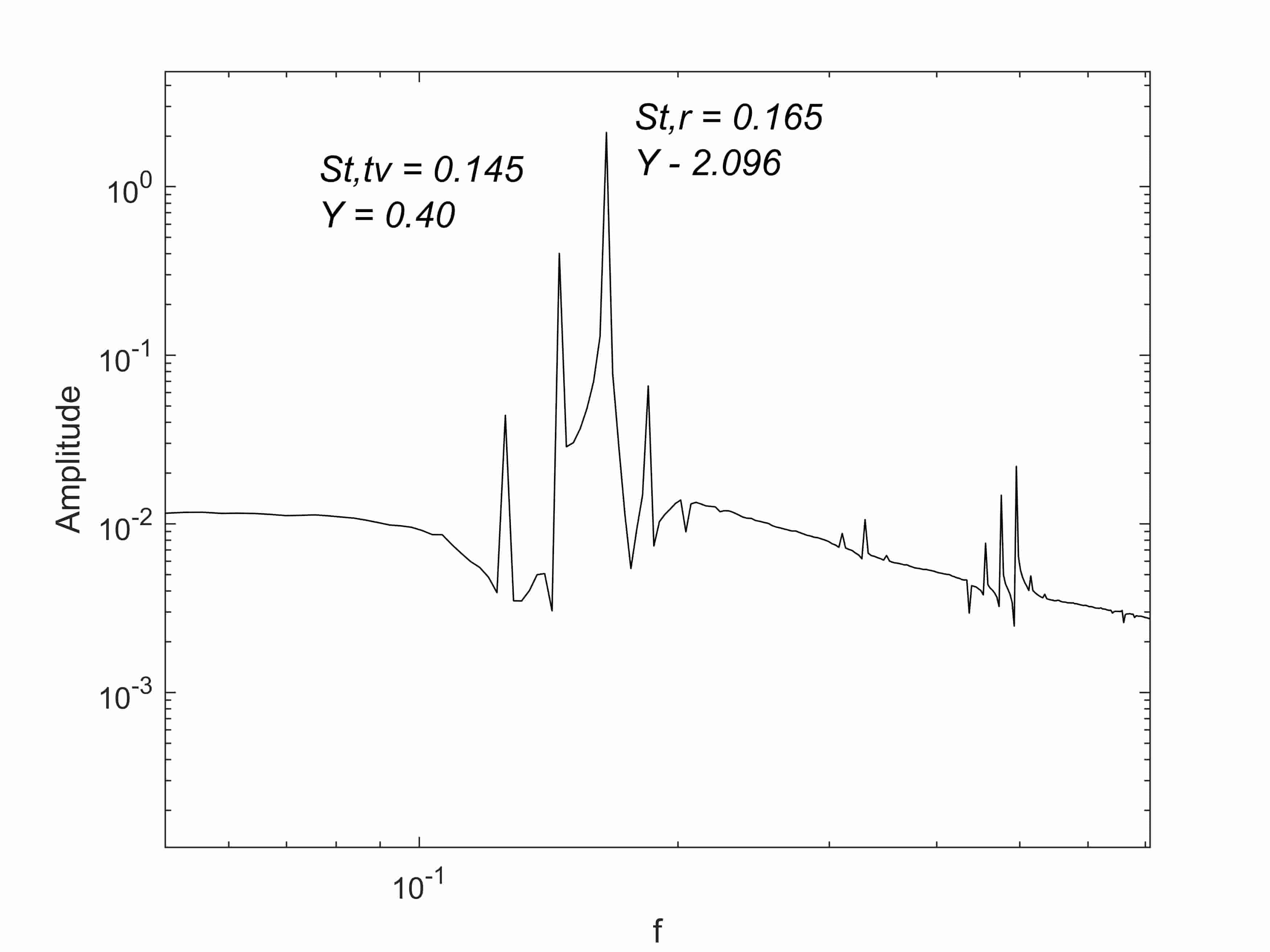} 
        \caption{$f_{tr}=0.9$,$St_{r}=0.165$}
        \label{fig:_1,3_}
        \end{subfigure}
        \hfill
    \begin{subfigure}[b]{6.2cm}
        \includegraphics[scale=0.055]{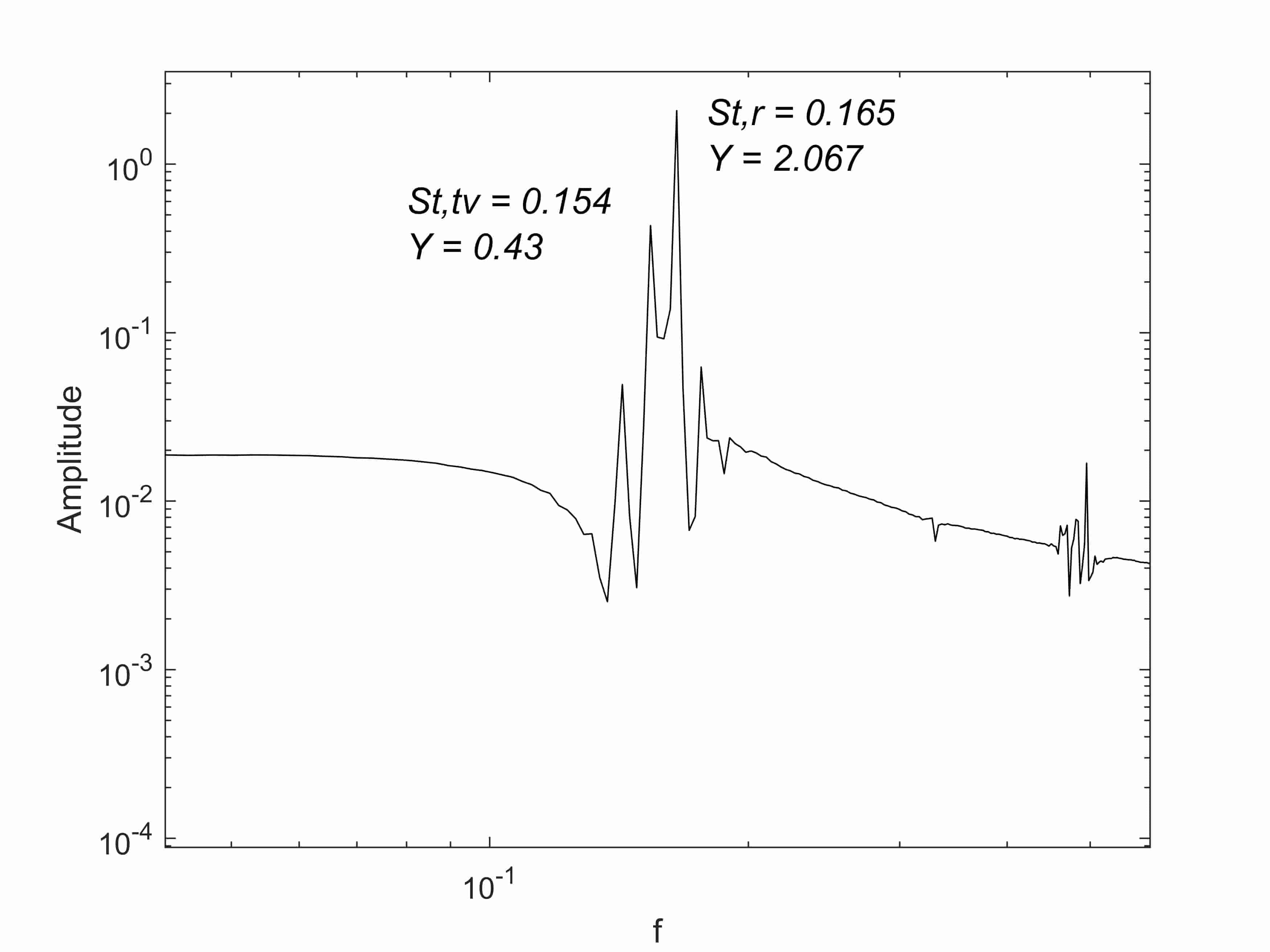} 
        \caption{$f_{tr}=0.95$,$St_{r}=0.165$}
        \label{fig:_1,4_}
    \end{subfigure}
        \hfill
    \begin{subfigure}[b]{6.2cm}
        \includegraphics[scale=0.055]{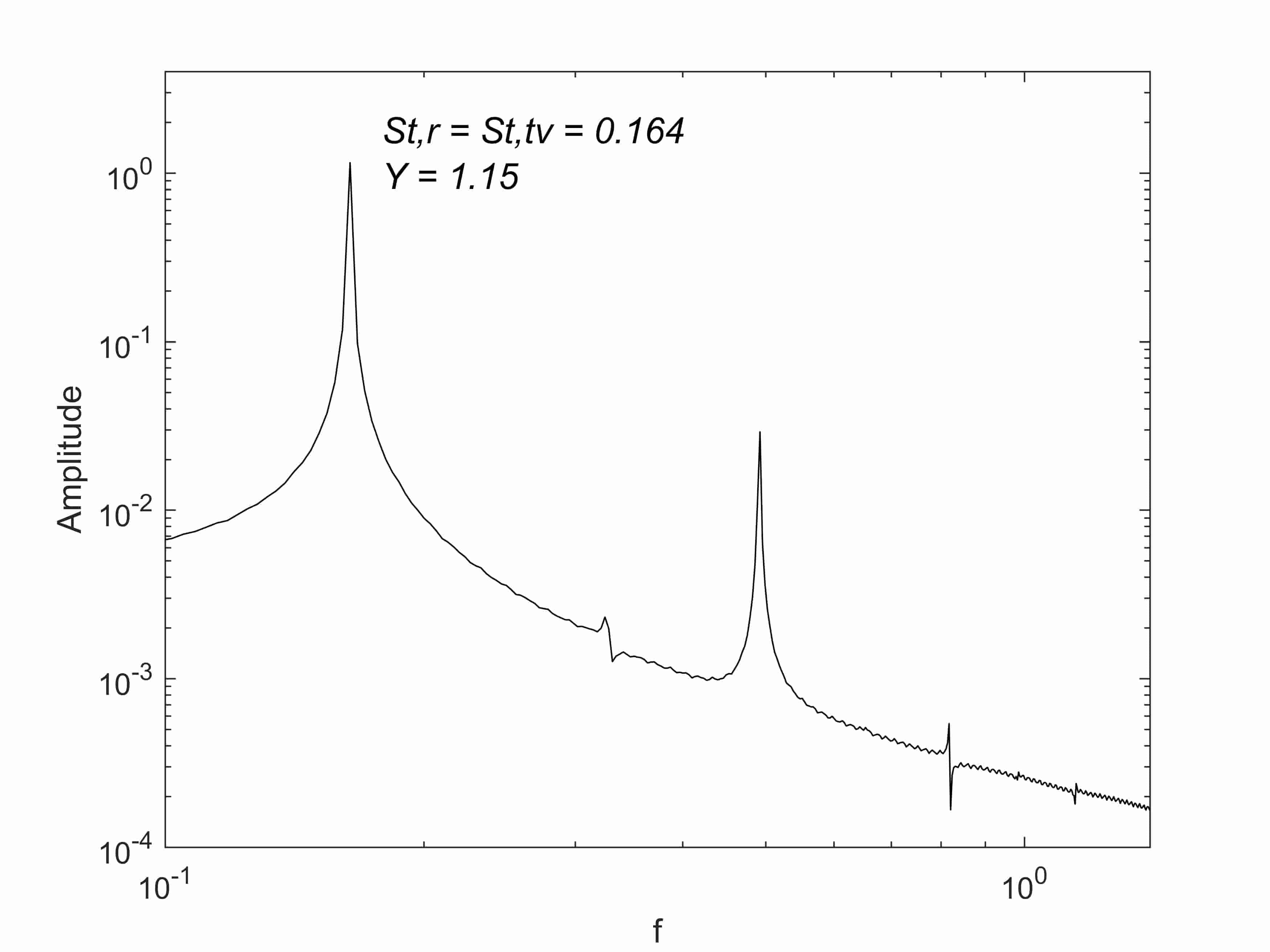} 
        \caption{$f_{tr} = 1.0$,$St_{r} = 0.165$}
        \label{fig:Broad Peak}
    \end{subfigure}
    \hfill
    \begin{subfigure}[b]{6.2cm}
        \includegraphics[scale=0.055]{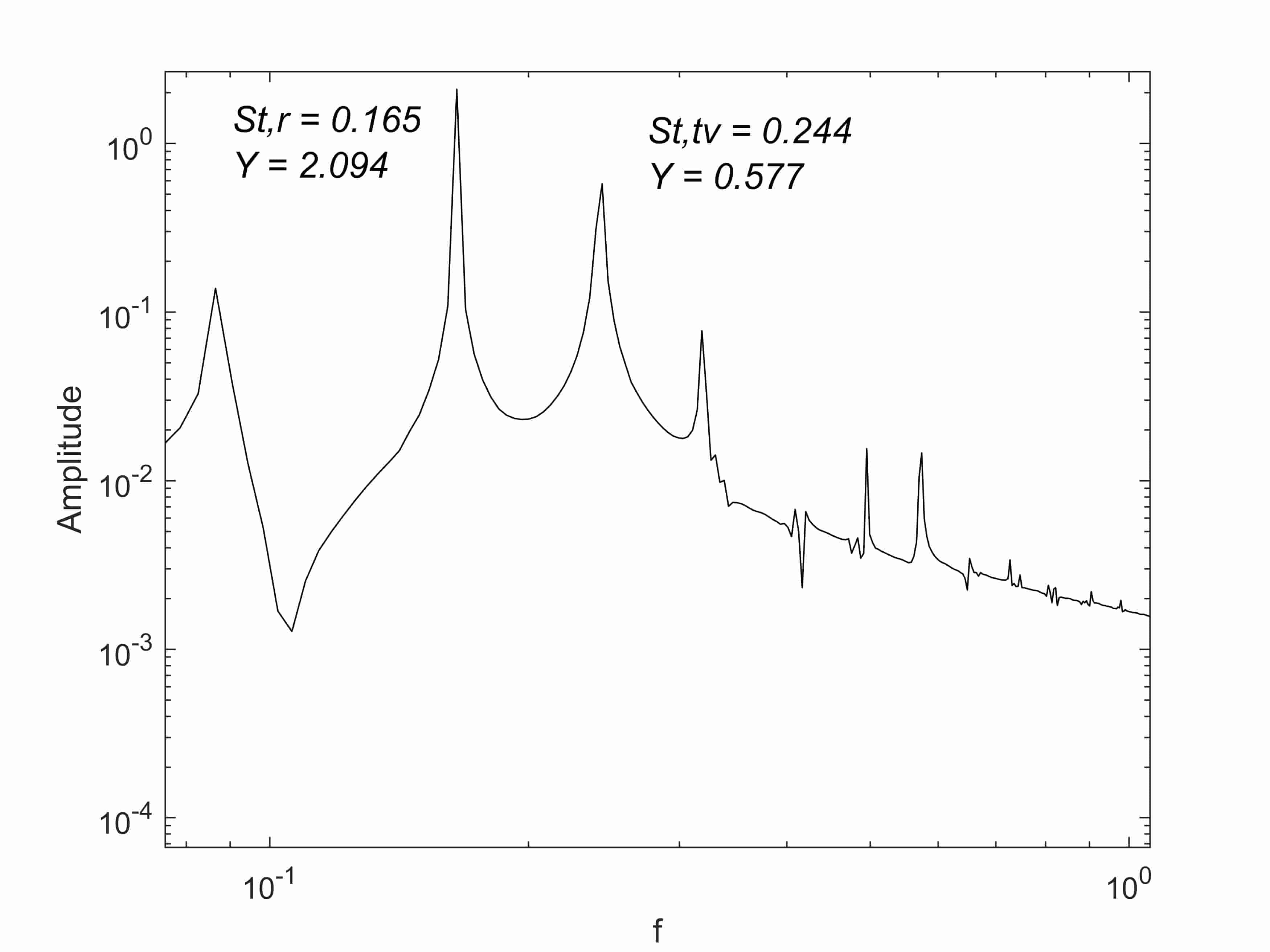} 
        \caption{$f_{tr} = 1.5$,$St_{r} = 0.165$}
        \label{fig:_1,5_}
    \end{subfigure}
    \hfill
    \begin{subfigure}[b]{6.2cm}
        \includegraphics[scale=0.055]{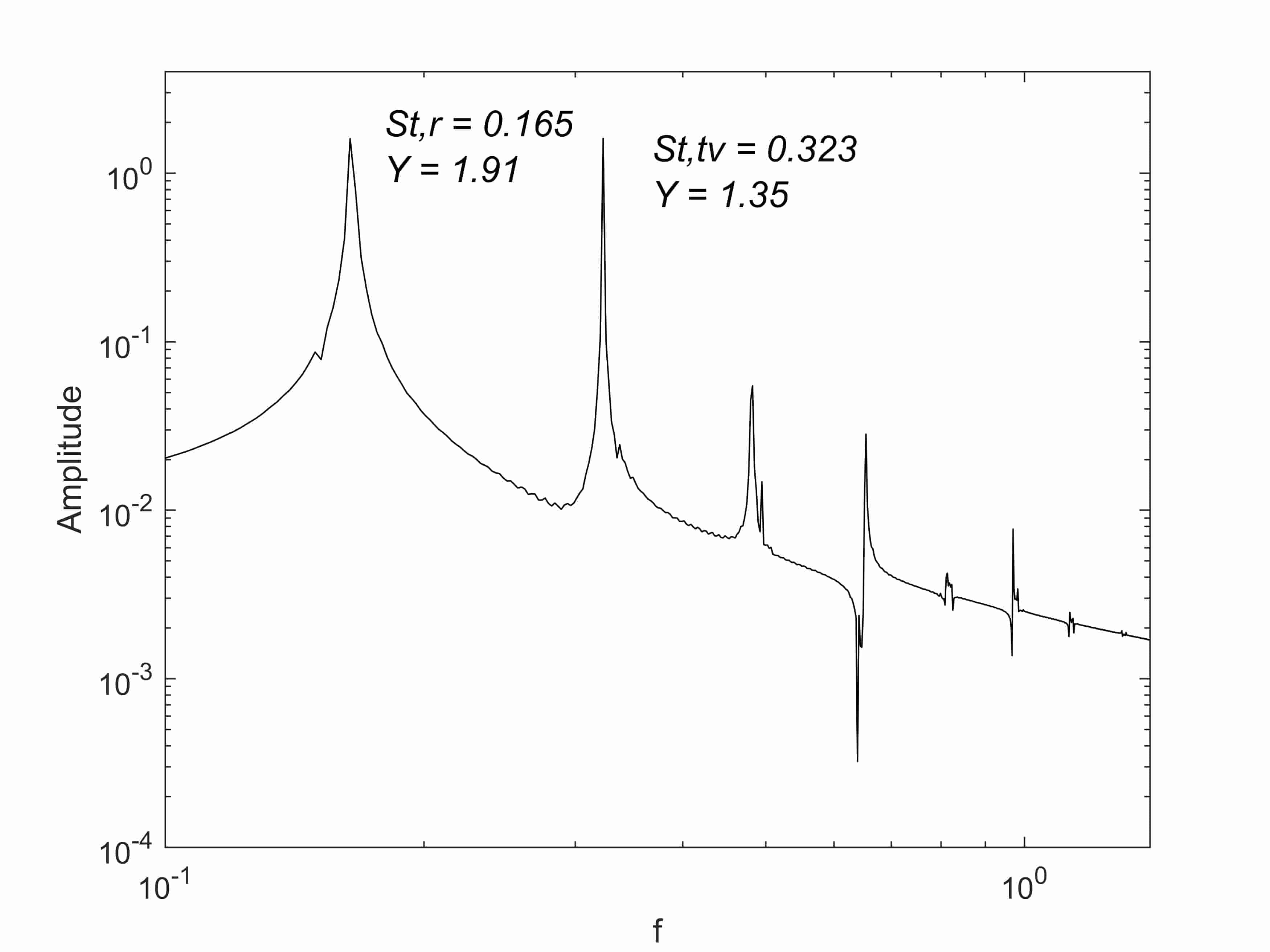} 
        \caption{$f_{tr} = 2.0$,$St_{r} = 0.165$}
        \label{fig:_1,1_}
    \end{subfigure}
    \hfill
    \begin{subfigure}[b]{6.2cm}
        \centering
        \includegraphics[scale=0.055]{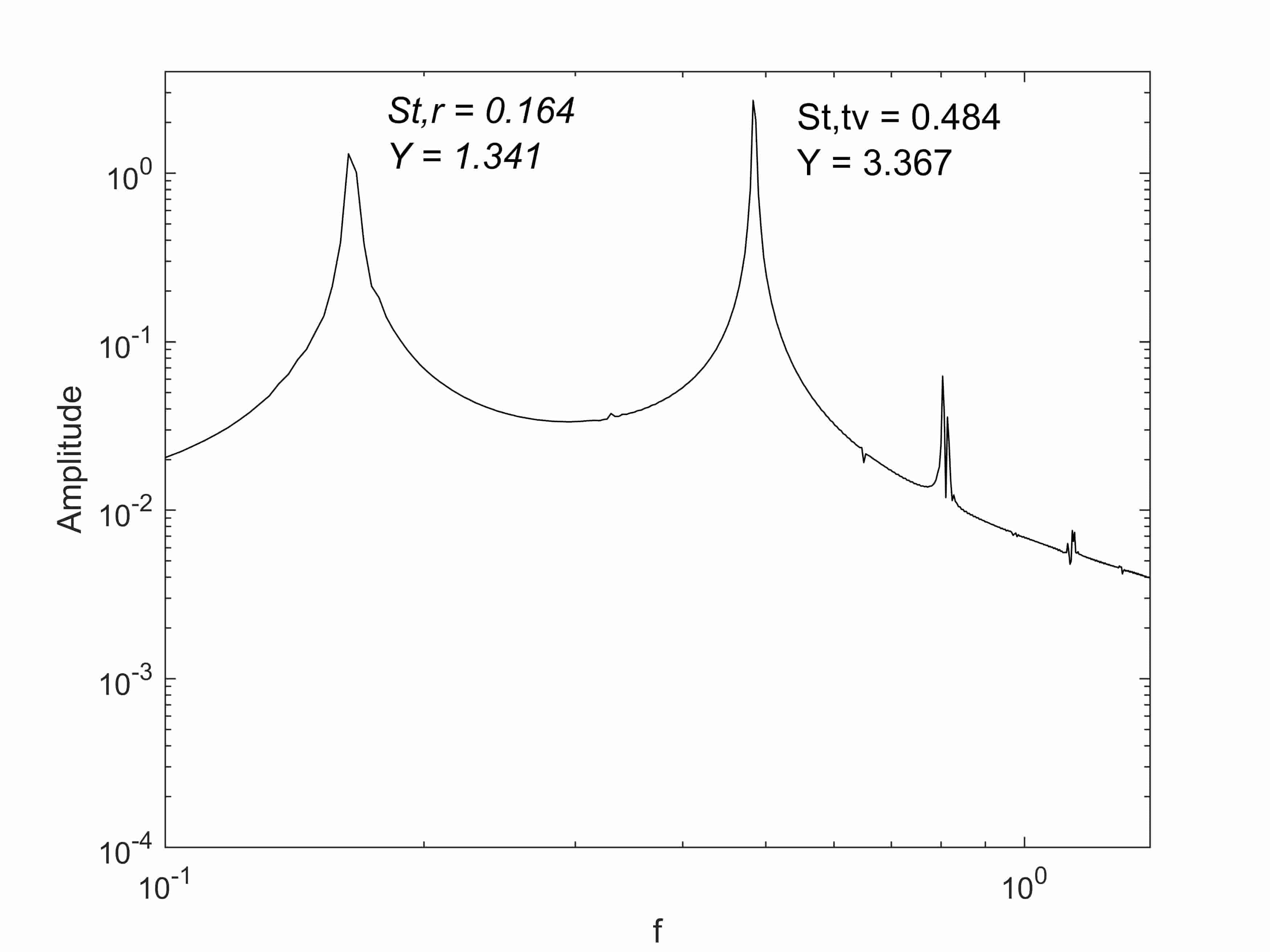} 
        \caption{$f_{tr} = 3.0$,$St_{r} = 0.165$}
        \label{fig:_1,2_}
    \end{subfigure}
    \caption{Frequency Characteristics of Lift at various transverse oscillation frequencies (Mode 1)}
    \end{figure}

    At frequencies lower than $f_{tr} = 1.0$ from Fig \ref{fig:_1,3_} and \ref{fig:_1,4_}, the lift amplitude (noted from \ref{fig:_1,3_Cl} and \ref{fig:_1,4_Cl}) contribution  attributable to frequencies $St_{tv}$ is observed to be roughly $16\%$ and that to $St_{r}$ close to $83\%$.  
    For $f_{tr}=1.0$, as seen from Fig \ref{fig:Broad Peak}, a single dominant broad peak appears consisting of all the three frequencies (all close to the $St_{n}$ of the system). 
    At frequencies higher than $f_{tr}=1.0$ from Fig \ref{fig:_1,5_}, \ref{fig:_1,1_} and \ref{fig:_1,2_}, the transverse oscillation contribution to the lift was observed to be higher. Contribution to lift due to  transverse oscillations was found to be as high as $36\%$ with rotary oscillation and vortex shedding frequencies constituting $52\%$ of the lift amplitude as seen in Fig \ref{fig:_1,1_Cl}, at $f_{tr}=2.0$ from Fig \ref{fig:_1,1_}. As $f_{tr}$ increases further, the transverse oscillations dominate the lift contribution with their contribution as high as $60\%$ of the lift amplitude in Fig \ref{fig:_1,2_Cl}, as seen from Fig \ref{fig:_1,2_}.
    
    \afterpage{
    \begin{figure}[hp]
        \begin{subfigure}[b]{10cm}
        \centering
        \includegraphics[scale=0.05]{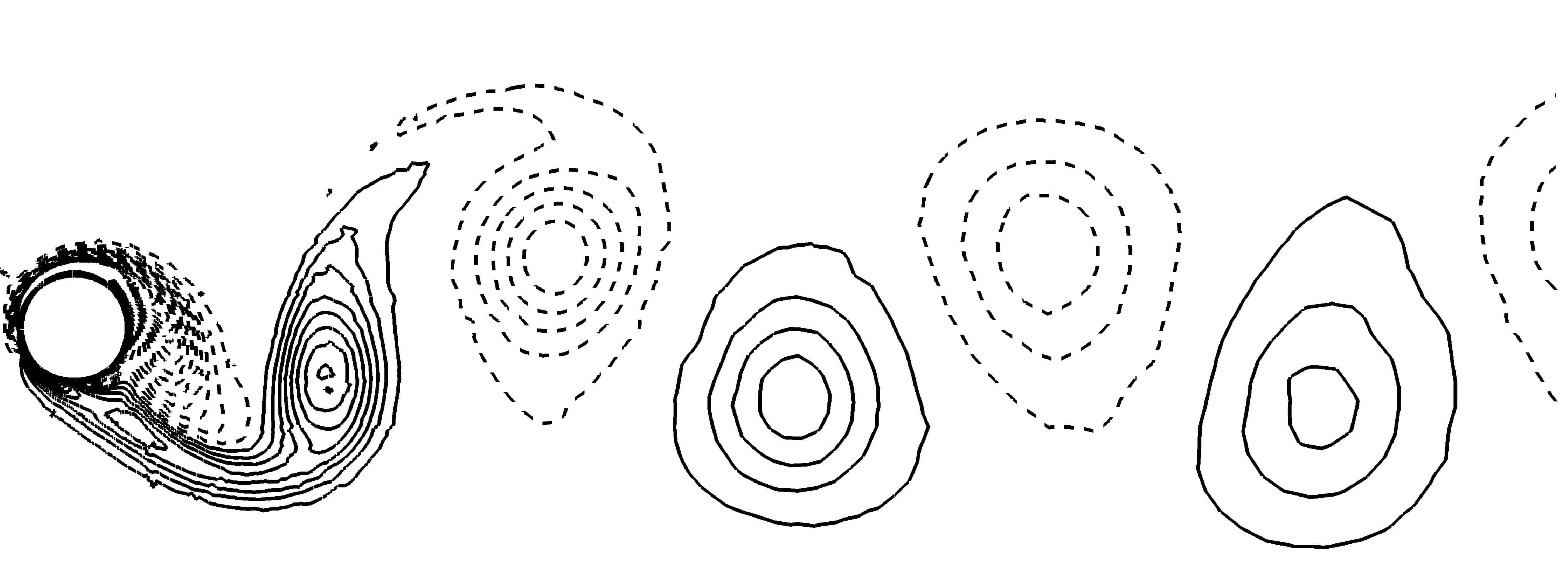}
        \caption{Mode 1}
        \label{fig:VSMode1}
        \end{subfigure}
        \hfill
        \begin{subfigure}[b]{10cm}
        \centering
        \includegraphics[scale=0.05]{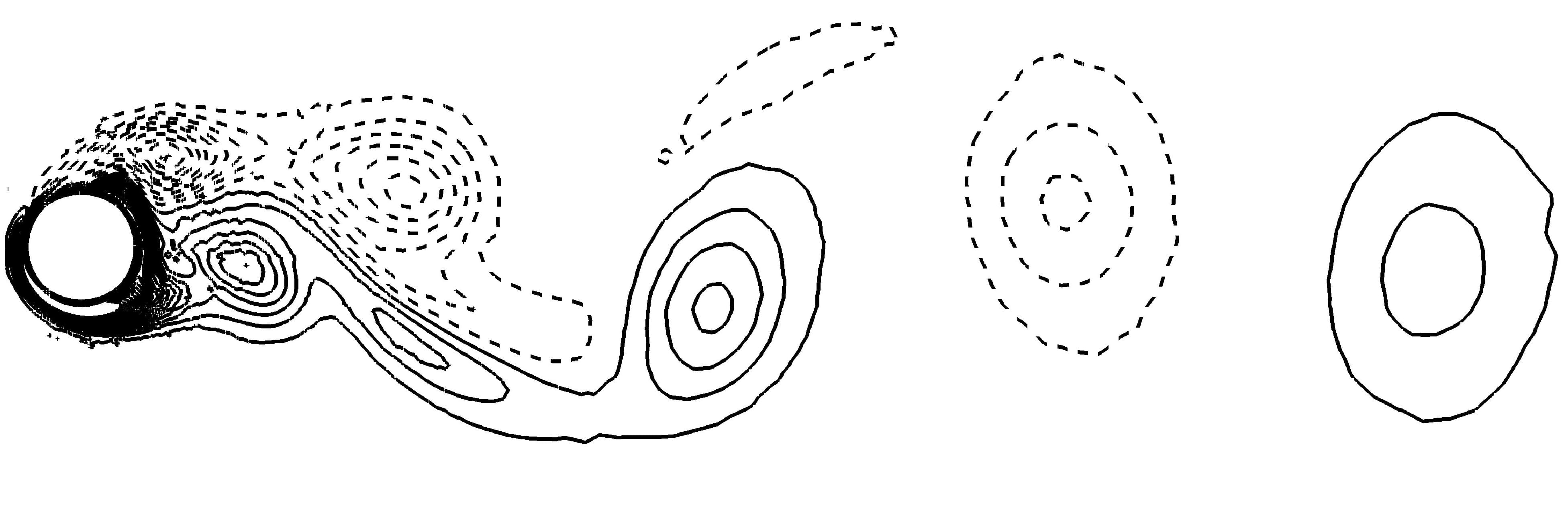}
        \caption{Mode 2}
        \label{fig:VSMode2}
        \end{subfigure}
        \hfill
        \begin{subfigure}[b]{10cm}
        \centering
        \includegraphics[scale=0.05]{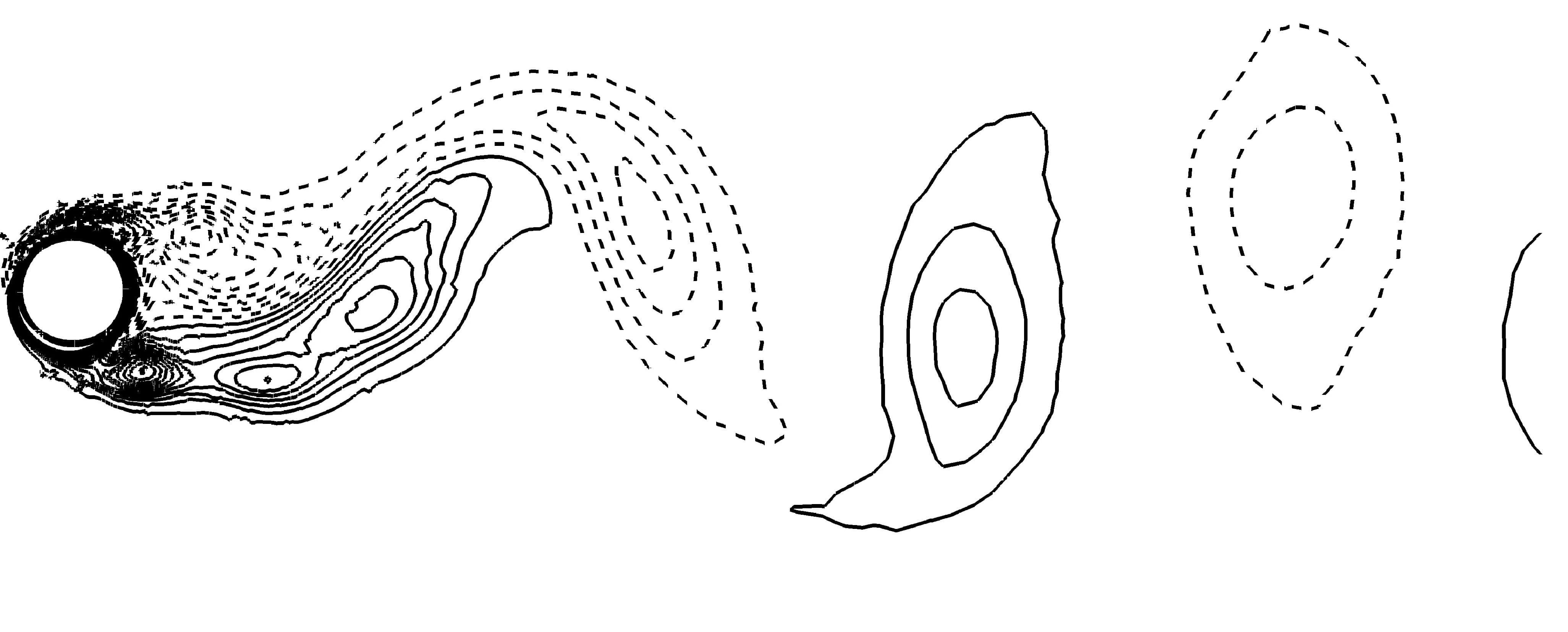}
        \caption{Mode 3}
        \label{fig:VSMode3}
        \end{subfigure}
        \hfill
        \begin{subfigure}[b]{10cm}
        \centering
        \includegraphics[scale=0.05]{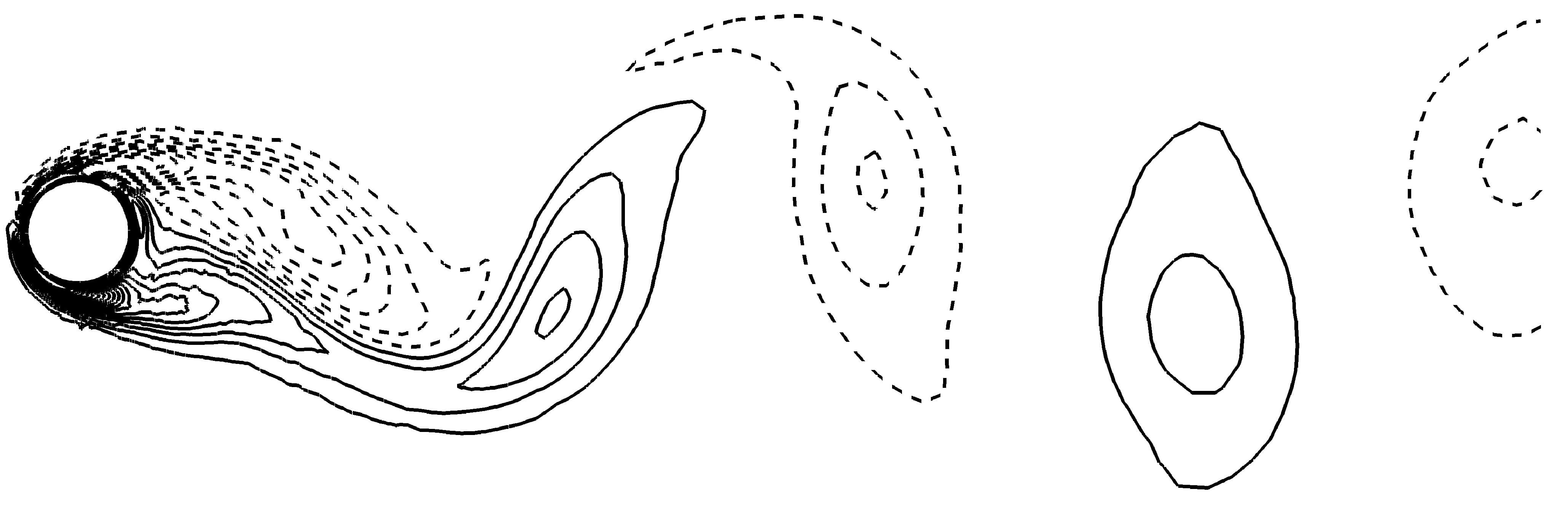}
        \caption{Mode 4}
        \label{fig:VSMode4}
        \end{subfigure}
    \caption{Distinct Vortex Shedding Modes}
    \end{figure}
    }
    
     Vortex shedding mode 2, with the rotary oscillation parameters $St_{r}=0.4$ and $A_{r}=2.0$, is characterized by the dissipation of vortices without coalescence downstream as given by Tokumaru and Dimotakis \cite{Tokumaru}. When transverse oscillations were imposed the vortex shedding pattern was seen to consist of elliptic vortices attached to circular vortices, which dissipated further downstream as seen from Fig (\ref{fig:VSMode2}). This mode was found to become less prominent as $f_{tr}$ was increased. Partial dissipation of vortices in the far downstream and the shedding of vortices downstream was observed at $f_{tr}=2.0$, a deviation from the usual characteristics of the vortex shedding mode.
    At $f_{tr}=0.9$ and $f_{tr}=0.95$ the frequency peaks for lift were observed to be prominent for both the $St_{tv}$ and $St_{r}$, as observed from Fig \ref{fig:_2,3_} and \ref{fig:_2,4_}, with approximately $50\%$ contribution from each towards the net lift amplitude generated (Fig  \ref{fig:_2,3_Cl} and \ref{fig:_2,4_Cl}). The reason behind these peaks belonging to the $St_{tv}$ and $St_{r}$ is due to the lock-in achieved by $St_{tv}$ with $St_{vs}$, leading to a common peak representing the two visible in Fig \ref{fig:_2,3_} and \ref{fig:_2,4_}, since in both these cases, $St_{tv}$ and $St_{n}$ lie in close proximity. Lift amplitude generated in this case came out to be significantly lower as compared to that of the earlier mode because of lock-in established by the $St_{r}$ with $St_{n}$ in mode 1.
    As the transverse oscillation frequency ratio was increased to $f_{tr}=1.5$ and $f_{tr}=2.0$, from Fig \ref{fig:_2,5_} and \ref{fig:_2,1_}, the transverse oscillation peaks similar to the previous mode discussed, were seen to account for majority of lift amplitude (Fig \ref{fig:_2,5_Cl} and \ref{fig:_2,1_Cl}) generation. The contributions towards lift generation (Fig \ref{fig:_2,1_Cl} and \ref{fig:_2,2_Cl}) from the $St_{tv}$ values corresponding to $f_{tr}=2.0$ and $f_{tr}=3.0$, from Fig \ref{fig:_2,1_} and \ref{fig:_2,2_}, were found to be $75-80\%$.

     The vortex shedding mode 3 is uniquely characterized by the synchronous vortex generation with the rotational forcing imposed. The smaller vortices generated merge with larger ones in the near wake to form a multi-polar vortex structure due to phase lag between the immediate and far wake. With rotary oscillation parameters $St_{r}=0.8$ and $A_{r}=2.0$, $f_{tr}$ was varied from 0.9 to 3.0 with $A_{tv} = 0.2$. This mode was found to become further enhanced when the cylinder was subjected to transverse oscillations since vortices generated by the rotary and transverse oscillations merged better due to greater phase lag between the shear layers and far wake. At $f_{tr}=0.9$ and $0.95$ the frequencies $St_{tv}$ and $St_{r}$ both impart $50\%$ to the net lift magnitude (Fig \ref{fig:_3,3_Cl} and \ref{fig:_3,4_Cl}) as seen from the frequency characteristics in Fig \ref{fig:_3,3_} and \ref{fig:_3,4_}, similar to the observations in Mode 2. 
    As frequency $St_{tv}$ increases to higher values when $f_{tr}=1.5$, $2.0$ and $3.0$, it was observed from numerical experiments that its contribution to lift amplitude (Fig \ref{fig:_3,5_Cl}, \ref{fig:_3,1_Cl} and \ref{fig:_3,2_Cl}) increases upto $70\%$, with the rotary oscillation frequency's contribution near $5-6\%$, as seen from Fig \ref{fig:_3,5_}, \ref{fig:_3,1_} and \ref{fig:_3,2_}.    
    
    Vortex shedding mode 4, with rotary oscillation parameters taken as $St_{r}=0.8$ and  $A_{r}=0.45$, can be characterized by the generation of small scale vortices existing in the shear layers near the surface of the cylinder as seen in Fig. \ref{fig:VSMode4}. The primary difference seen between the vortex shedding mode 3 and 4 is the synchronized vortex generation with rotary oscillation in the former. The vortex shedding pattern in the downstream wake for mode 4 resembles that of flow past a stationary cylinder. From the numerical experiments conducted, two major peaks emerge in the lift frequency characteristics at $f_{tr}=0.9$ and $0.95$ due to lock-in between the frequencies $St_{tv}$ and $St_{n}$ ($St_{tv}$ and $St_{n}$ again being in close proximity), as seen in Fig. \ref{fig:_4,3_} and \ref{fig:_4,4_}. The contribution of transverse and rotary oscillation towards lift generation was found to be equally distributed as seen in the second and third modes earlier.  
    Similar trends are observed at high transverse frequency ratios $f_{tr}=2.0$ and $3.0$ from Fig. \ref{fig:_4,1_} and \ref{fig:_4,2_}, where the contribution of the same was observed to be $98\%$. The reason for this being that $A_{r}$ taken was significantly lower than that of the rest of the modes and $A_{tv}$ was taken to be the same.
    
    \newpage
    
   \begin{figure}[hp]
        \begin{subfigure}[b]{6.2cm}
        \includegraphics[scale=0.04]{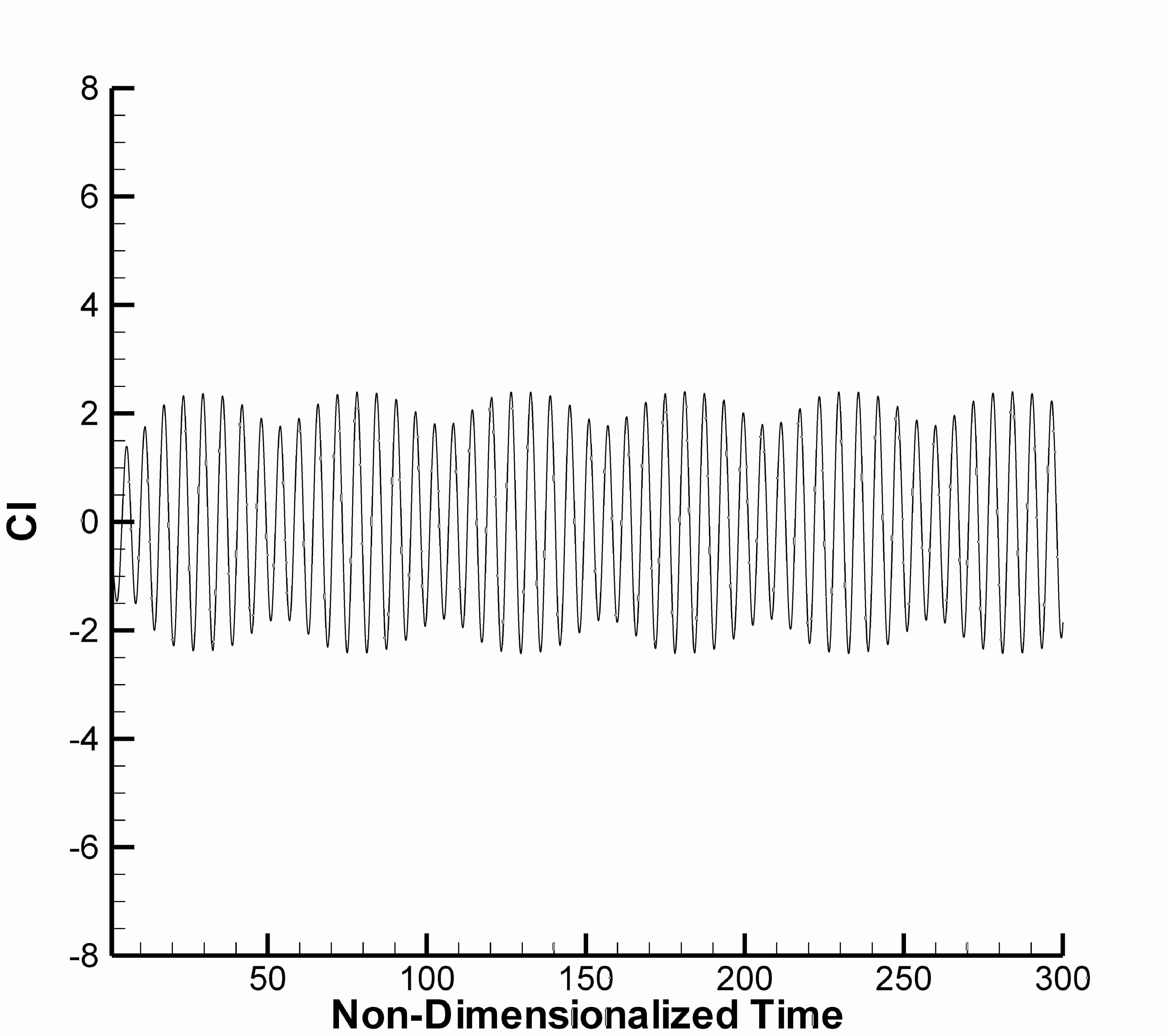} 
        \caption{$f_{tr}=0.9$,$St_{r}=0.165$}
        \label{fig:_1,3_Cl}
        \end{subfigure}
        \hfill
    \begin{subfigure}[b]{6.2cm}
        \includegraphics[scale=0.04]{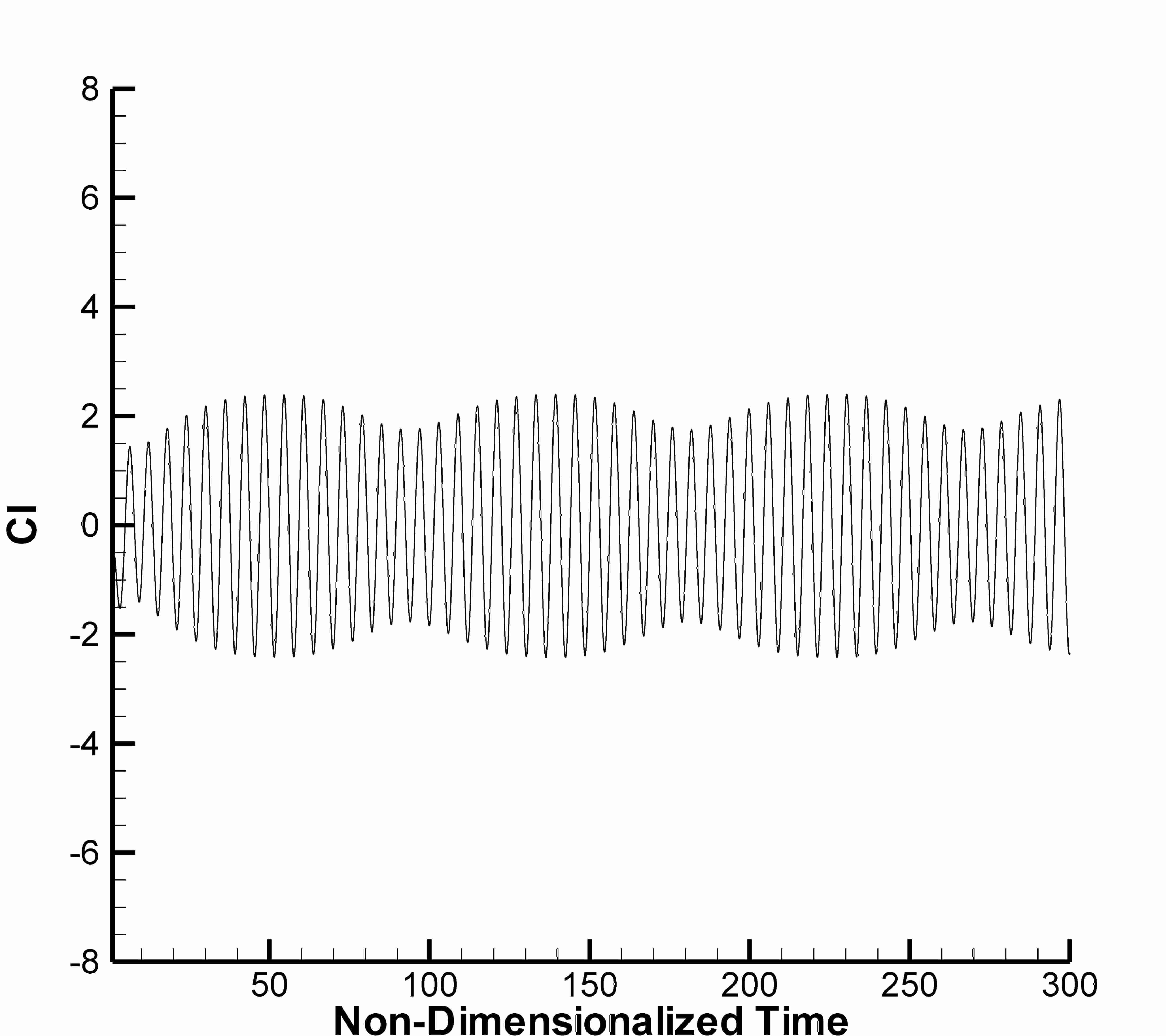} 
        \caption{$f_{tr}=0.95$,$St_{r}=0.165$}
        \label{fig:_1,4_Cl}
    \end{subfigure}
    \hfill
    \begin{subfigure}[b]{6.2cm}
        \includegraphics[scale=0.04]{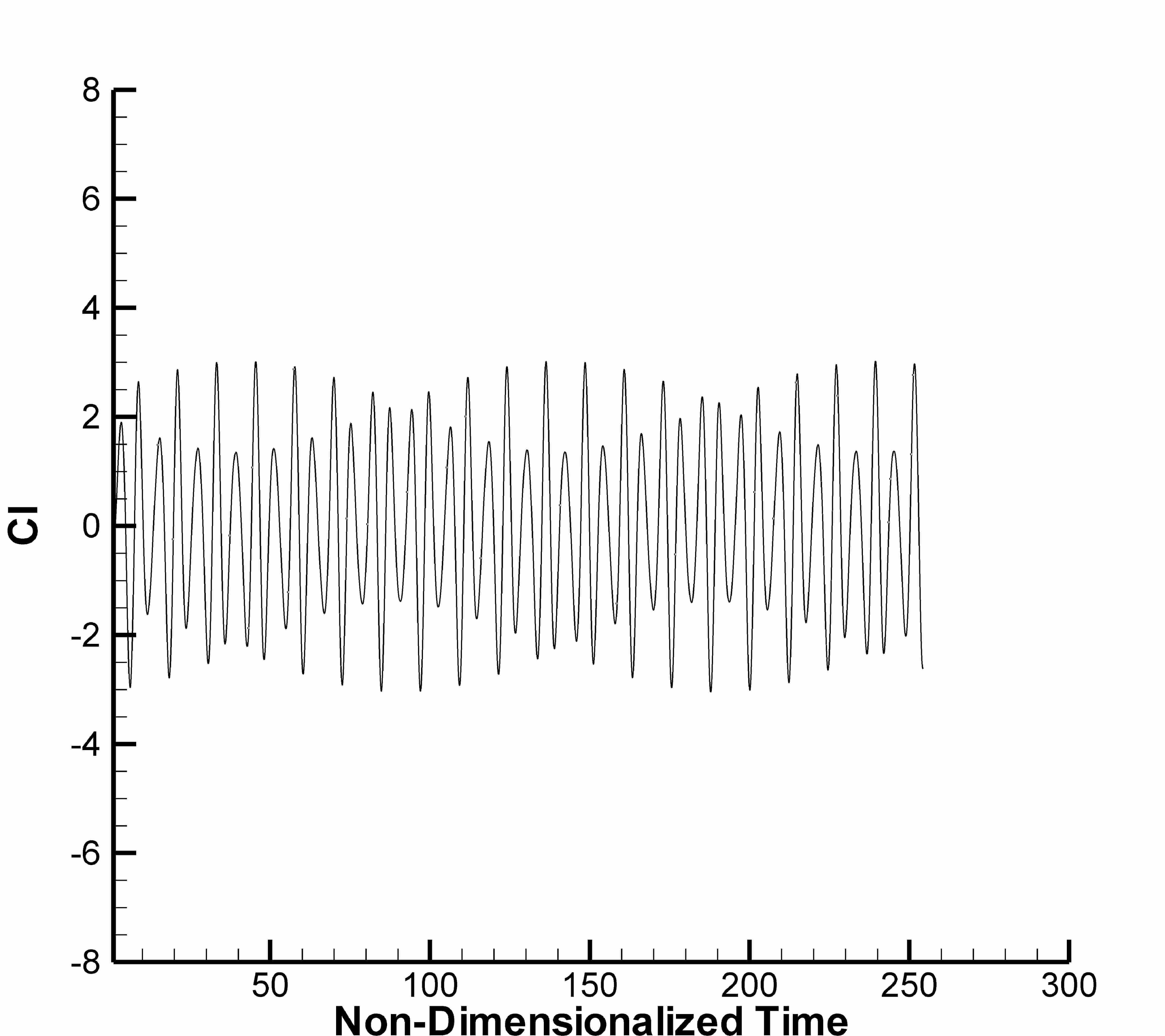} 
        \caption{$f_{tr} = 1.5$,$St_{r} = 0.165$}
        \label{fig:_1,5_Cl}
    \end{subfigure}
    \hfill
    \begin{subfigure}[b]{6.2cm}
        \includegraphics[scale=0.04]{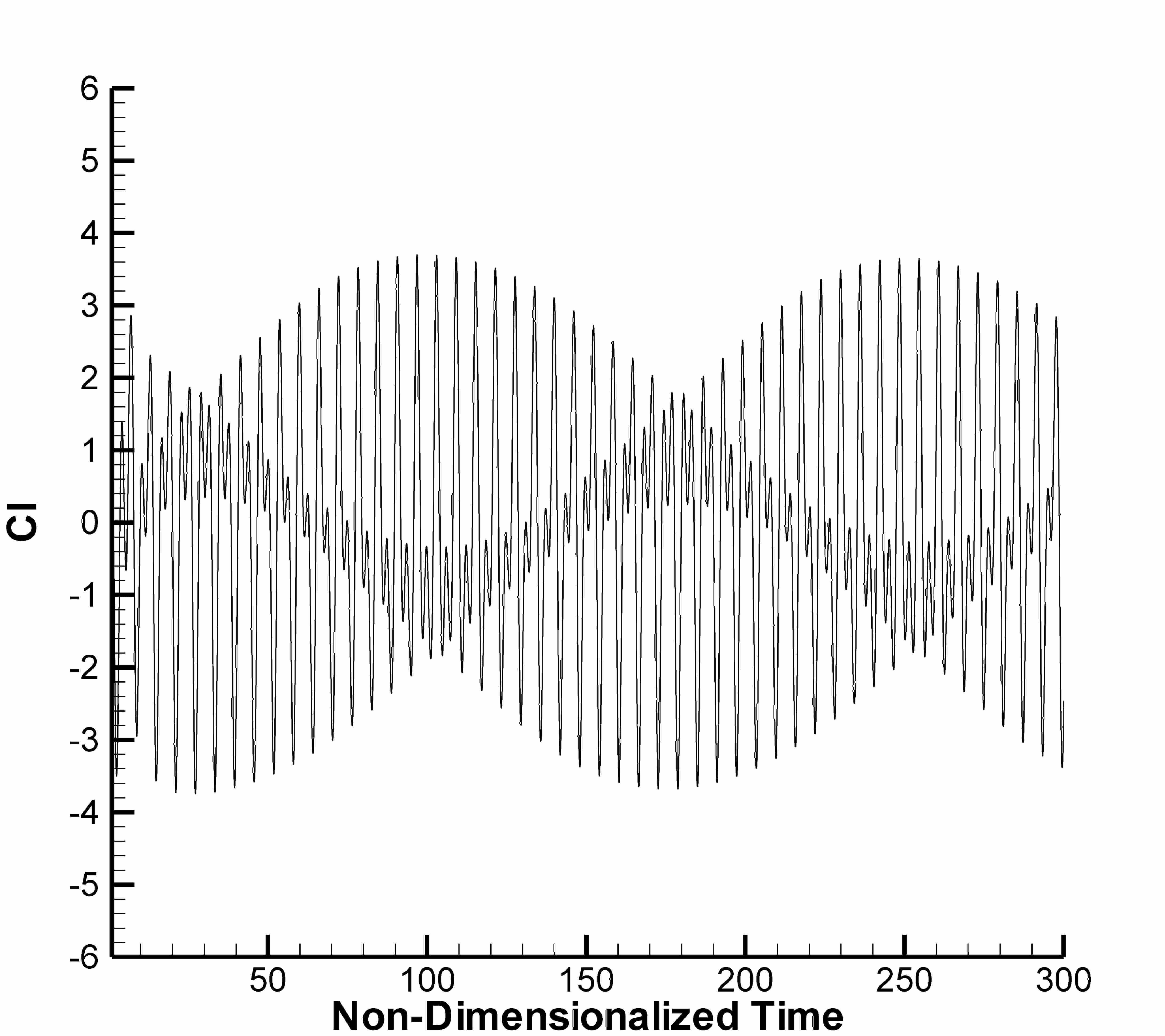} 
        \caption{$f_{tr} = 2.0$,$St_{r} = 0.165$}
        \label{fig:_1,1_Cl}
    \end{subfigure}
    \hfill
    \begin{subfigure}[b]{6.2cm}
        \centering
        \includegraphics[scale=0.04]{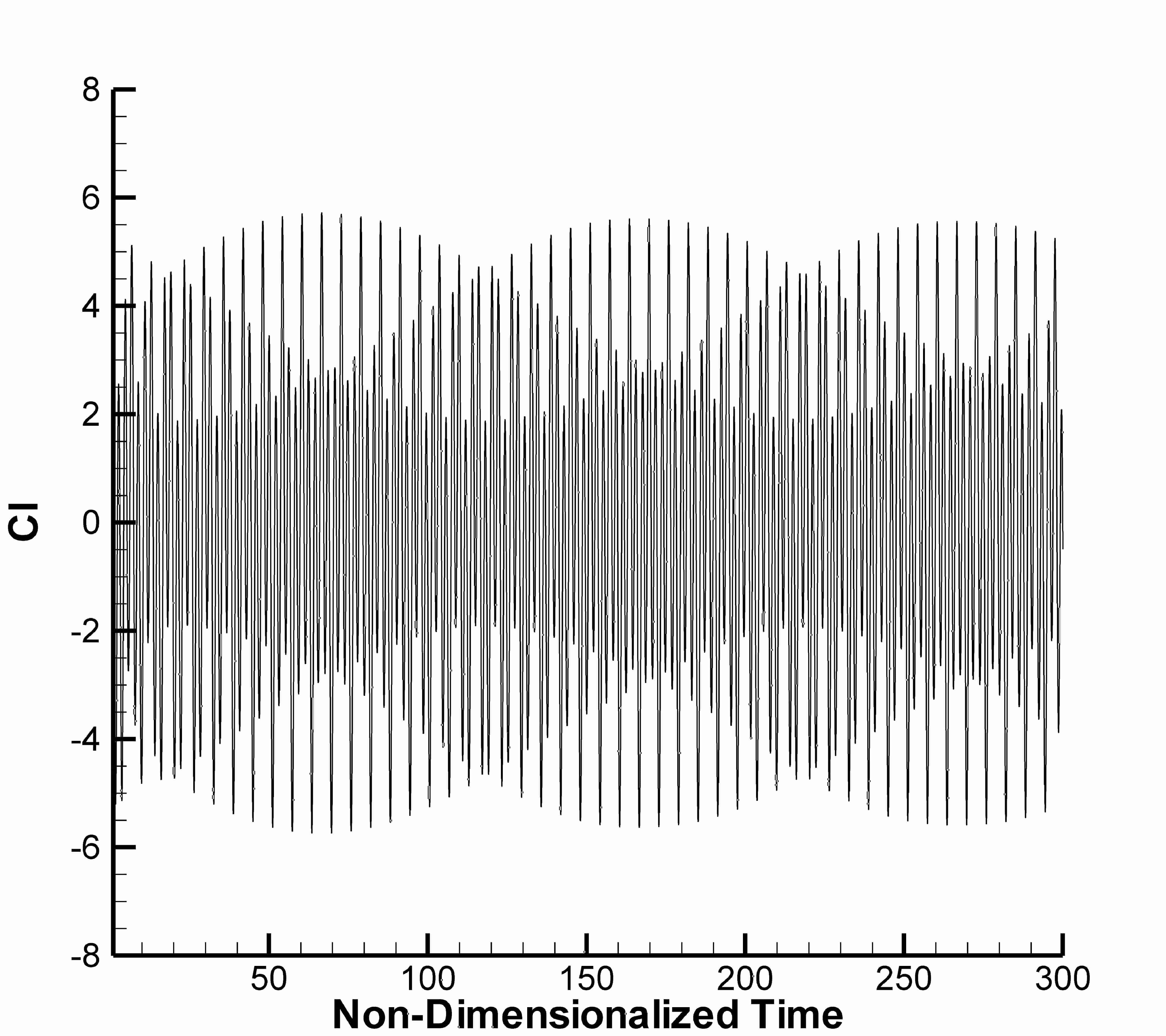} 
        \caption{$f_{tr} = 3.0$,$St_{r} = 0.165$}
        \label{fig:_1,2_Cl}
    \end{subfigure}
    \caption{Lift Coefficient at various transverse oscillation frequencies (Mode 1)}
    \end{figure}
    
    \begin{figure}[hp]
        \begin{subfigure}[b]{6.2cm}
        \includegraphics[scale=0.055]{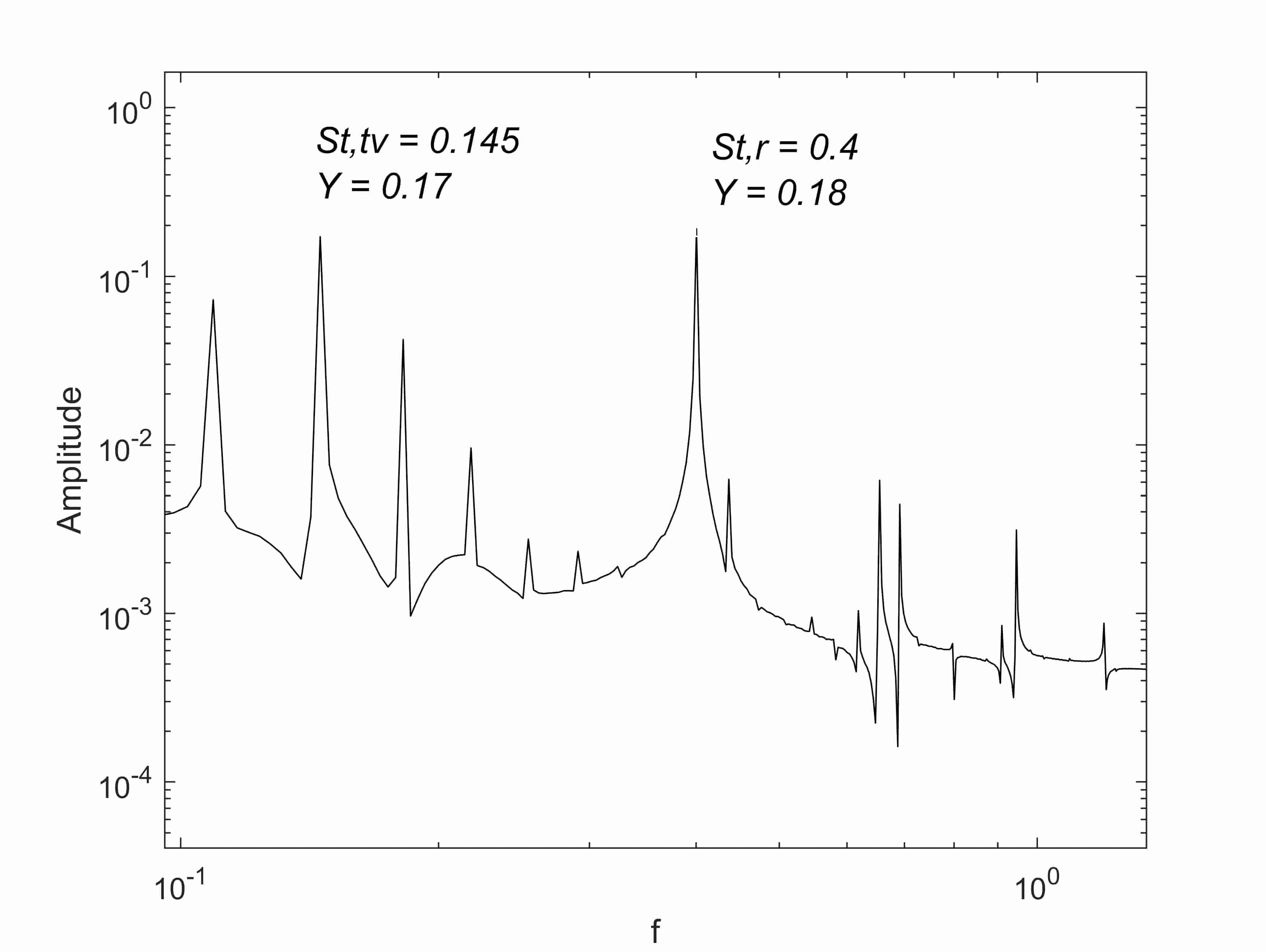} 
        \caption{$f_{tr}=0.9$,$St_{r}=0.3$}
        \label{fig:_2,3_}
        \end{subfigure}
        \hfill
    \begin{subfigure}[b]{6.2cm}
        \includegraphics[scale=0.055]{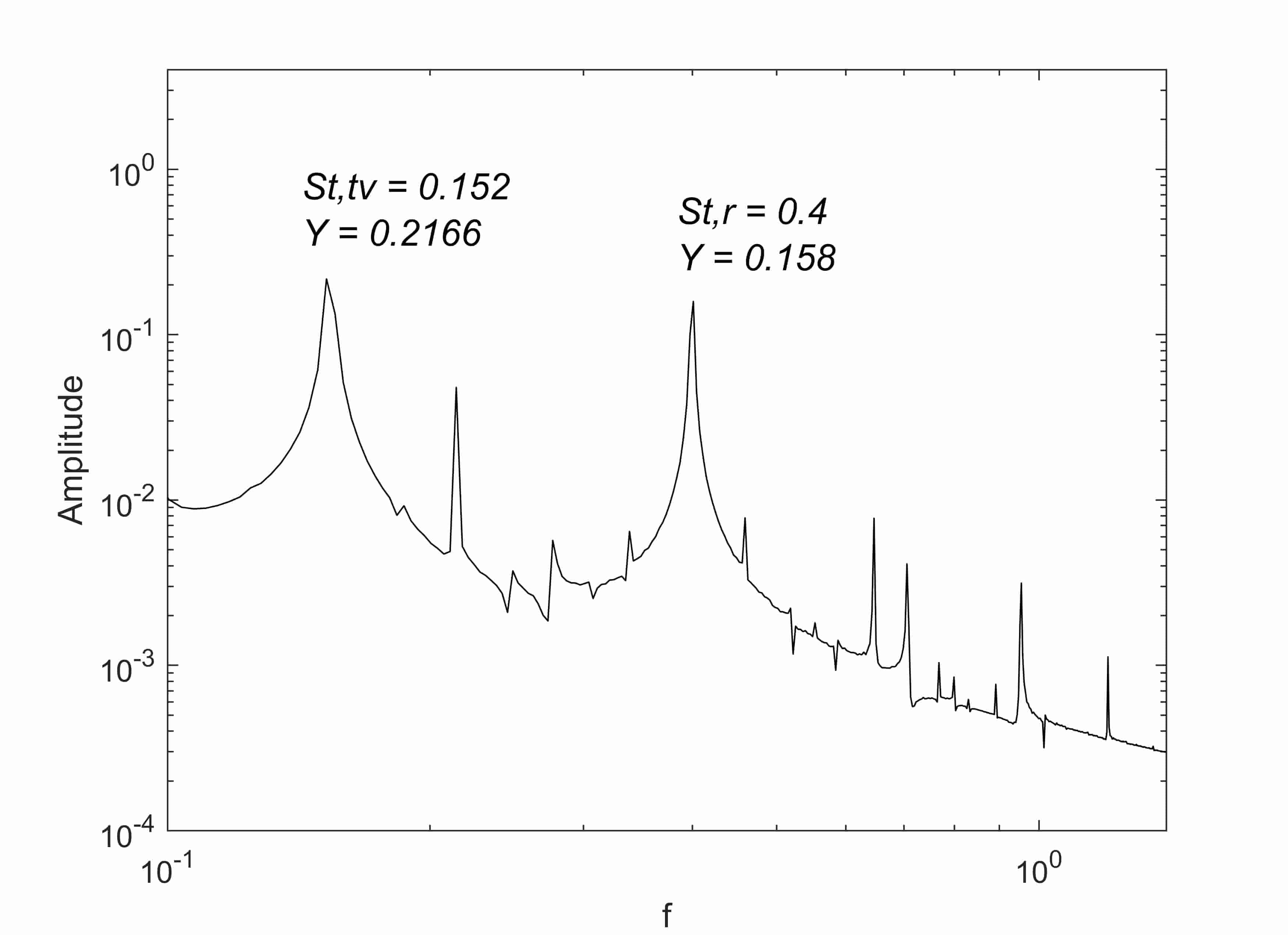} 
        \caption{$f_{tr}=0.95$,$St_{r}=0.4$}
        \label{fig:_2,4_}
    \end{subfigure}
        \hfill
    \begin{subfigure}[b]{6.2cm}
        \includegraphics[scale=0.055]{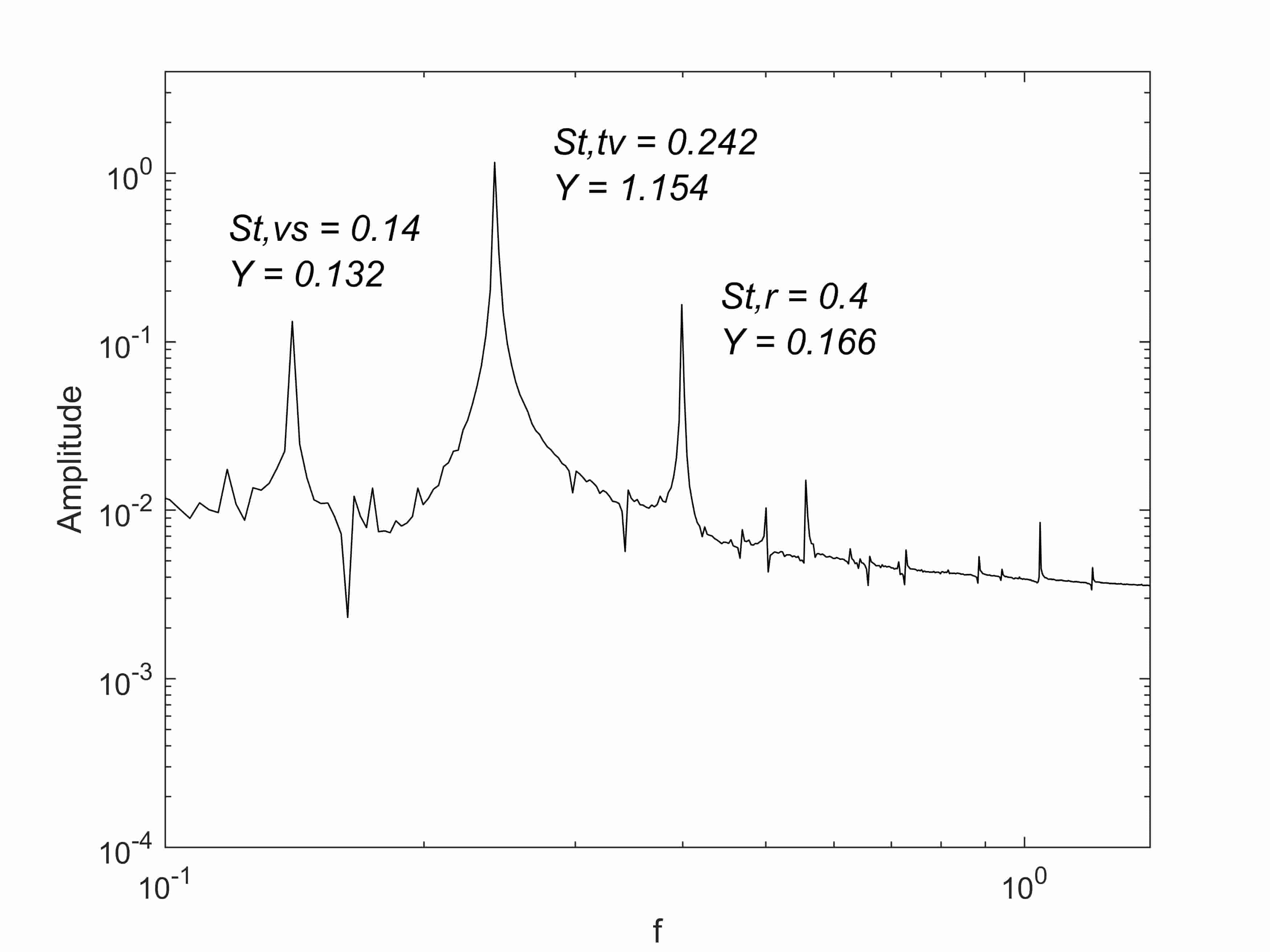} 
        \caption{$f_{tr} = 1.5$,$St_{r} = 0.4$}
        \label{fig:_2,5_}
    \end{subfigure}
    \hfill
    \begin{subfigure}[b]{6.2cm}
        \includegraphics[scale=0.055]{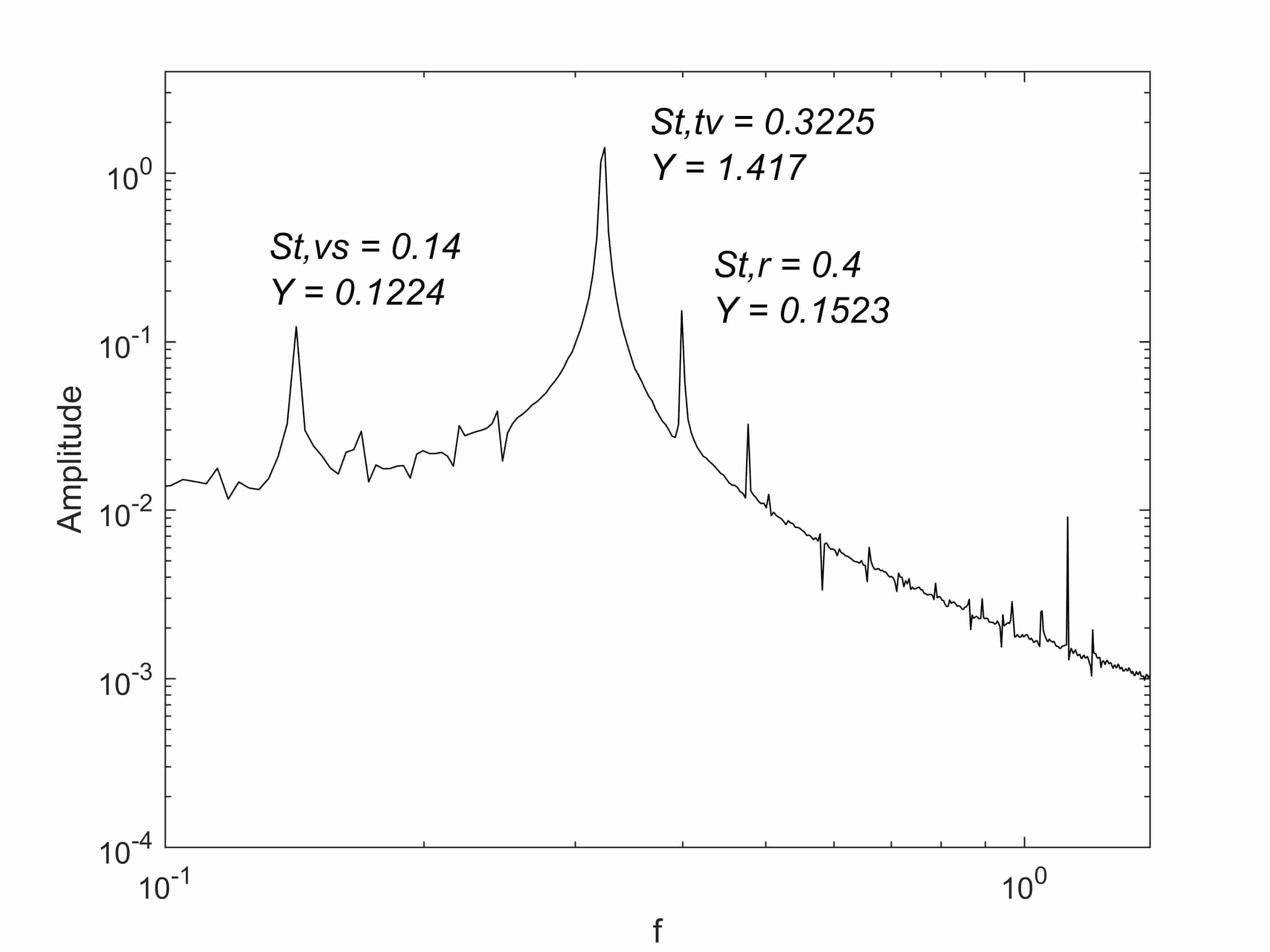} 
        \caption{$f_{tr} = 2.0$,$St_{r} = 0.4$}
        \label{fig:_2,1_}
    \end{subfigure}
    \hfill
    \begin{subfigure}[b]{6.2cm}
        \centering
        \includegraphics[scale=0.055]{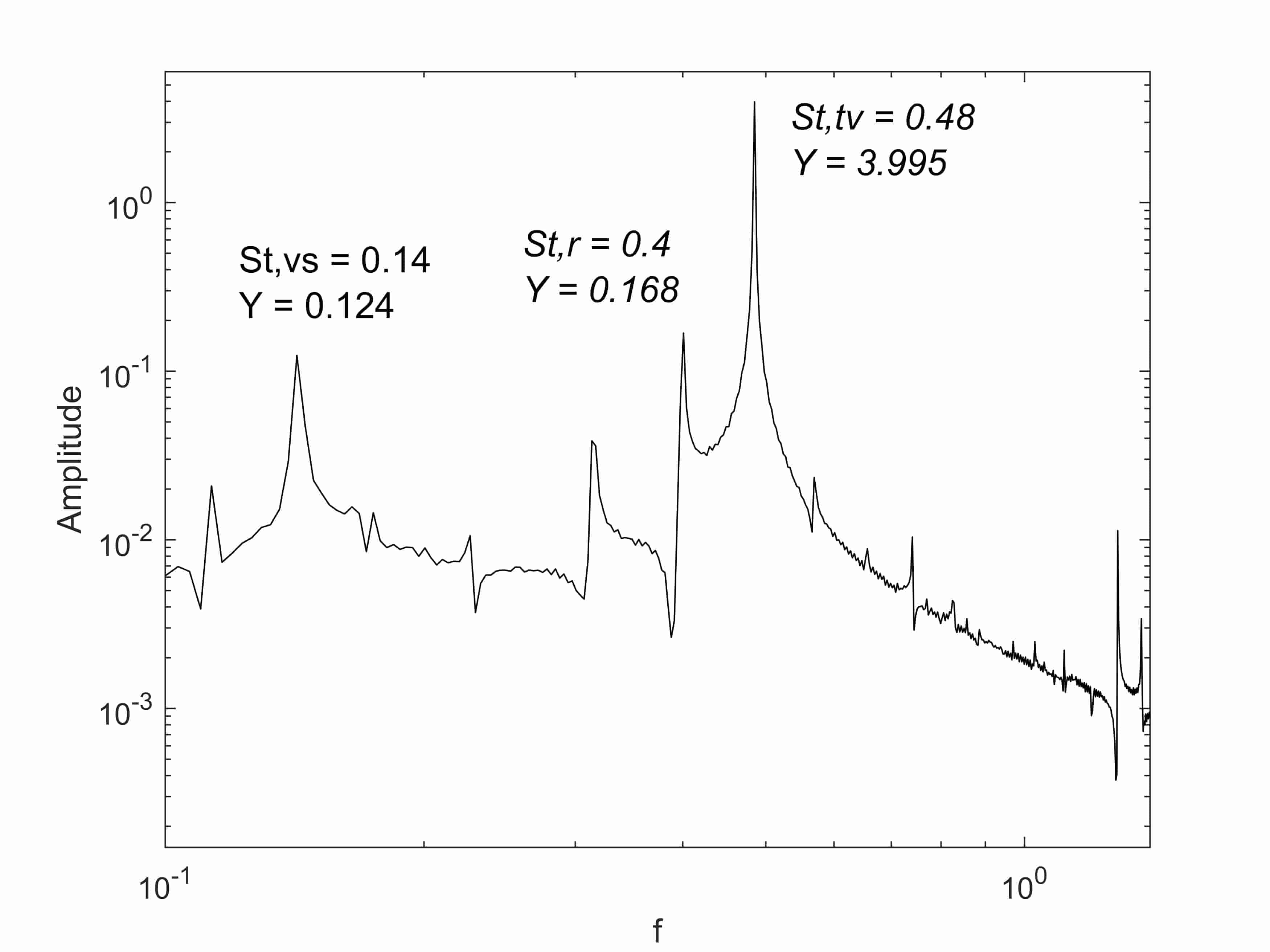} 
        \caption{$f_{tr} = 3.0$,$St_{r} = 0.4$}
        \label{fig:_2,2_}
    \end{subfigure}
    \caption{Frequency Characteristics of Lift at various transverse oscillation frequencies (Mode 2)}
    \end{figure}
    
     \begin{figure}[hp]
        \begin{subfigure}[b]{6.2cm}
        \includegraphics[scale=0.04]{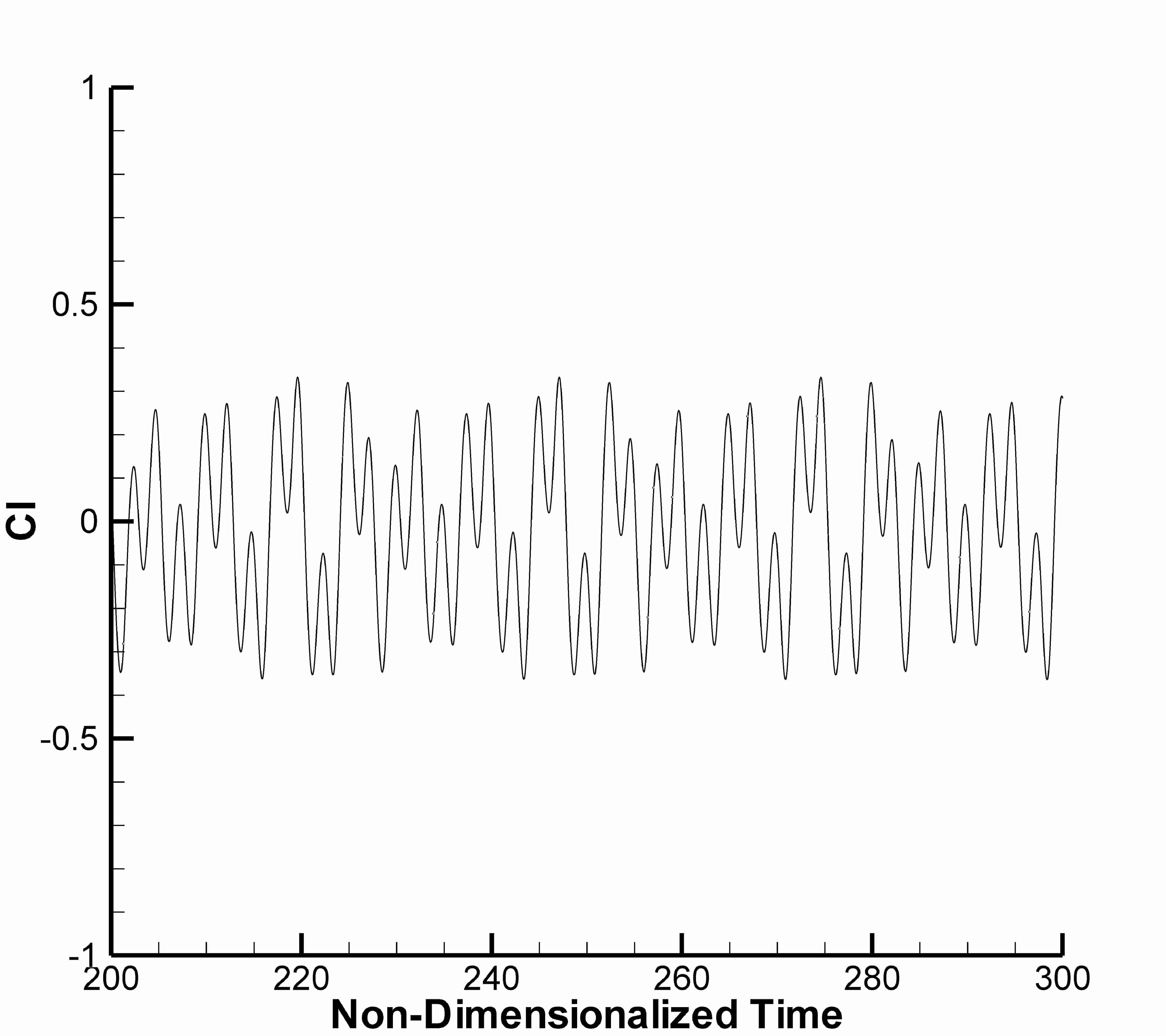} 
        \caption{$f_{tr}=0.9$,$St_{r}=0.4$}
        \label{fig:_2,3_Cl}
        \end{subfigure}
        \hfill
    \begin{subfigure}[b]{6.2cm}
        \includegraphics[scale=0.04]{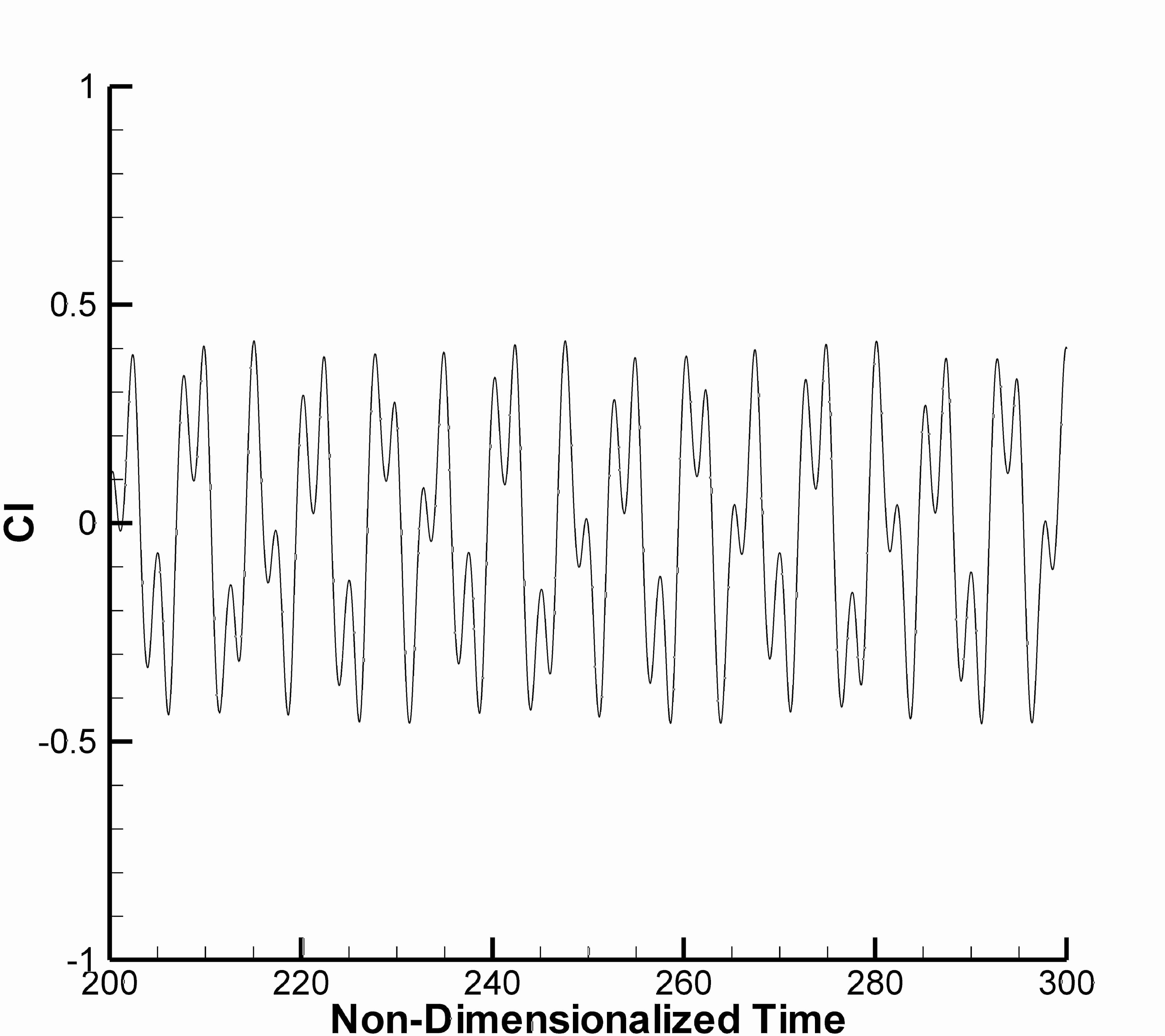} 
        \caption{$f_{tr}=0.95$,$St_{r}=0.4$}
        \label{fig:_2,4_Cl}
    \end{subfigure}
        \hfill
    \begin{subfigure}[b]{6.2cm}
        \includegraphics[scale=0.04]{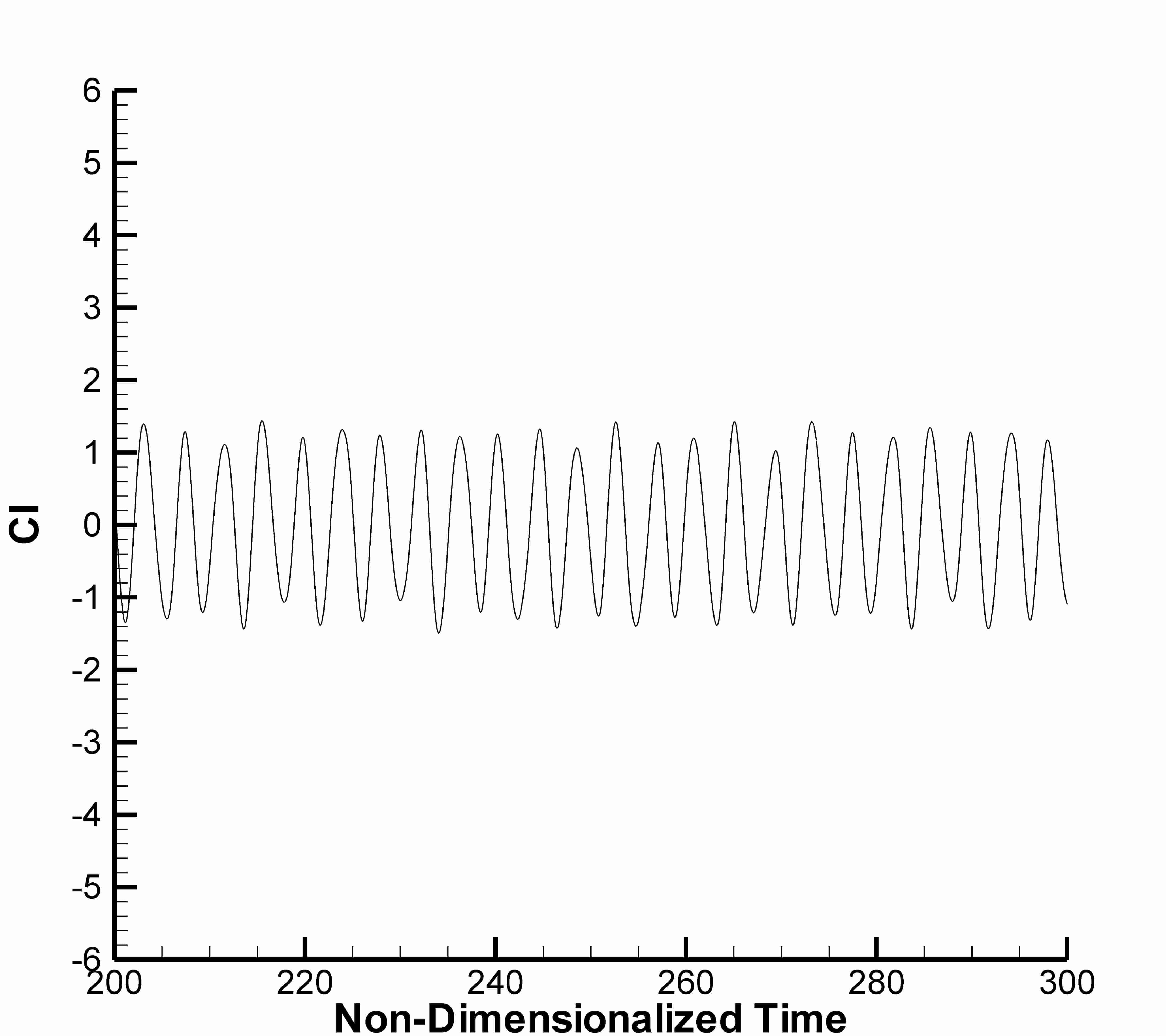} 
        \caption{$f_{tr} = 1.5$,$St_{r} = 0.4$}
        \label{fig:_2,5_Cl}
    \end{subfigure}
    \hfill
    \begin{subfigure}[b]{6.2cm}
        \includegraphics[scale=0.04]{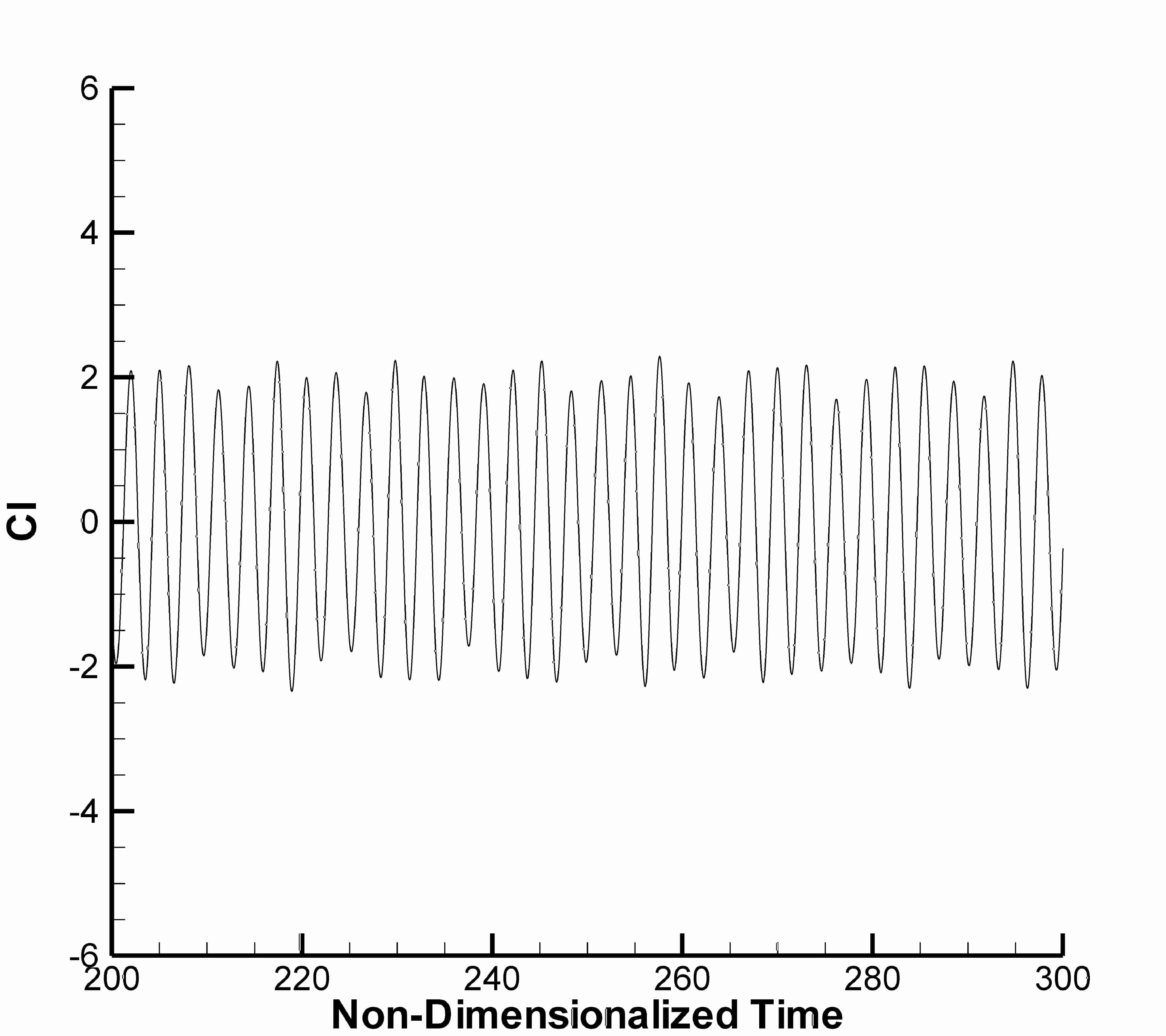} 
        \caption{$f_{tr} = 2.0$,$St_{r} = 0.4$}
        \label{fig:_2,1_Cl}
    \end{subfigure}
    \hfill
    \begin{subfigure}[b]{6.2cm}
        \centering
        \includegraphics[scale=0.04]{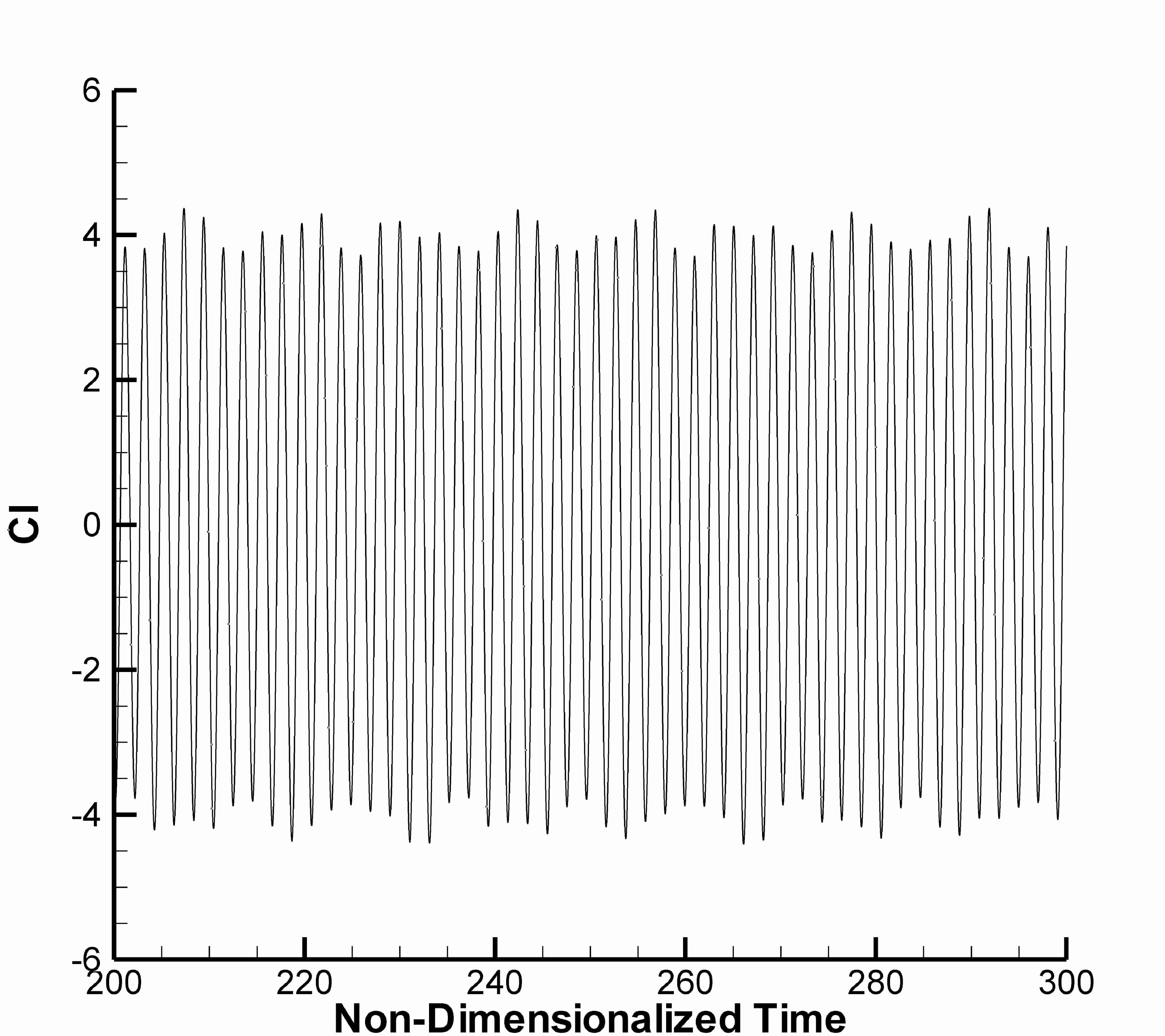} 
        \caption{$f_{tr} = 3.0$,$St_{r} = 0.4$}
        \label{fig:_2,2_Cl}
    \end{subfigure}
    \caption{Lift Coefficient at various transverse oscillation frequencies (Mode 2)}
    \end{figure}
    
    \begin{figure}[hp]
        \begin{subfigure}[b]{6.2cm}
        \includegraphics[scale=0.055]{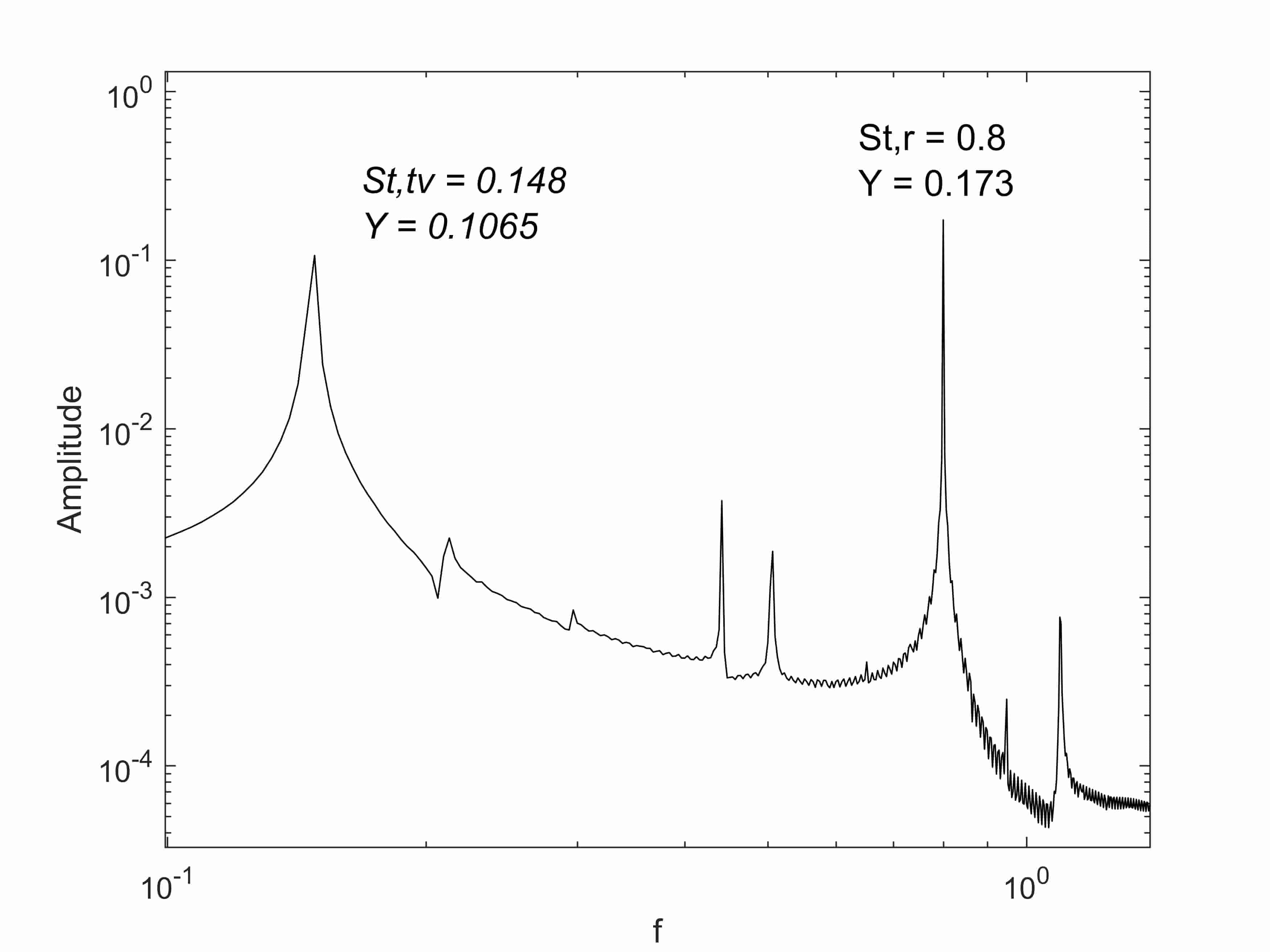} 
        \caption{$f_{tr}=0.9$,$St_{r}=0.8$}
        \label{fig:_3,3_}
        \end{subfigure}
        \hfill
    \begin{subfigure}[b]{6.2cm}
        \includegraphics[scale=0.055]{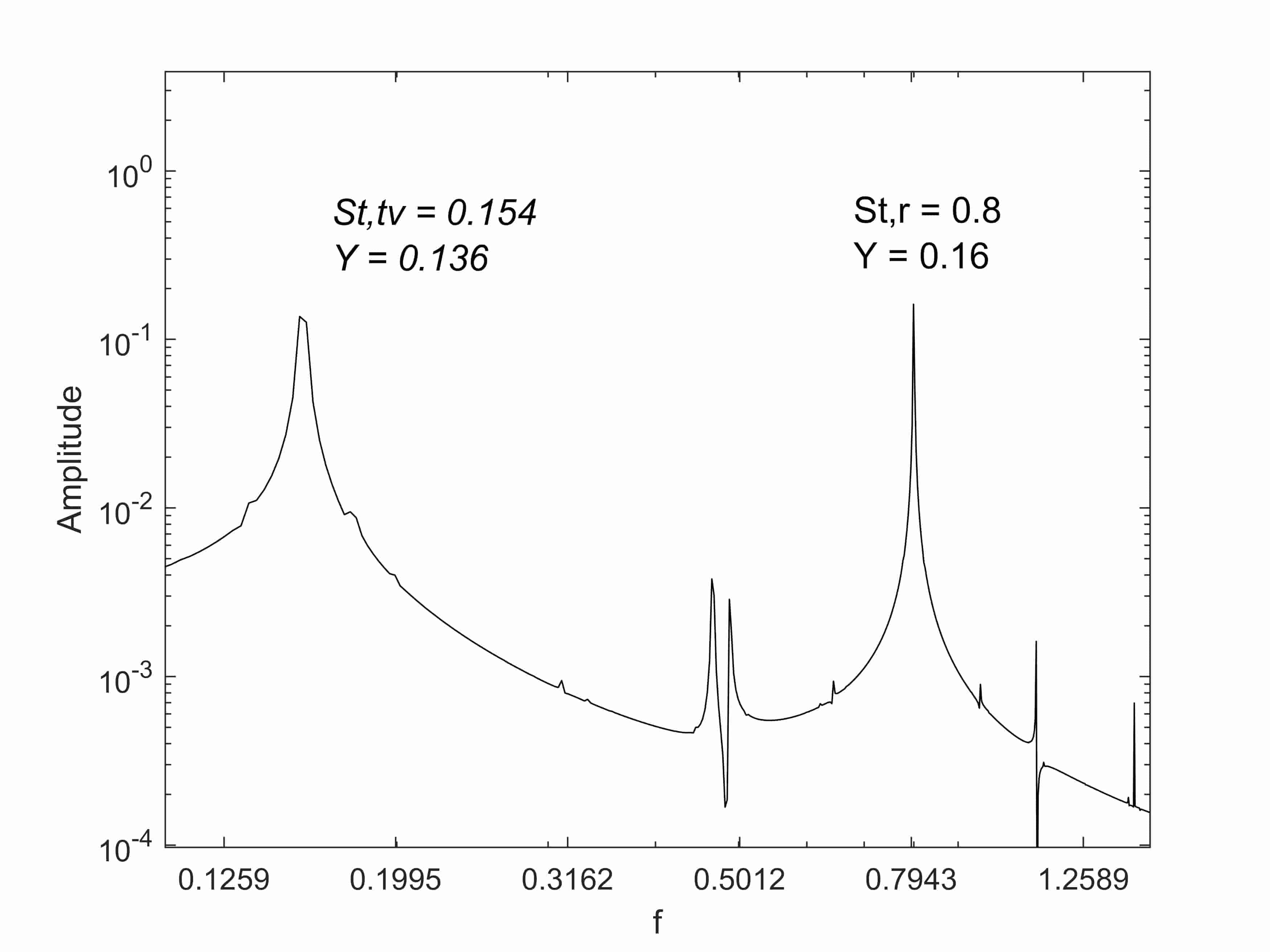} 
        \caption{$f_{tr}=0.95$,$St_{r}=0.8$}
        \label{fig:_3,4_}
    \end{subfigure}
        \hfill
    \begin{subfigure}[b]{6.2cm}
        \includegraphics[scale=0.055]{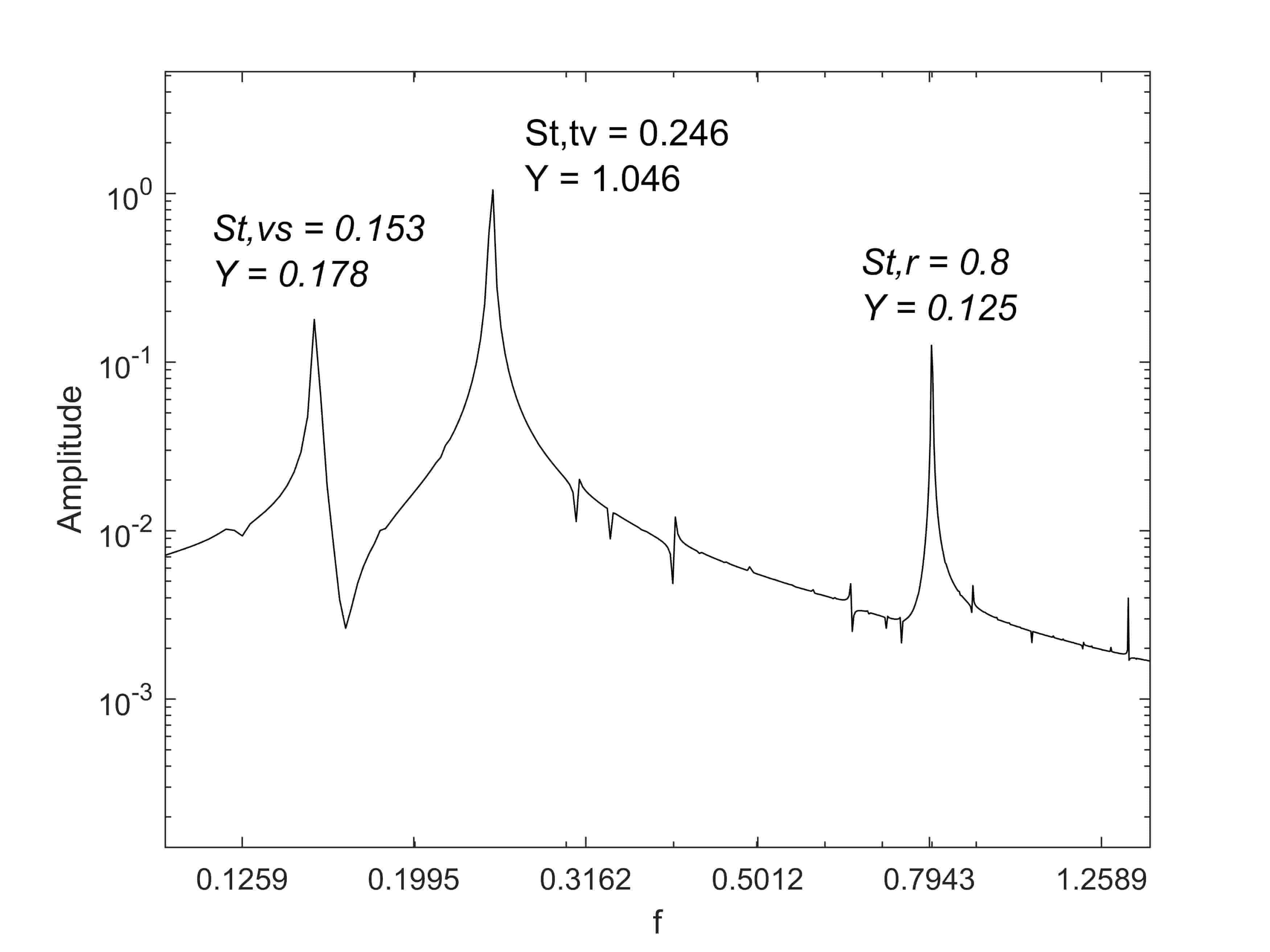} 
        \caption{$f_{tr} = 1.5$,$St_{r} = 0.8$}
        \label{fig:_3,5_}
    \end{subfigure}
    \hfill
    \begin{subfigure}[b]{6.2cm}
        \includegraphics[scale=0.055]{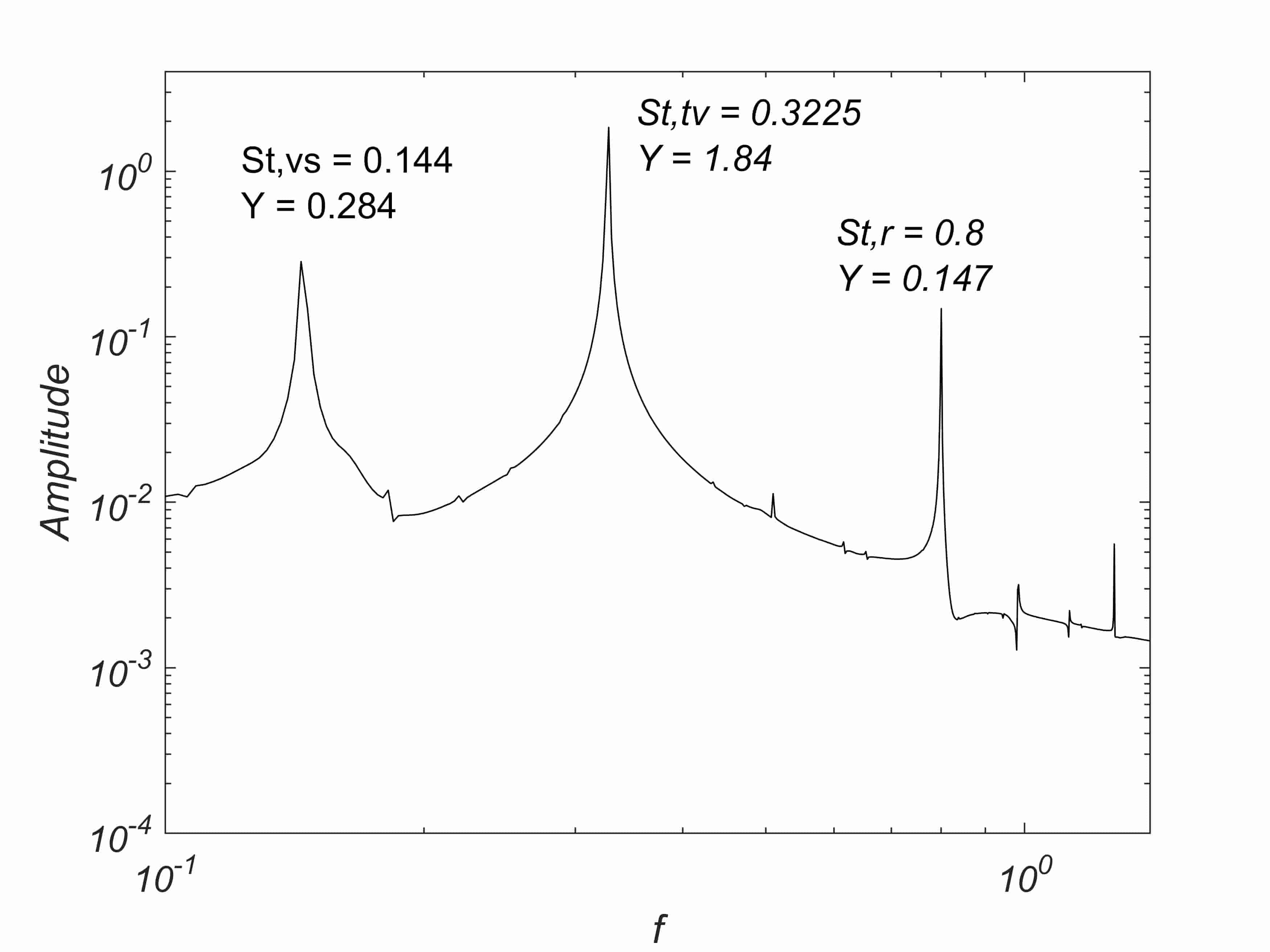} 
        \caption{$f_{tr} = 2.0$,$St_{r} = 0.8$}
        \label{fig:_3,1_}
    \end{subfigure}
    \hfill
    \begin{subfigure}[b]{6.2cm}
        \centering
        \includegraphics[scale=0.055]{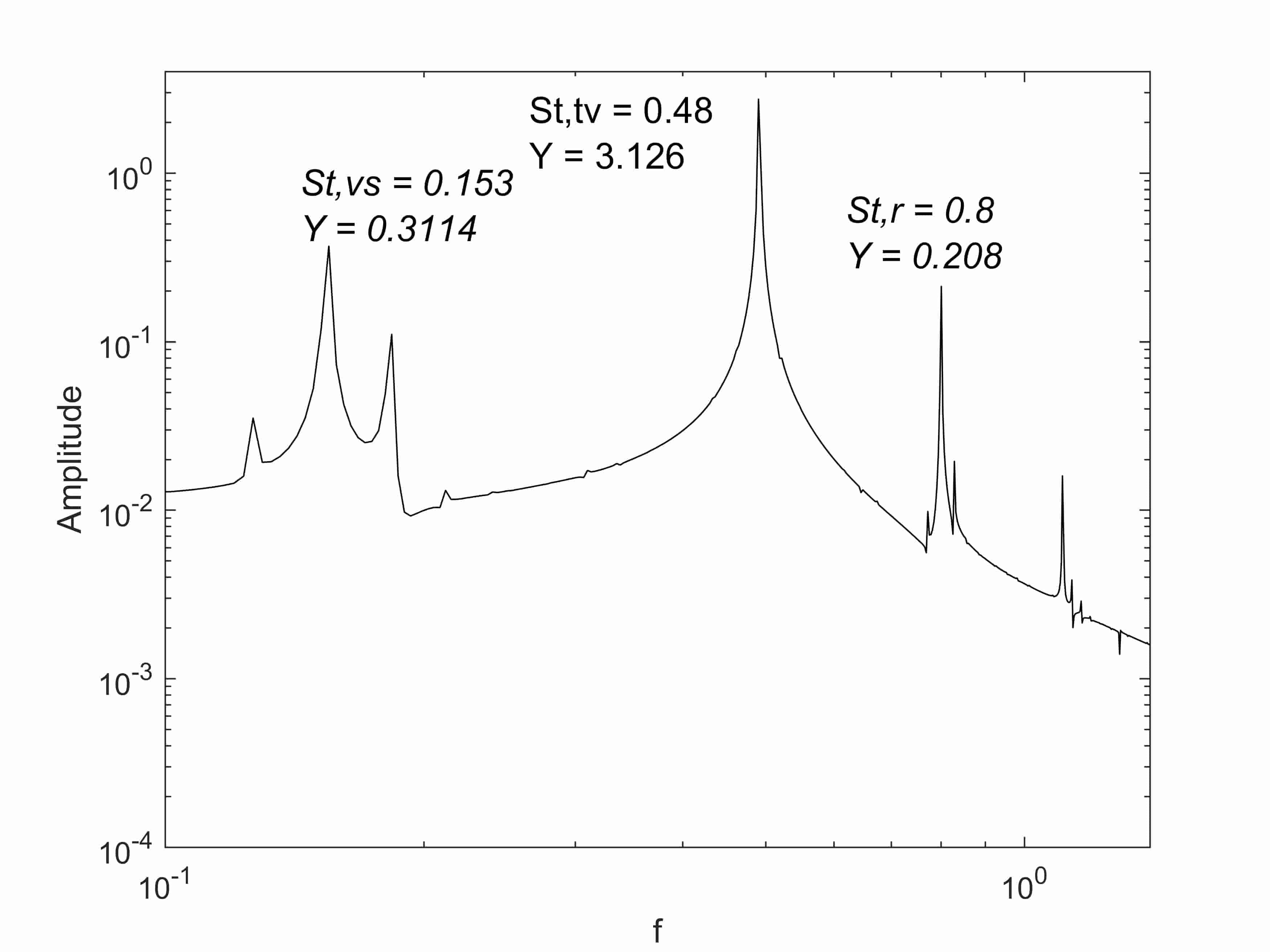} 
        \caption{$f_{tr} = 3.0$,$St_{r} = 0.8$}
        \label{fig:_3,2_}
    \end{subfigure}
    \caption{Frequency Characteristics of Lift at various transverse oscillation frequencies (Mode 3)}
    \end{figure}

    \begin{figure}[hp]
        \begin{subfigure}[b]{6.2cm}
        \includegraphics[scale=0.04]{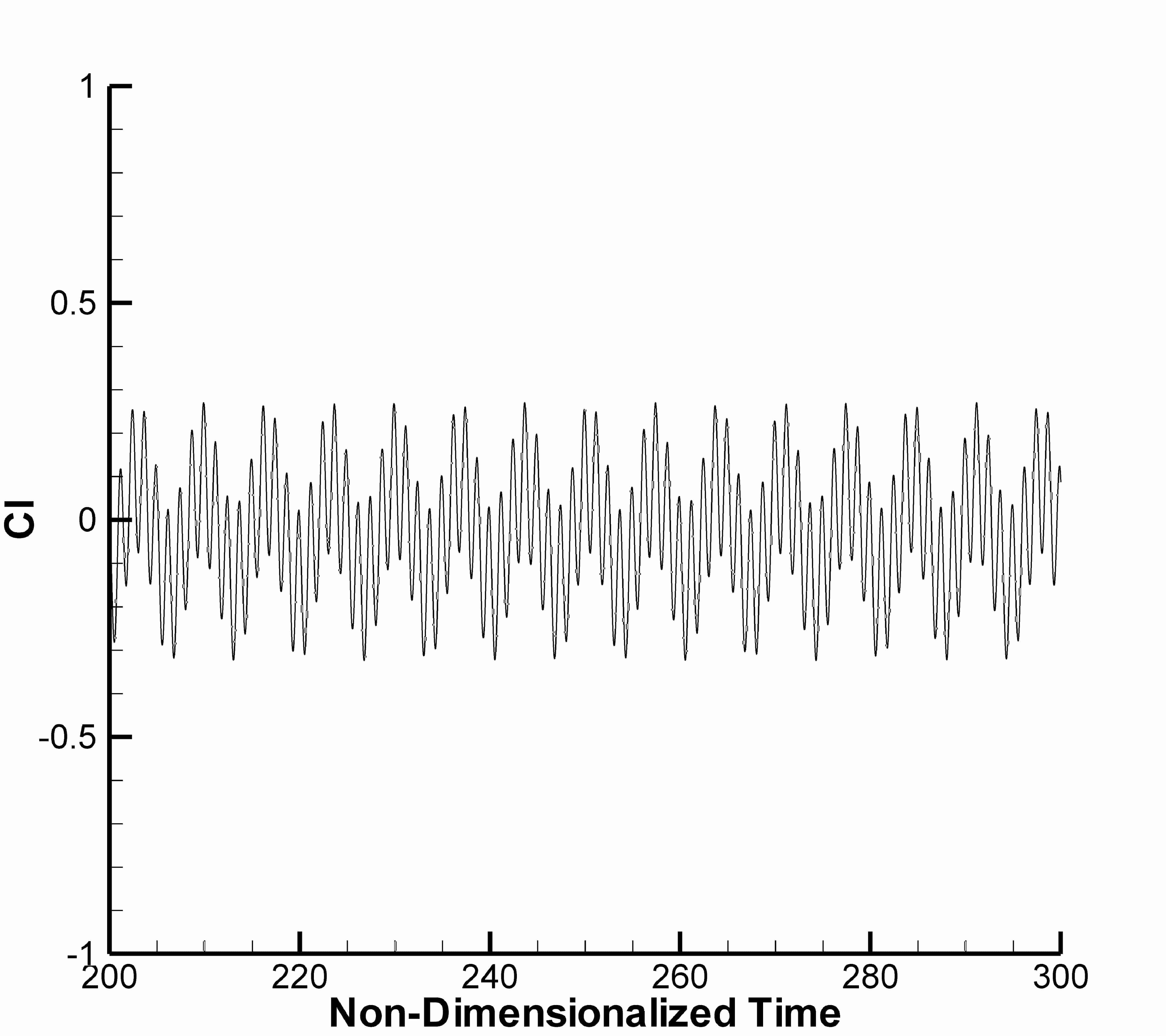} 
        \caption{$f_{tr}=0.9$,$St_{r}=0.8$}
        \label{fig:_3,3_Cl}
        \end{subfigure}
        \hfill
    \begin{subfigure}[b]{6.2cm}
        \includegraphics[scale=0.04]{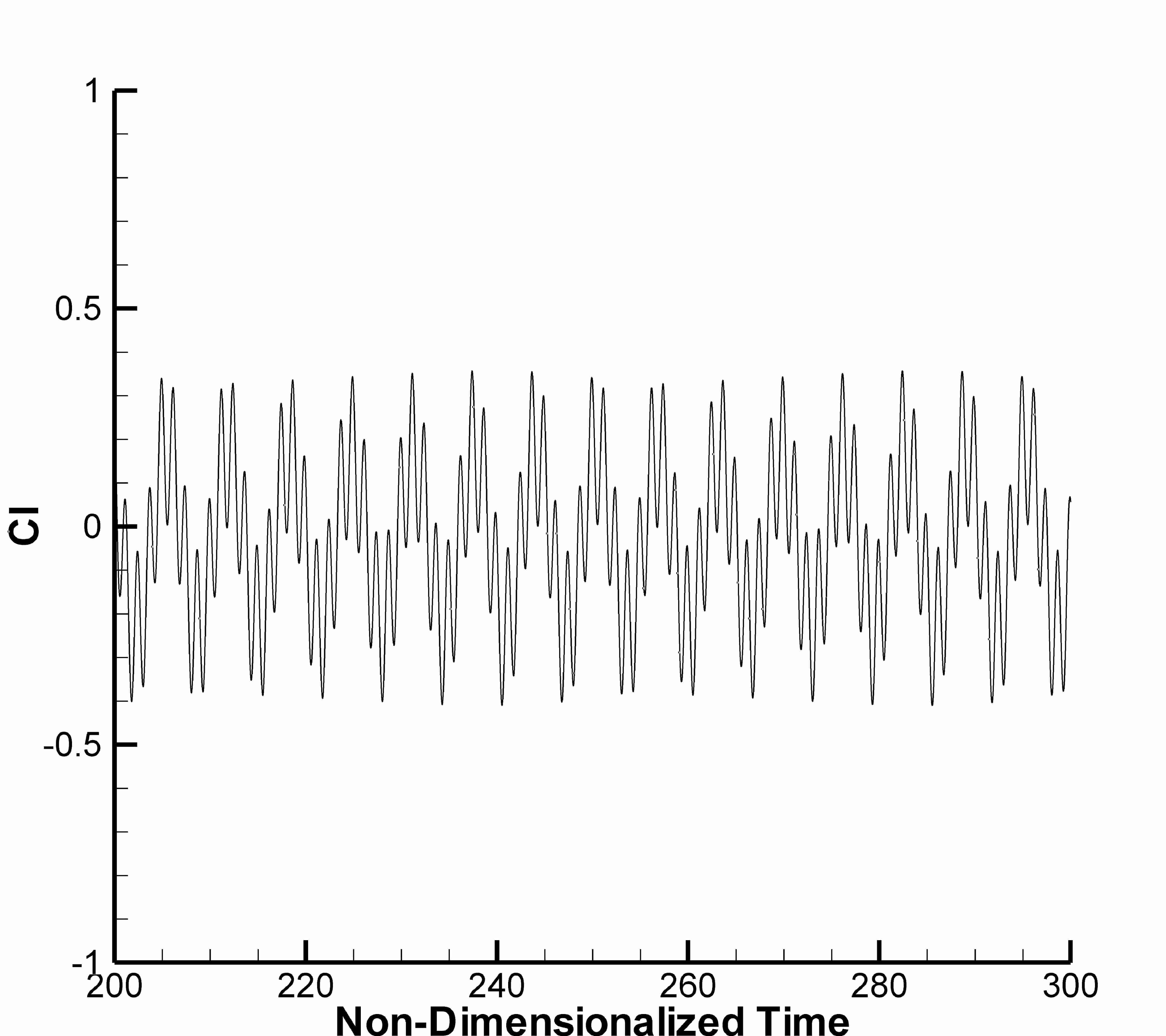} 
        \caption{$f_{tr}=0.95$,$St_{r}=0.8$}
        \label{fig:_3,4_Cl}
    \end{subfigure}
        \hfill
    \begin{subfigure}[b]{6.2cm}
        \includegraphics[scale=0.04]{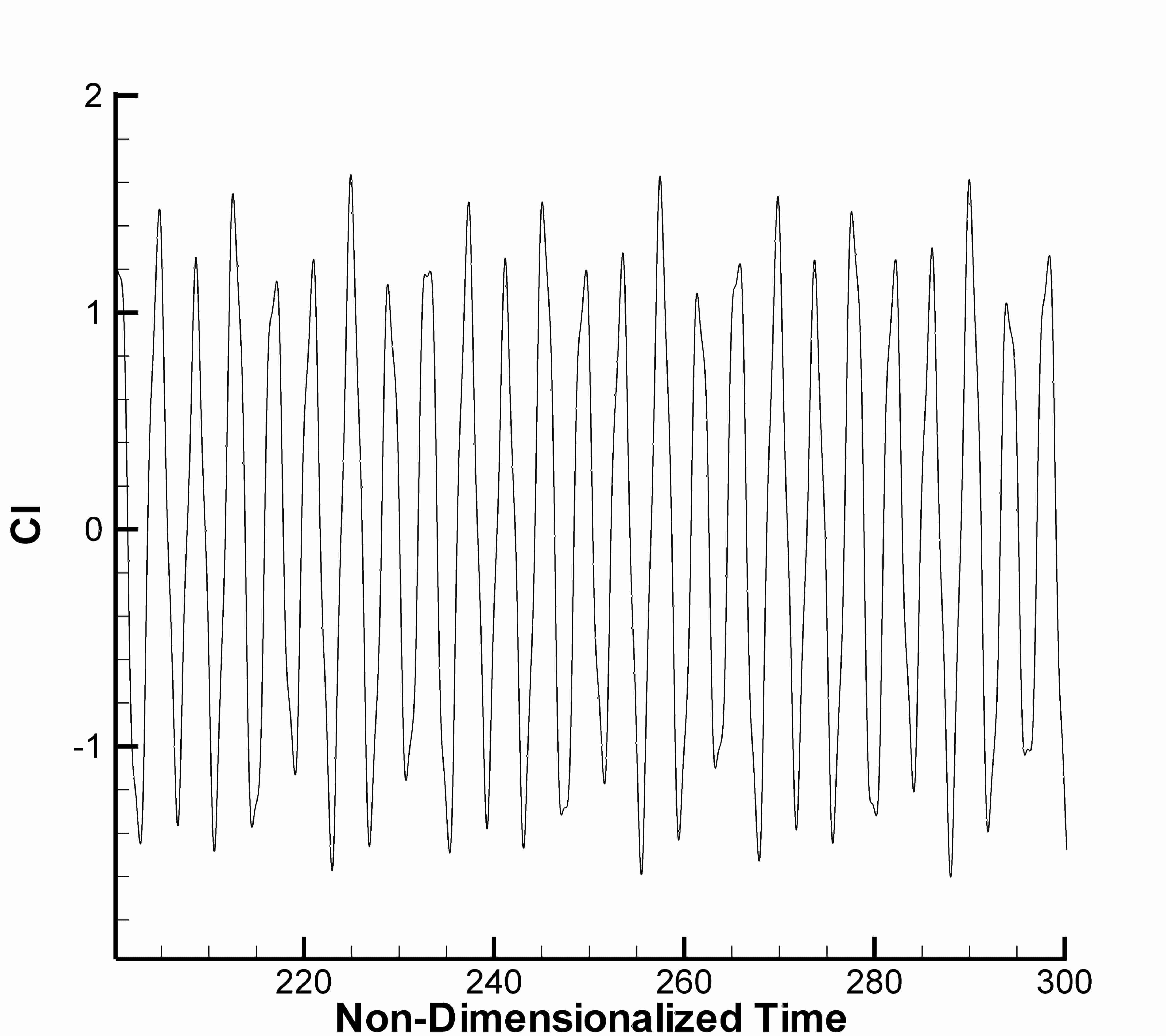} 
        \caption{$f_{tr} = 1.5$,$St_{r} = 0.8$}
        \label{fig:_3,5_Cl}
    \end{subfigure}
    \hfill
    \begin{subfigure}[b]{6.2cm}
        \includegraphics[scale=0.04]{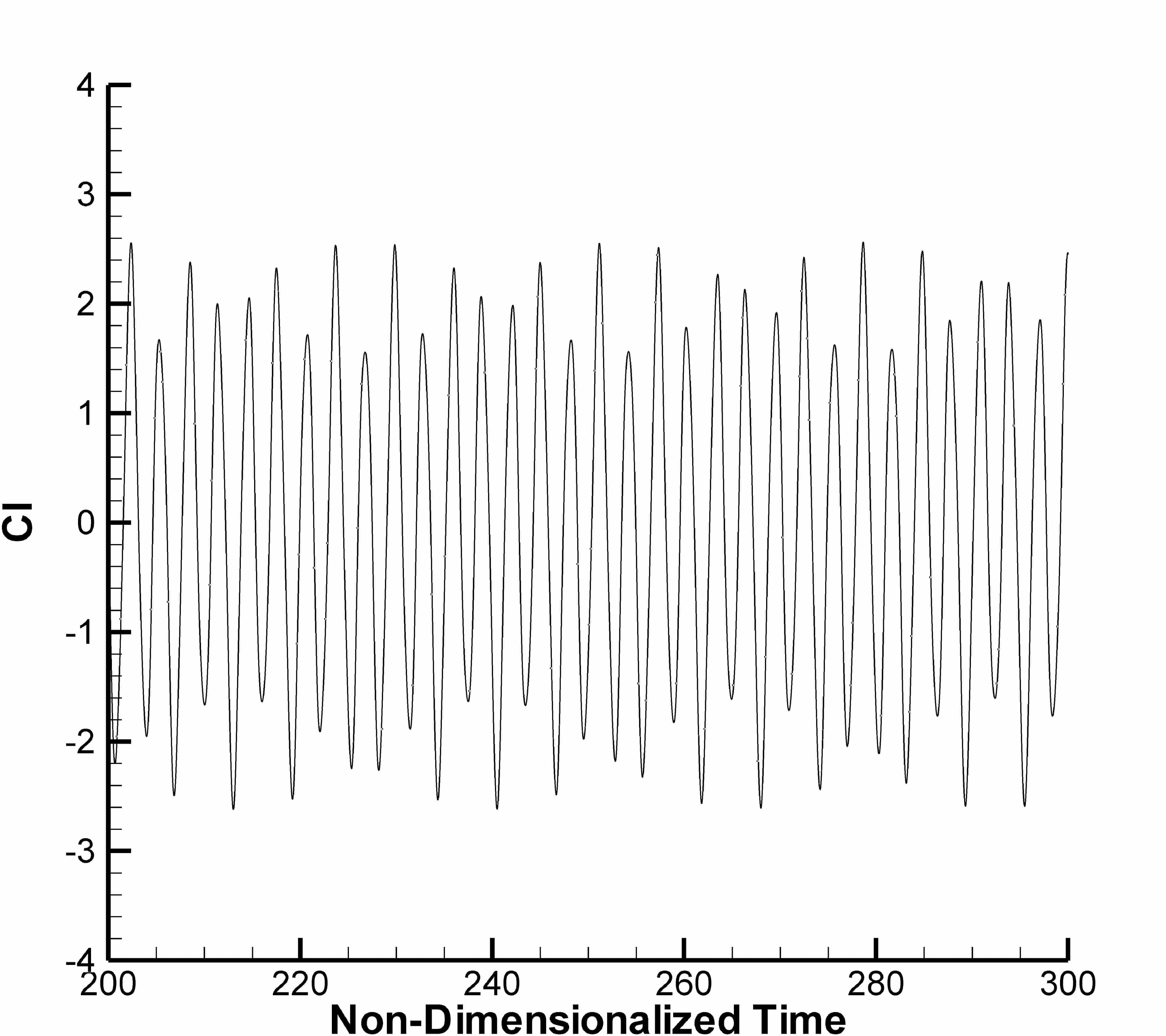} 
        \caption{$f_{tr} = 2.0$,$St_{r} = 0.8$}
        \label{fig:_3,1_Cl}
    \end{subfigure}
    \hfill
    \begin{subfigure}[b]{6.2cm}
        \centering
        \includegraphics[scale=0.04]{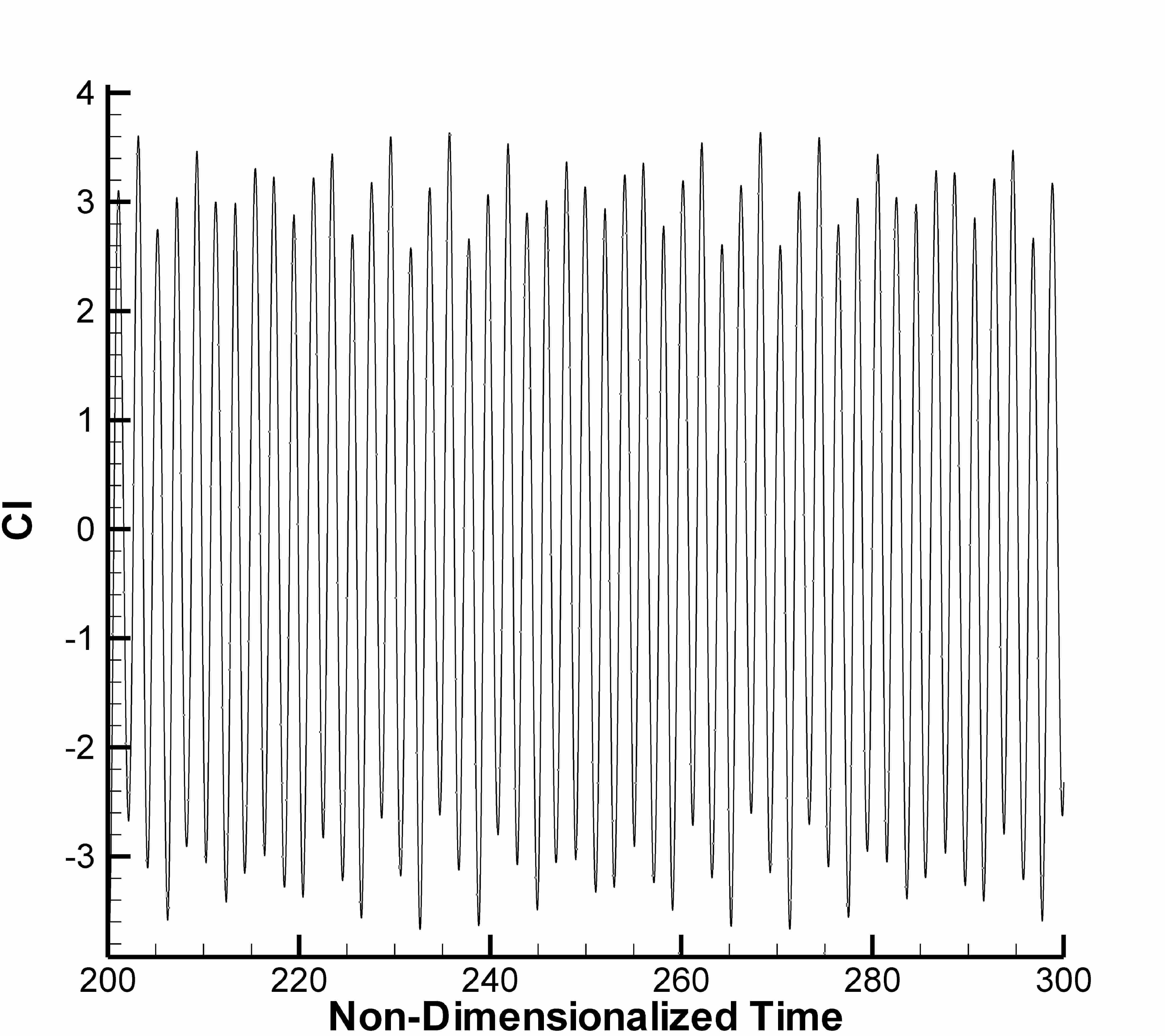} 
        \caption{$f_{tr} = 3.0$,$St_{r} = 0.8$}
        \label{fig:_3,2_Cl}
    \end{subfigure}
    \caption{ Lift Coefficient at various transverse oscillation frequencies (Mode 3)}
    \end{figure}
    
    \begin{figure}[hp]
        \begin{subfigure}[b]{6.2cm}
        \includegraphics[scale=0.055]{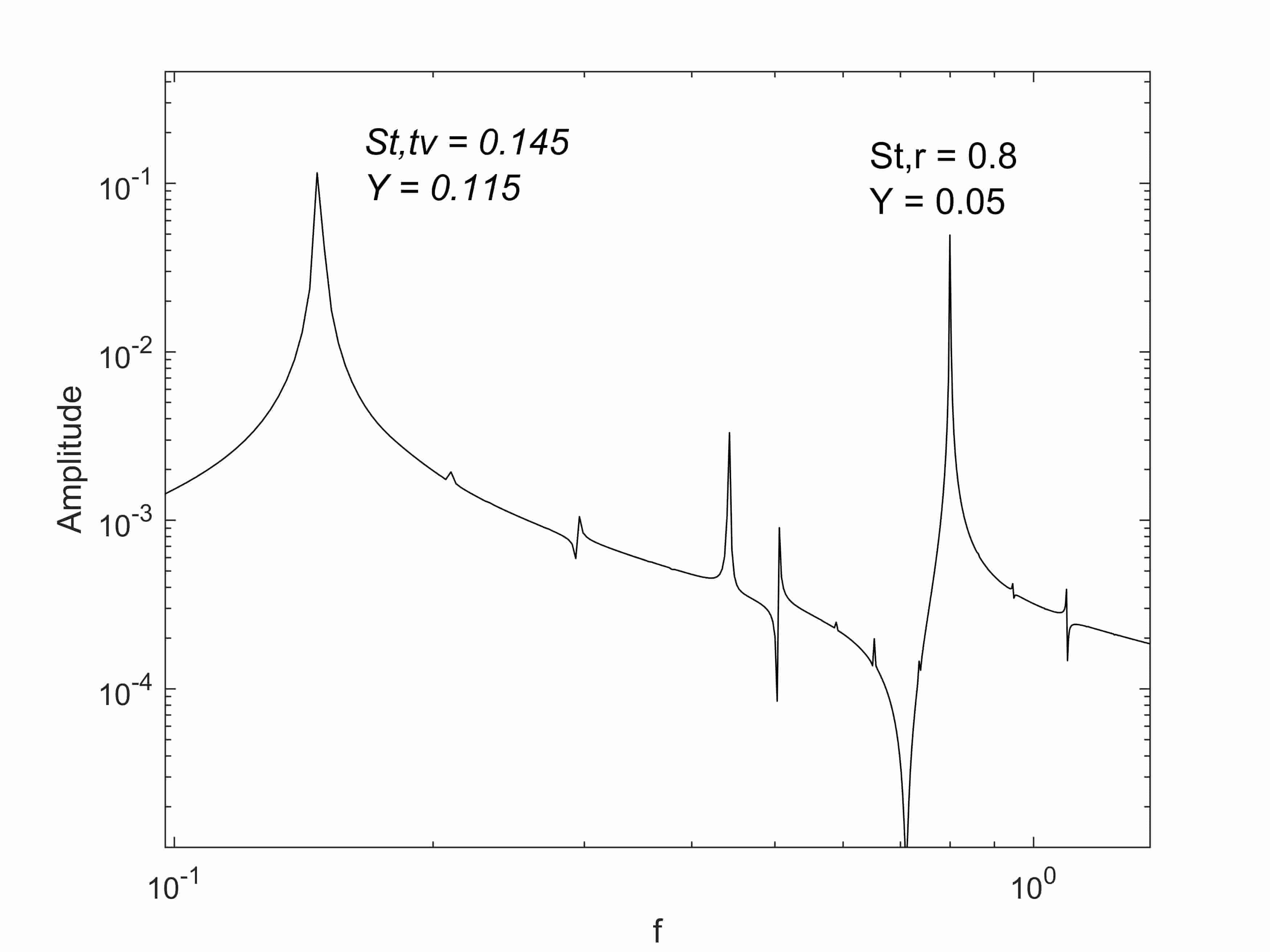} 
        \caption{$f_{tr}=0.9$,$St_{r}=0.8$}
        \label{fig:_4,3_}
        \end{subfigure}
        \hfill
    \begin{subfigure}[b]{6.2cm}
        \includegraphics[scale=0.055]{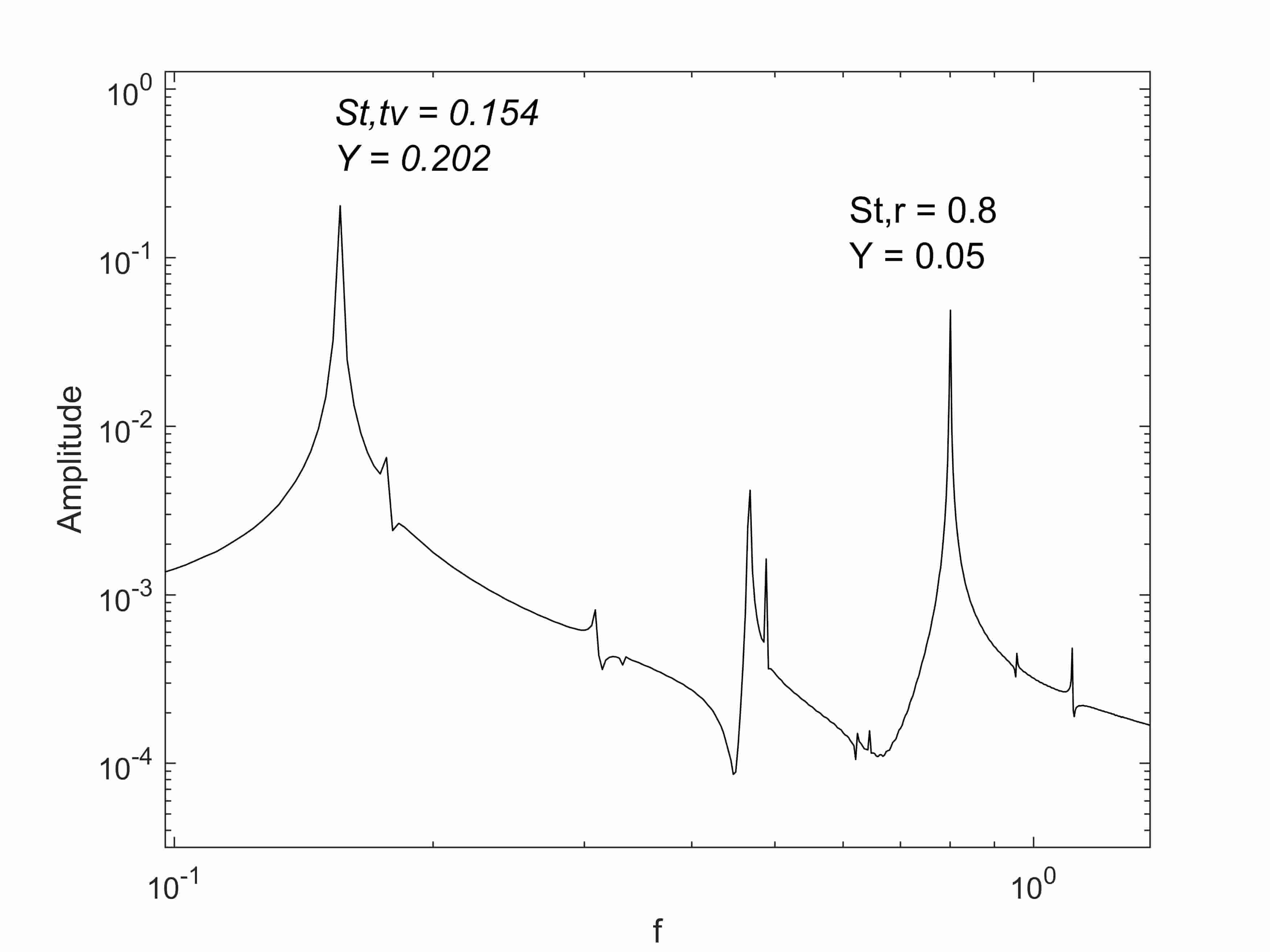} 
        \caption{$f_{tr}=0.95$,$St_{r}=0.8$}
        \label{fig:_4,4_}
    \end{subfigure}
        \hfill
    \begin{subfigure}[b]{6.2cm}
        \includegraphics[scale=0.055]{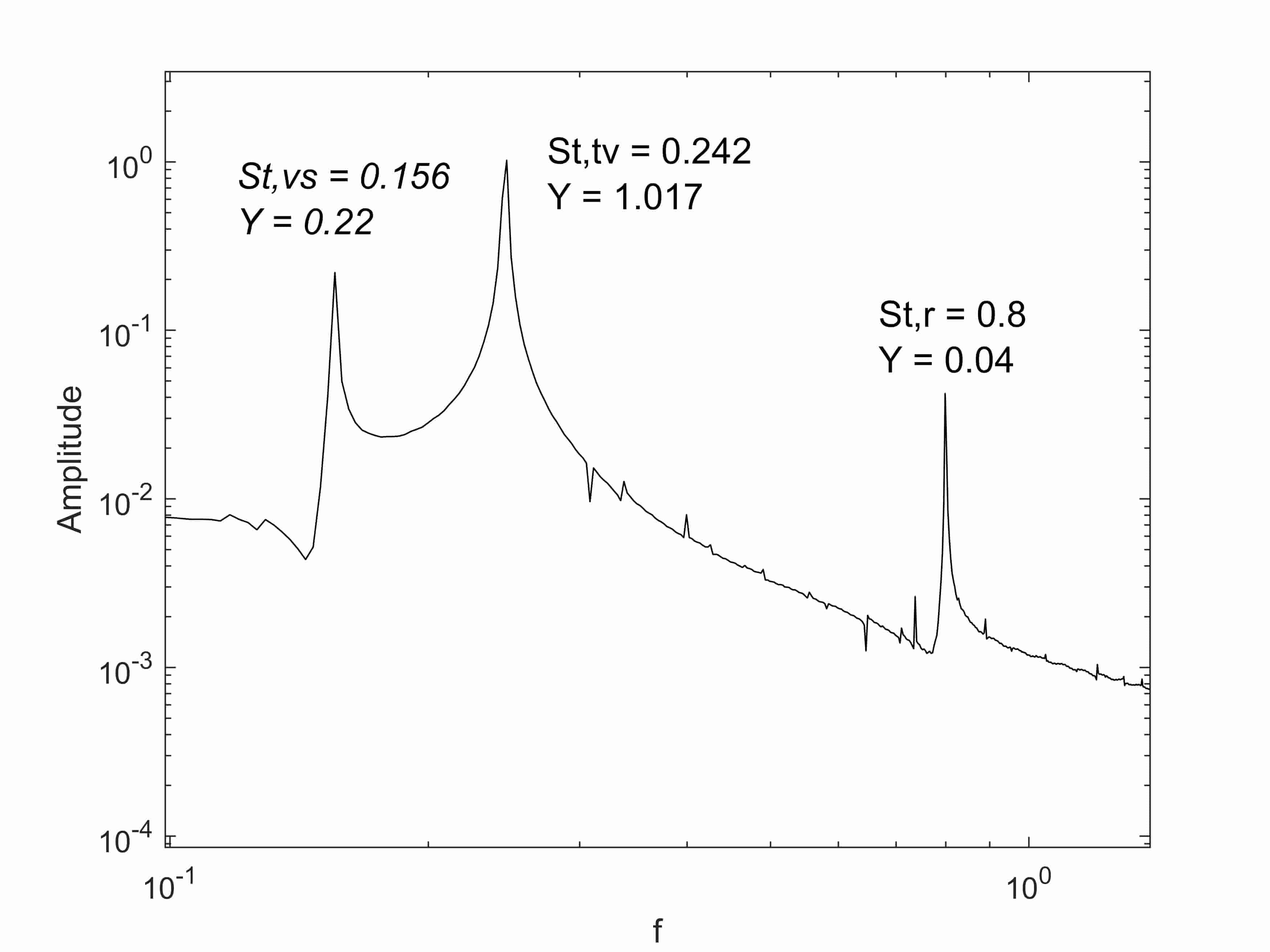} 
        \caption{$f_{tr} = 1.5$,$St_{r} = 0.8$}
        \label{fig:_4,5_}
    \end{subfigure}
    \hfill
    \begin{subfigure}[b]{6.2cm}
        \includegraphics[scale=0.055]{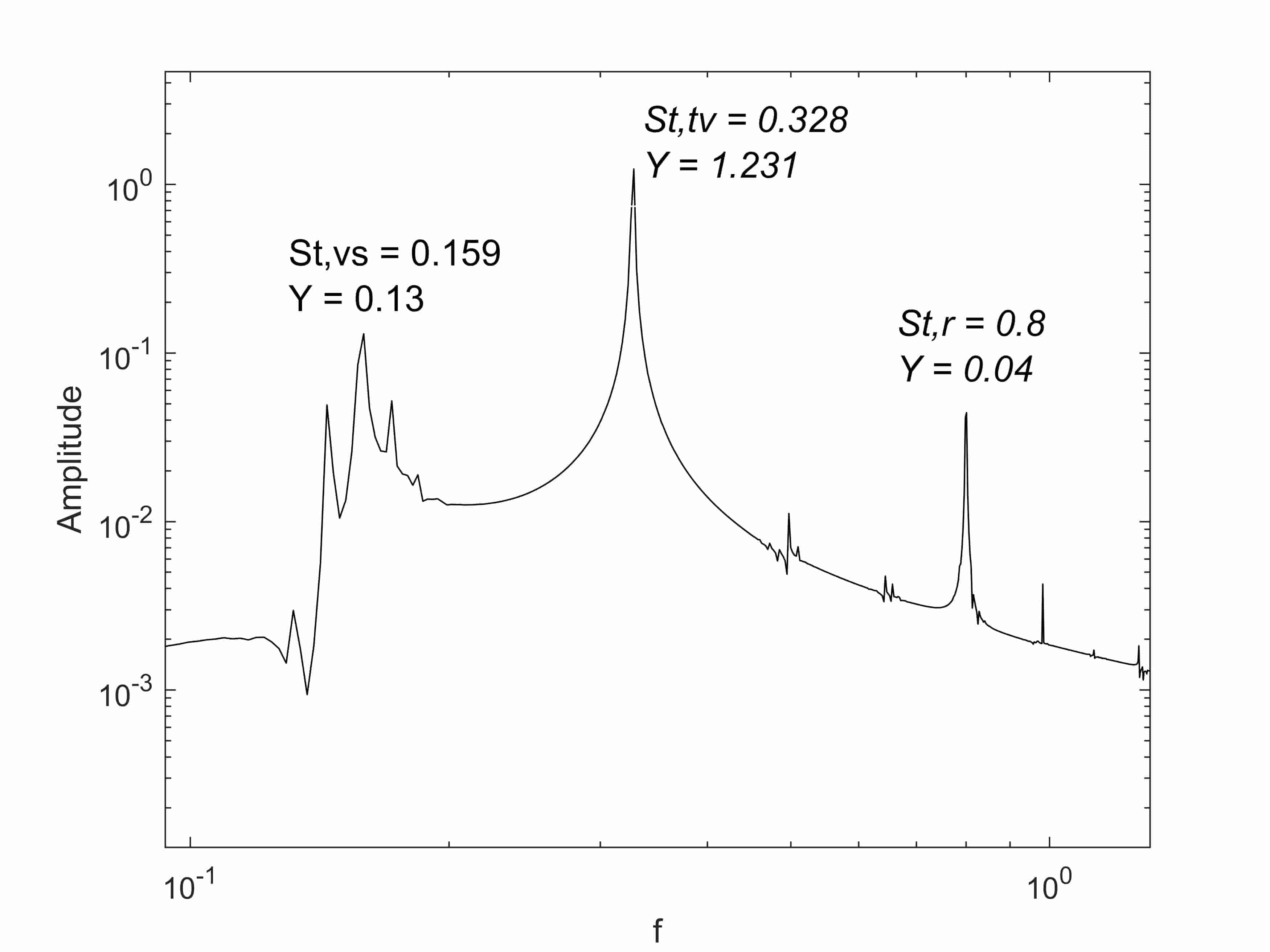} 
        \caption{$f_{tr} = 2.0$,$St_{r} = 0.8$}
        \label{fig:_4,1_}
    \end{subfigure}
    \hfill
    \begin{subfigure}[b]{6.2cm}
        \centering
        \includegraphics[scale=0.055]{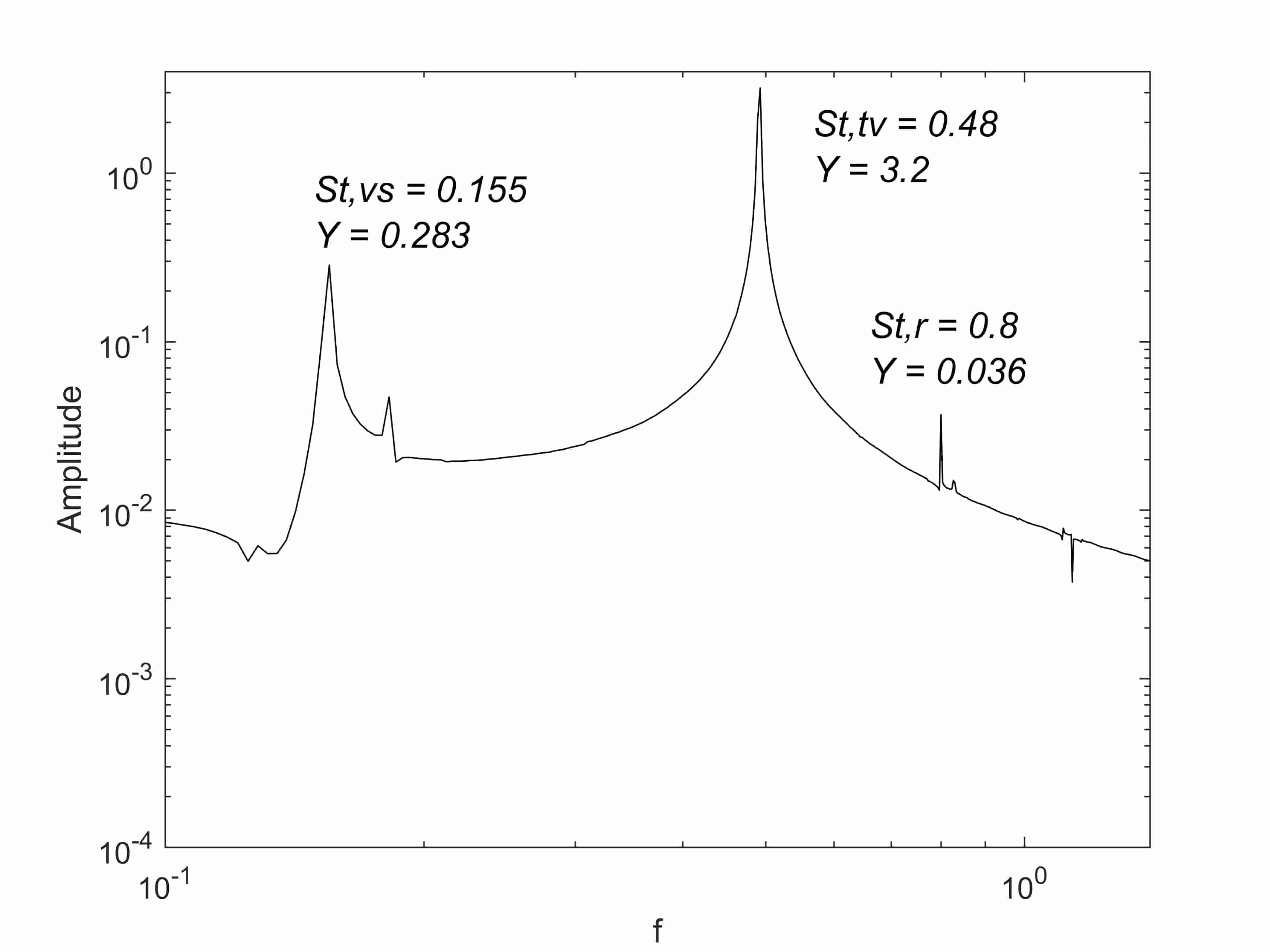} 
        \caption{$f_{tr} = 3.0$,$St_{r} = 0.8$}
        \label{fig:_4,2_}
    \end{subfigure}
    \caption{Frequency Characteristics of Lift at various transverse oscillation frequencies (Mode 4)}
    \end{figure}
    
     \begin{figure}[hp]
        \begin{subfigure}[b]{6.2cm}
        \includegraphics[scale=0.04]{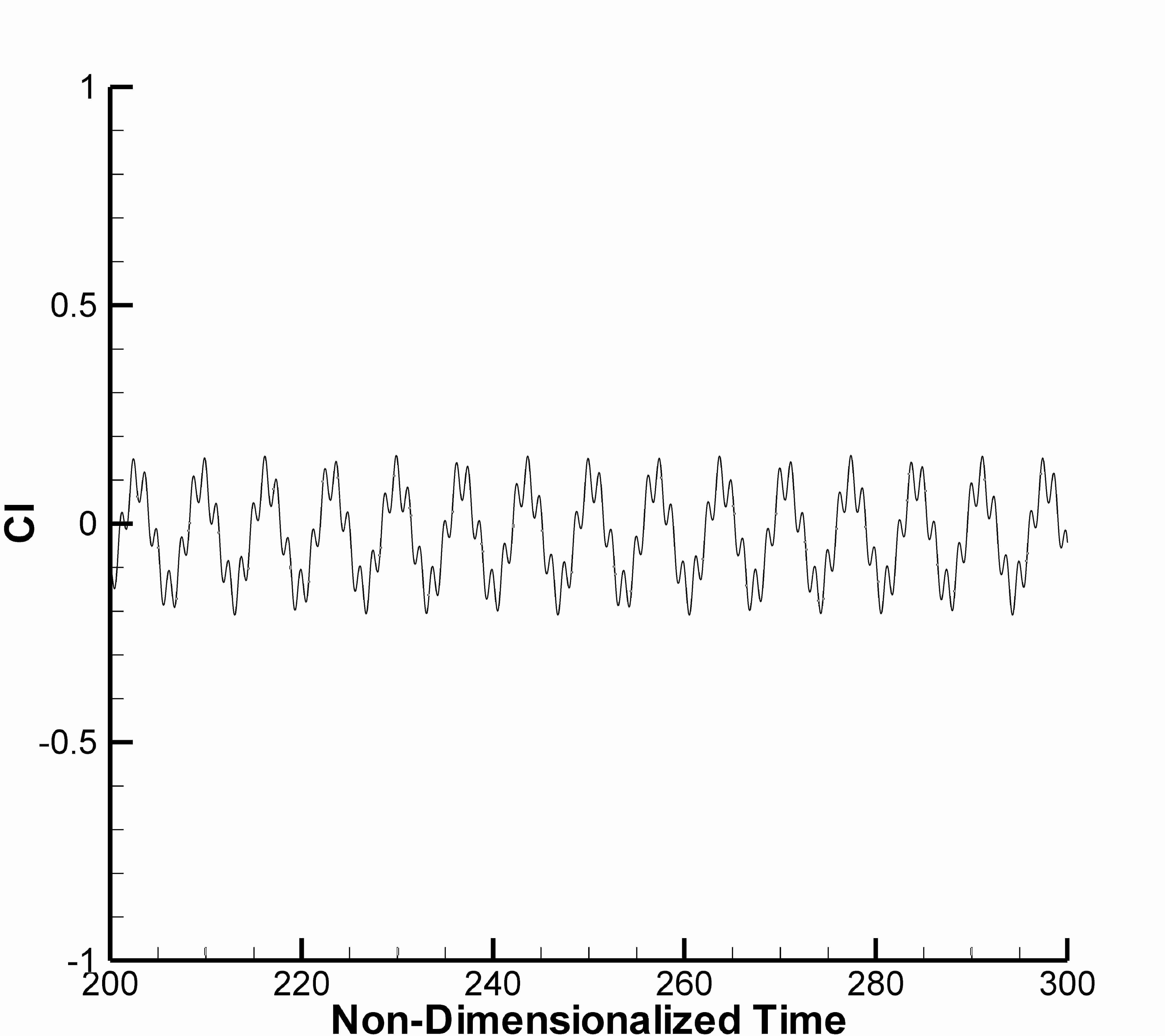} 
        \caption{$f_{tr}=0.9$,$St_{r}=0.8$}
        \label{fig:_4,3_Cl}
        \end{subfigure}
        \hfill
    \begin{subfigure}[b]{6.2cm}
        \includegraphics[scale=0.04]{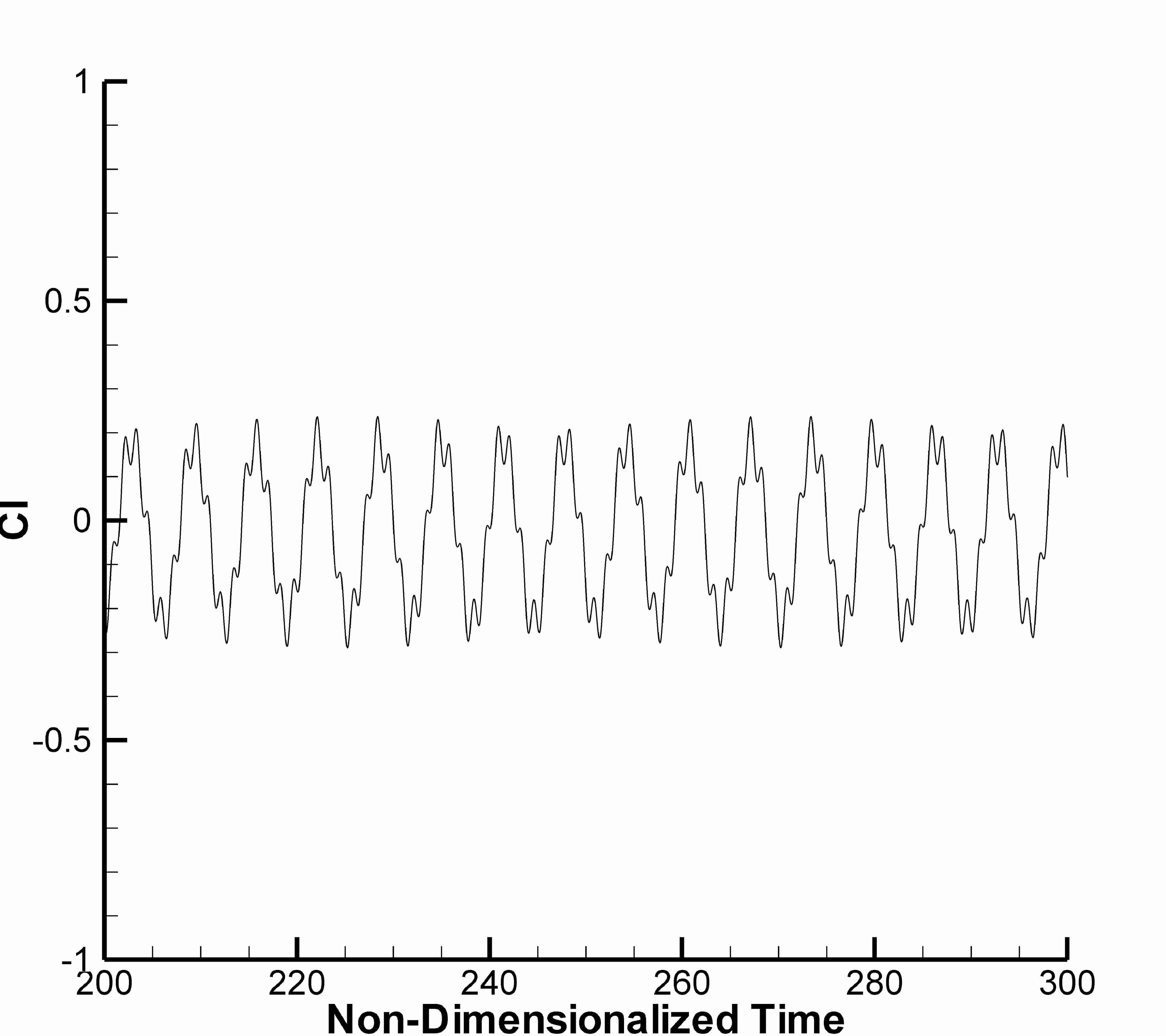} 
        \caption{$f_{tr}=0.95$,$St_{r}=0.8$}
        \label{fig:_4,4_Cl}
    \end{subfigure}
        \hfill
    \begin{subfigure}[b]{6.2cm}
        \includegraphics[scale=0.04]{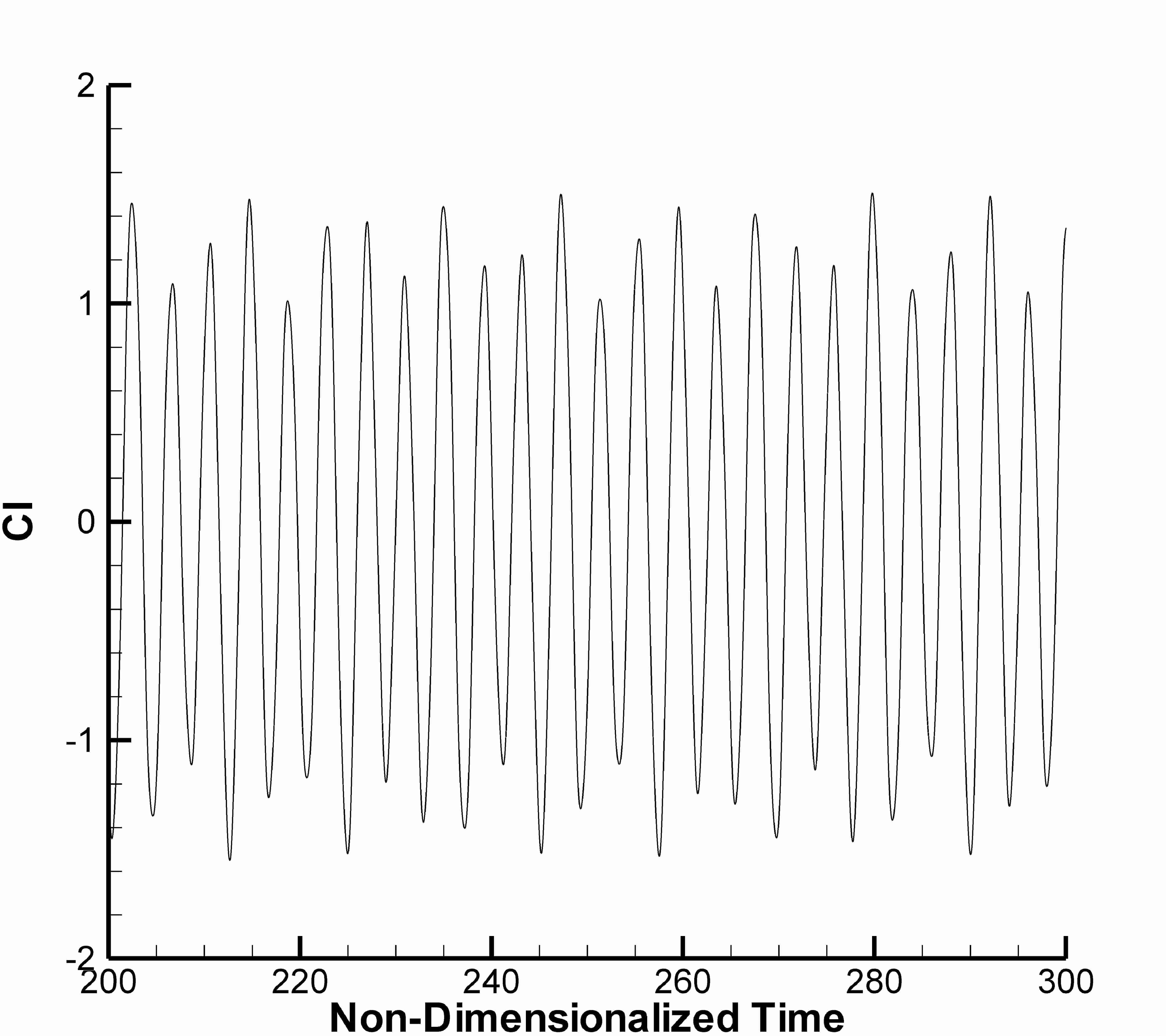} 
        \caption{$f_{tr} = 1.5$,$St_{r} = 0.8$}
        \label{fig:_4,5_Cl}
    \end{subfigure}
    \hfill
    \begin{subfigure}[b]{6.2cm}
        \includegraphics[scale=0.04]{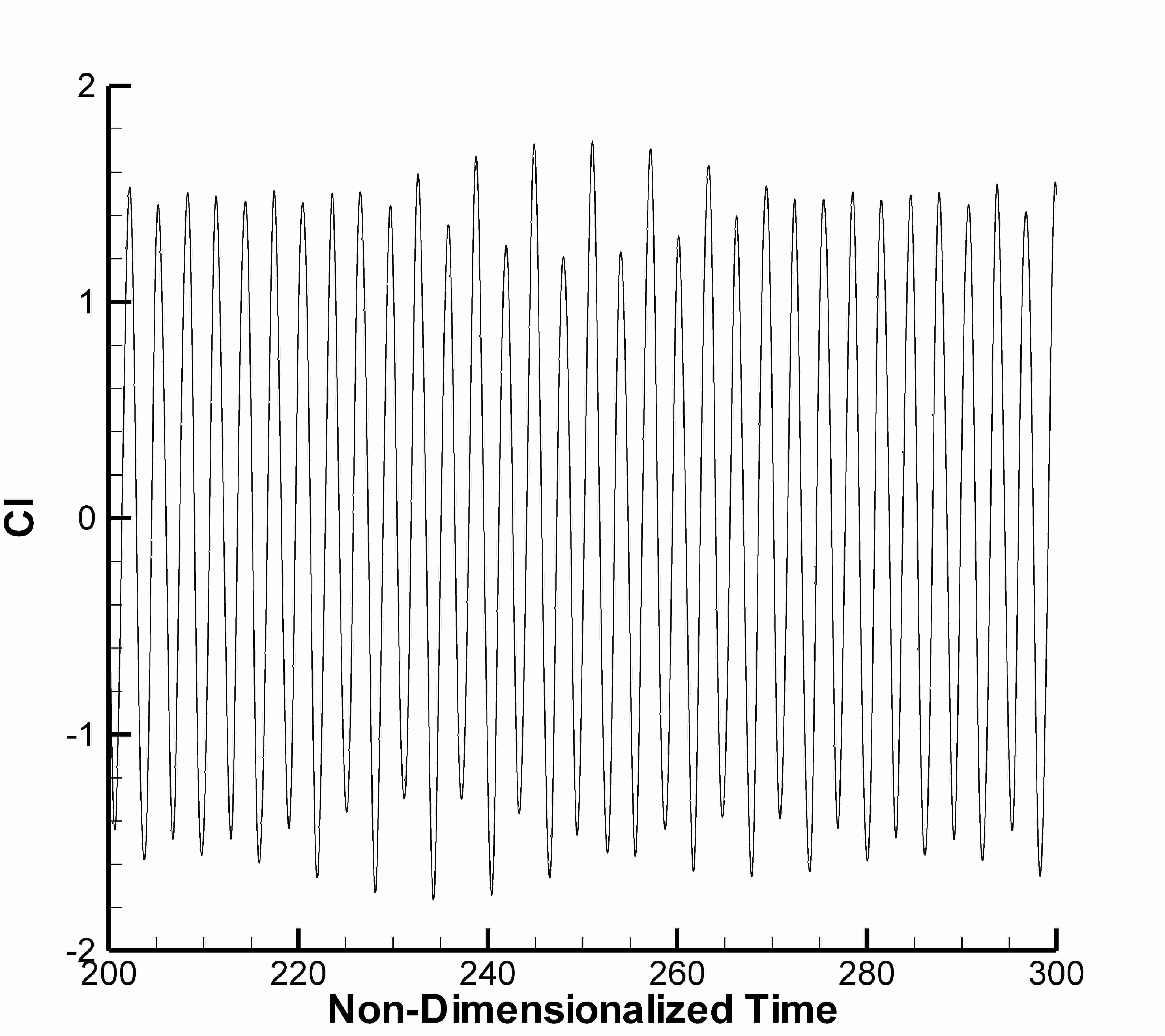} 
        \caption{$f_{tr} = 2.0$,$St_{r} = 0.8$}
        \label{fig:_4,1_Cl}
    \end{subfigure}
    \hfill
    \begin{subfigure}[b]{8cm}
        \centering
        \includegraphics[scale=0.04]{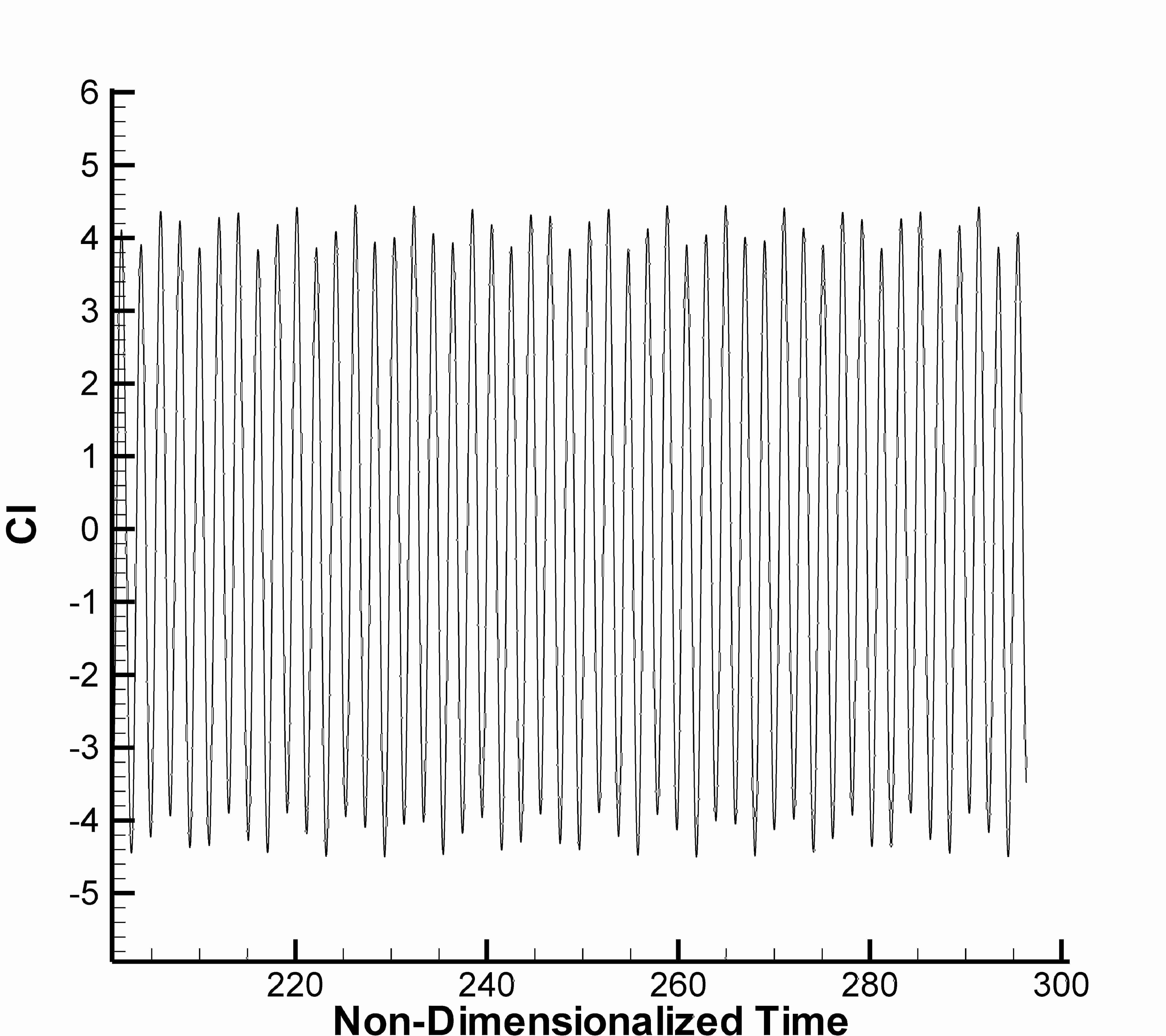} 
        \caption{$f_{tr} = 3.0$,$St_{r} = 0.8$}
        \label{fig:_4,2_Cl}
    \end{subfigure}
    \caption{Lift Coefficient at various transverse oscillation frequencies (Mode 4)}
    \end{figure}

\clearpage
\subsection{Variation of aerodynamic coefficients due to transverse oscillations for various vortex shedding modes}
~\\
For vortex shedding frequencies belonging to mode 2 and 3, similar variation of $\overline{c}_{d}$ and $c_{l,rms}$ was observed. From Fig. \ref{fig:Mode2} and \ref{fig:Mode3}, upon reaching transverse frequency ratios near $f_{tv} = 1.0$, $\overline{c}_{d}$ increases and reaches a maximum as also observed in works of Singh et al \cite{Singh} and Pham et al \cite{Pham}. For frequencies beyond this synchronization region, $\overline{c}_{d}$ reduces and increases monotonously for higher transverse oscillation frequencies past near $f_{tv} = 1.25$. For Mode 4, from Fig. \ref{fig:Mode4}, the increase in $\overline{c}_{d}$ was observed near $f_{tv} = 1.5$ onwards. This phenomenon was distinct in the latter due to the dominating effect of forced transverse oscillations over rotational oscillations, resulting from the lower rotational oscillation amplitude taken in this mode when compared with all other modes. $c_{l,rms}$ monotonously increases for greater values of $f_{tv}$ for modes 2,3 and 4. For vortex shedding mode 4, the same trend following the previous vortex shedding modes 2 and 3 was observed, with descent in $\overline{c}_{d}$ followed by a monotonous increase for frequencies beyond the region. For mode 1, from Fig. \ref{fig:Mode1} contrary to the all other modes, $\overline{c}_{d}$ and $c_{l,rms}$ decrease in the synchronization region near $f_{tv} = 1.0$ followed further by a monotonous increase. 

\section{Conclusion}
The developed ALE-CFR scheme on validation yielded good agreement for independent transverse and rotational oscillations imposed on the cylinder. The scheme was then employed to solve the complex flow past a cylinder oscillating with both such modes and interesting phenomena were also observed. The vortex shedding modes were found to retain their distinguishable characteristics in the near wake, even after forced transverse oscillations were imposed. The vertical length of the vortex shedding area was observed to increase as a result of transverse oscillations. Vortex shedding mode 3 appeared to become more prominent as the transverse oscillation frequency was increased, due to a greater phase lag between the near and far wake given in Singh et al \cite{Singh}.
 Similar to the phenomenon reported by Cheng et al \cite{Cheng}, the lock-in occurred when $St_{r} \approx St_{n}$ as seen for mode 1, and $St_{tv} \approx St_{n}$ when transverse frequency ratios $St_{r}=0.9$ and $0.95$ for all the modes were taken. 
The peaks in the frequency characteristics for lift coefficient show that the dominant frequencies in the non lock-in regions correspond to the transverse, rotary oscillation and the vortex shedding frequencies, with the first two dominating.
The majority of lift contribution was accounted for by the transverse oscillations as we go to higher frequencies, even when $A_{tv}$ taken was $10\%$ of that of $A_{r}$. In the lock-in regions however, the lift was found to remain largely affected by the oscillation frequency chosen close to the natural vortex shedding frequency. For modes 2, 3 and 4, $\overline{c}_{d}$ was found to increase near $f_{tv}=1.0$ and then further reduce and continued to remain moderately variable for higher transverse oscillation frequencies. For mode 1, $c_{l,rms}$ and $\overline{c}_{d}$ decreased in the frequency region near $f_{tv}=1.0$ and then continued to increase for higher transverse oscillation frequencies. The reason for such a variation remains a topic to be explored in future studies.     

\afterpage{
\begin{figure}[h]
    \begin{subfigure}[b]{6.2cm}
    \centering
    \includegraphics[scale=0.04]{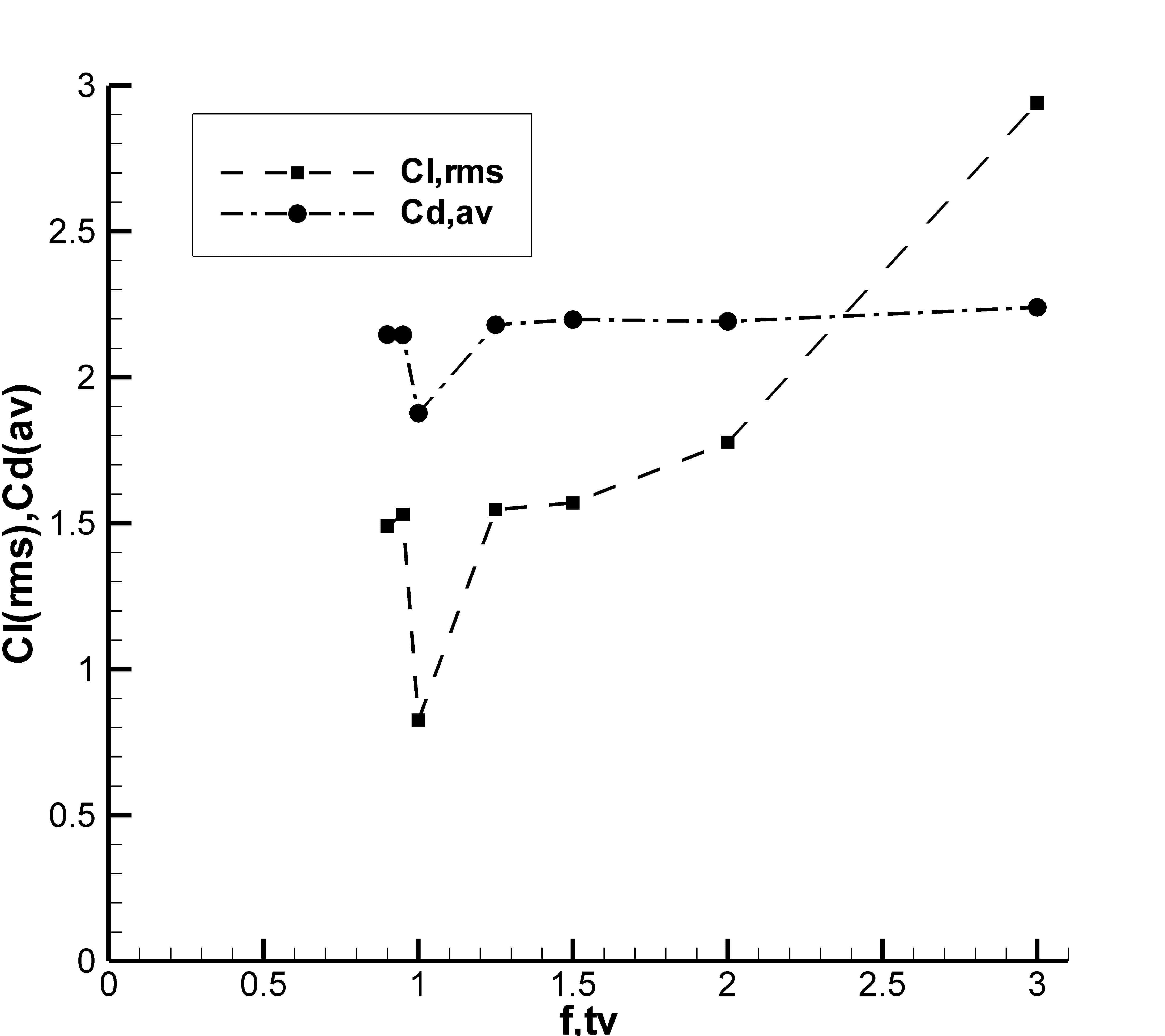}
    \caption{Mode1}
    \label{fig:Mode1}
    \end{subfigure}
    \hfill
    \begin{subfigure}[b]{6.2cm}
    \centering
    \includegraphics[scale=0.04]{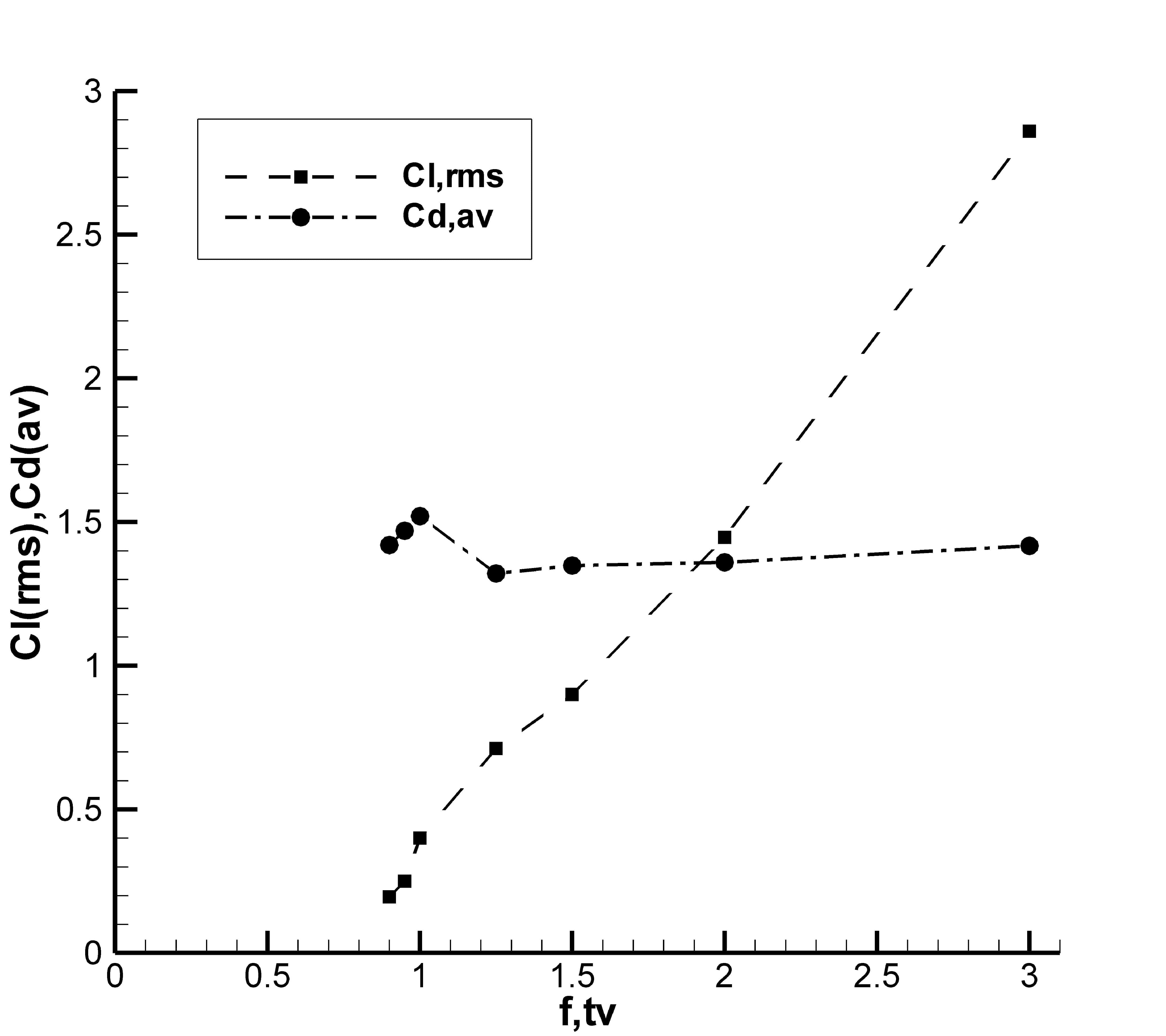}
    \caption{Mode2}
    \label{fig:Mode2}
    \end{subfigure}
    \hfill
    \begin{subfigure}[b]{6.2cm}
    \centering
    \includegraphics[scale=0.04]{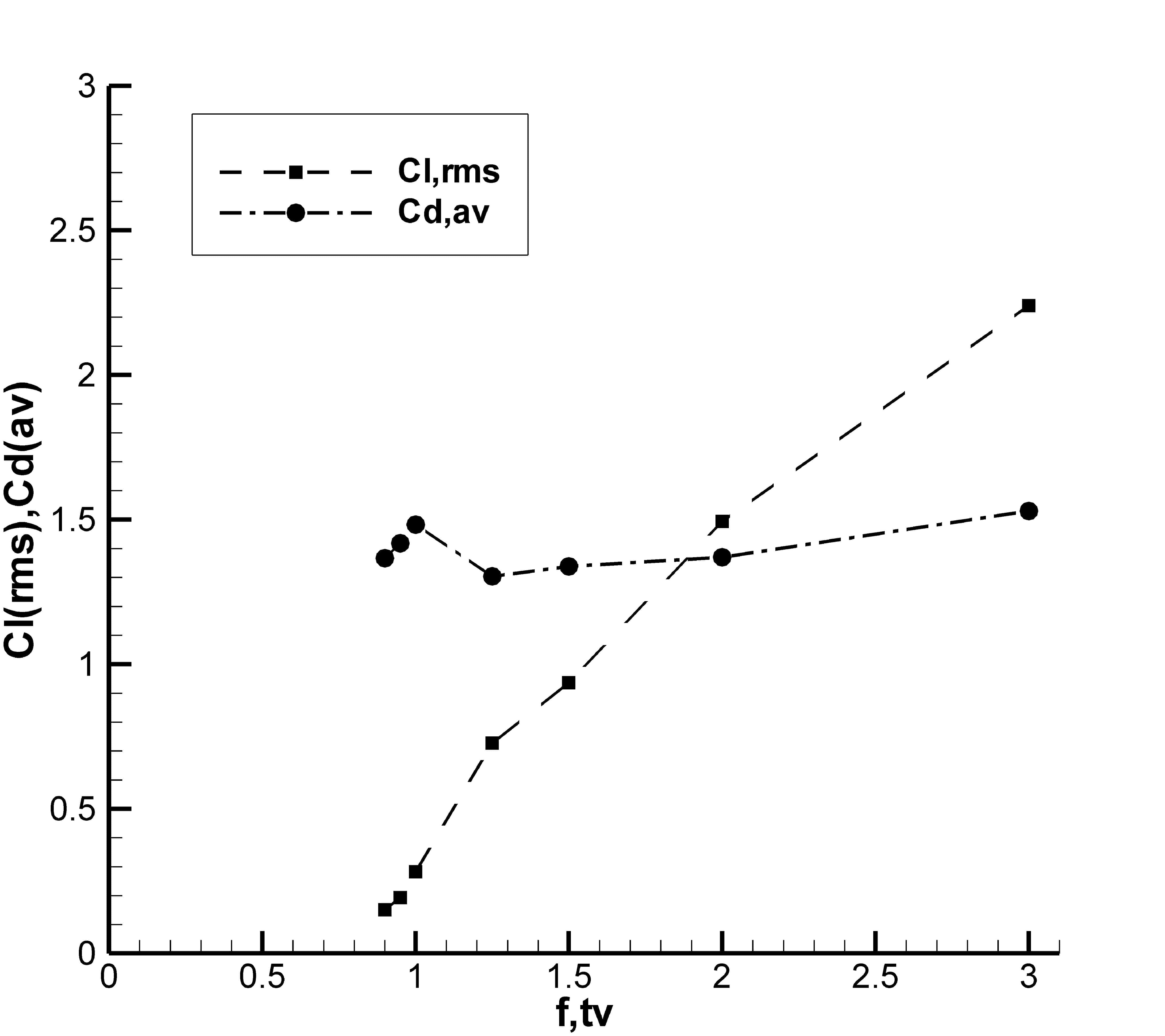}
    \caption{Mode3}
    \label{fig:Mode3}
    \end{subfigure}
    \hfill
    \begin{subfigure}[b]{6.2cm}
    \centering
    \includegraphics[scale=0.04]{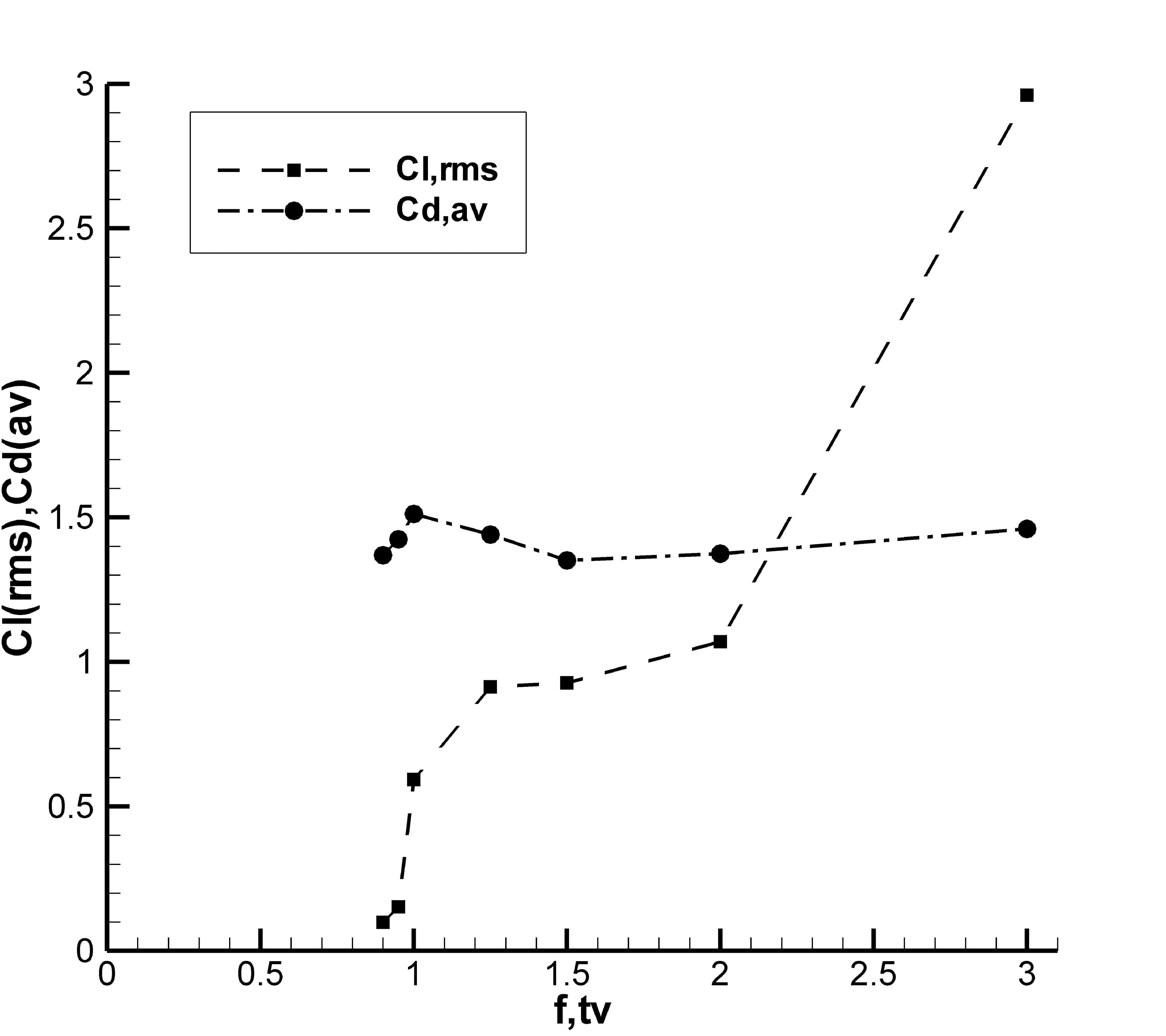}
    \caption{Mode4}
    \label{fig:Mode4}
    \end{subfigure}
    \caption{Variation of aerodynamic coefficients for different transverse oscillation frequencies}
\end{figure}
}

\afterpage{

}

\end{document}